\documentclass[preprint,10pt,3p]{elsarticle}

\usepackage{graphicx}
\usepackage{amssymb}
\usepackage{paralist}
\usepackage{amsmath}
\usepackage{float}
\usepackage{multirow}
\usepackage[table]{xcolor}
\usepackage{tikz}
\usepackage{csquotes}
\biboptions{sort&compress}
\usepackage{rotating}

\usepackage{url}
\usepackage{doi}
\usepackage{color}
\usepackage{graphicx}
\usepackage{subcaption}
\usepackage{bbm}
\usepackage{nomencl}
\makenomenclature

\allowdisplaybreaks

\hypersetup{
	colorlinks   = true, 
	urlcolor     = blue, 
	linkcolor    = blue, 
	citecolor    = blue 
}

\journal{Transportation Research Part C: Emerging Technologies}
\begin{document}
\begin{frontmatter}


\title{Developing 3D Virtual Safety Risk Terrain for UAS Operations in Complex Urban Environments}



\author{Zhenyu Gao\footnote{Postdoctoral Fellow, Department of Aerospace Engineering and Engineering Mechanics, The University of Texas at Austin}, John-Paul Clarke\footnote{Professor and Ernest Cockrell, Jr. Memorial Chair in Engineering, The University of Texas at Austin}, Javid Mardanov\footnote{Ph.D. Student, School of Aeronautics and Astronautics, Purdue University}, and Karen Marais\footnote{Professor and Associate Head, School of Aeronautics and Astronautics, Purdue University}}

\address{The University of Texas at Austin, Austin, TX 78712, United States}

\address{Purdue University, West Lafayette, IN 47907, United States}

\begin{abstract}

Unmanned Aerial Systems (UAS), an integral part of the Advanced Air Mobility (AAM) vision, are capable of performing a wide spectrum of tasks in urban environments. The societal integration of UAS is a pivotal challenge, as these systems must operate harmoniously within the constraints imposed by regulations and societal concerns. In complex urban environments, UAS safety has been a perennial obstacle to their large-scale deployment. To mitigate UAS safety risk and facilitate risk-aware UAS operations planning, we propose a novel concept called \textit{3D virtual risk terrain}. This concept converts public risk constraints in an urban environment into 3D exclusion zones that UAS operations should avoid to adequately reduce risk to Entities of Value (EoV). To implement the 3D virtual risk terrain, we develop a conditional probability framework that comprehensively integrates most existing basic models for UAS ground risk. To demonstrate the concept, we build risk terrains on a Chicago downtown model and observe their characteristics under different conditions. We believe that the 3D virtual risk terrain has the potential to become a new routine tool for risk-aware UAS operations planning, urban airspace management, and policy development. The same idea can also be extended to other forms of societal impacts, such as noise, privacy, and perceived risk. 

\end{abstract}


\end{frontmatter}



\section{Introduction}\label{sec:intro}


Advanced Air Mobility (AAM) is a novel air transport concept that integrates multiple transformational technologies, such as electric aircraft, small drones, and automated air traffic management, into the existing transportation and service system. As envisioned by National Aeronautics and Space Administration (NASA), AAM will enable the movement of people and cargo more effectively, especially in currently underserved local and regional settings\footnote{Advanced Air Mobility Mission Overview, NASA, \url{https://www.nasa.gov/aam/overview/}, Last Updated: Jun 23, 2022}. The main AAM vehicle concepts include Electric Vertical Take-Off \& Landing (eVTOL) aircraft, Electric Conventional Take-Off \& Landing (eCTOL) aircraft, and Unmanned Aerial Systems (UAS). The former two concepts are indispensable for Urban Air Mobility (UAM)~\citep{garrow2021urban}, a subset of AAM that focuses on sustainable air mobility technologies that will operate and transport passengers or cargo at lower altitudes in urban environments. The UAS is highly capable in a diverse set of tasks such as goods delivery, emergency services, public safety, infrastructure inspection, agriculture monitoring, meteorological research, and aerial photography and videography. By expanding the current services into a new dimension -- the sky, AAM is likely to become an integral part of the future urban infrastructure system. Nevertheless, the integration of AAM into the existing urban environment is still a challenging task~\citep{bauranov2021designing}. The societal impacts of the system are among the key challenges that are decisive in the existence and large-scale deployment of AAM. A sustainable AAM system must operate harmoniously within the constraints imposed by public concerns such as noise pollution~\citep{bian2021assessment,gao2023noise}, emissions~\citep{gao2022multi}, safety~\citep{wei2023risk,lin2020failure}, privacy~\citep{ding2022routing}, and equity~\citep{bennaceur2022passenger,chin2023protocol}. Consequently, the societal/community integration of aerospace systems has become a focal research topic in recent years~\citep{vascik2018analysis,gao2022probabilistic,gao2022statistics,yunus2023efficient,nassi2021sok}. 

UAS safety has been a perennial challenge, and one of the principal barriers to the large-scale deployment of UASs, especially in complex urban environments. Relevant literature from the communities of UAV design, controls and robotics, reliability engineering, and transportation can be classified into two broad categories: models for evaluating UAS safety risks, and protocols for mitigating UAS safety risks. The former involves several specific risk models such as failure models, recovery models, impact location models, stress models, exposure models, and harm/damage models~\citep{washington2017review}. The latter comprises aspects such as collision avoidance algorithms, regulations and policies, and risk-aware operations planning. When considering UAS (and other forms of AAM) operations in complex urban environments, airspace management is the kernel of operations planning and policy making. An urban airspace defines volumes in the 3D space where an UAS is allowed to operate. The `no-fly' zones are areas where flying is prohibited due to urban topographies such as buildings, and societal impacts such as noise, privacy, and safety. On the flight safety side, although the UAS community has made concrete progress towards risk-aware trajectory planning in recent years~\citep{lin2020failure,pang2022uav,primatesta2020ground}, the existing works either (1) only consider trajectory planning in the 2D domain and not in a 3D urban environment, or (2) require a repetitive process to identify a flight path while minimizing Third Party Risk (TPR). In addition, to the best of our knowledge, no published work has investigated how TPR or ground risk could affect airspace management in complex urban environments. In this work, we propose the novel idea of 3D virtual risk terrain for UAS operations planning in complex urban environments. The core idea is to convert risk constraints in an urban environment into 3D exclusion zones that UAS operations should avoid to adequately reduce
risk to the Entities of Value (EoV) in the urban space. In our view, the 3D virtual risk terrain has three advantages over other existing methods for risk-aware UAS operations planning:
\begin{enumerate}
    \item \textit{It enables efficient UAS trajectory planning.} A combination of the 3D virtual risk terrain and the physical urban terrain defines an overall acceptable fly zone for an UAS to operate. This turns the original 3D trajectory optimization problem into a much more straightforward terrain avoidance problem, which can be solved by a non-repetitive trajectory generation process. 
    \item \textit{It can be extended to other societal impacts.} The same concept can be applied to create `no-fly' zones for other public acceptance factors such as community noise and privacy. The union of some or all of these `no-fly' zones will generate an overall acceptable fly zone, where UAS operations can circumvent/limit multiple or all societal impacts of the system.
    \item \textit{It can serve as a guideline to policy makers.} By specifying the spatial and temporal variation of minimum clearance distance to be maintained from people and properties, this new tool can play a significant role in the development of safety regulations for UAS operations. It can also be used by aviation airworthiness authorities as a basis for determining the reliability and equipment requirements for the UAS to operate within a certain space in an urban environment. 
\end{enumerate}

In this paper, we develop an integrated risk-based approach for generating 3D virtual risk terrain in urban environment, which facilitates risk-aware UAS trajectory planning. Specifically, this work bridges two categories of relevant literature. It combines most types of UAS safety risk sub-models to generate protocols for mitigating UAS safety risks. The integrated approach considers a total of seven sub-models from four areas: systems failure, third-party information, urban topography, and safety requirements to achieve the overarching objective. In addition, the framework is flexible in accommodating different probabilistic models, accounting for uncertainty, and capturing temporal dependencies in third-party exposure. Overall, we summarize our four primary contributions as follows:
\begin{enumerate}
    \item \textit{Proposing the novel concept of 3D virtual risk terrain for UAS operations.} This concept translates TPR or ground risk considerations into acceptable fly zones and provides a new angle for risk-aware UAS trajectory planning and urban airspace management. Based on our observations, a similar concept has not appeared in the literature before. 
    \item \textit{Developing a holistic computational framework to generate 3D virtual risk terrain.} A review paper~\citep{washington2017review} summarized a total of seven types of sub-models for UAS ground risk. Whilst most relevant works in the literature have a limited coverage, this framework integrates all seven sub-models (in a modified organization). A mathematical framework based on conditional probability connects the sub-models together.
    \item \textit{Conducting numerical examples and generating prototypes of the proposed concept.} Using a Chicago downtown area as the background, we generate prototypes of the 3D virtual risk terrains under different conditions. The results are presented using both data visualization and quantitative measures. This first set of prototypes serves to provide insights to the key patterns of the virtual risk terrain.
    \item \textit{Interacting with 3D virtual terrains from other societal impacts.} In a project sponsored by NASA, a companion work \citep{gao2023noise} has developed 3D virtual acoustic terrain for aerial vehicle trajectory planning with limited noise impacts. This work also presents results that combine the two virtual terrains for the same urban model and vehicle type.
\end{enumerate}

The remainder of the paper is organized as follows. Section~\ref{sec:background} reviews literature in two relevant streams and identifies the research gap. Section~\ref{sec:models} introduces the proposed overall approach and details of the sub-models. Section~\ref{sec:case} applies the proposed approach to a real-world 3D urban model to generate prototypes of the proposed concept. Section~\ref{sec:remarks} discusses the limitations and extensions of the study before Section~\ref{sec:conclusions} concludes the paper.

\section{Background}\label{sec:background}


\subsection{Modeling of UAS Safety Risk}

A holistic understanding of the risks UAS operations pose to people and property in urban and suburban environments is key to the development of UAS safety and airworthiness regulations. Airworthiness authorities, such as the Federal Aviation Administration (FAA)~\citep{faa2019uas} and the European Aviation Safety Authority (EASA)~\citep{european2015concept} have promoted the adoption of a risk-based approach to develop safety regulatory frameworks for UASs. Risk assessment typically consists of three steps: risk identification, risk analysis, and risk evaluation~\citep{iso2018risk}. Of the two primary UAS risk sources, collision risk and ground risk, this work focuses on the ground risk of UAS: the system's risk to people or structures on the ground due to system failure during operation. A comprehensive survey paper~\citep{washington2017review} identified seven basic models in the assessment of UAS ground risk: failure model, impact location model, recovery model, stress model, exposure model, incident stress model, and harm model. Detailed definitions and recent developments for these sub-models are provided in Section~\ref{sec:models}. Each of these seven models is a dedicated research area in its own right. For example, the impact location model is at the intersection of flight dynamics and probabilistic modeling; the exposure model is enabled by Geographic Information System (GIS) data; and the harm model is a branch of solid mechanics and biomechanics. Therefore, a comprehensive assessment of UAS safety risk is interdisciplinary in nature and should integrate the latest research outcomes from a variety of specialized research fields. 

The complexity in UAS safety risk modeling presents two dimensions of challenge/difficulty to this unique problem. The first dimension of challenge lies in the diversity of UASs. Compared to Conventional Piloted Aircraft (CPA), UASs consist of a more heterogeneous set of aerial vehicles (with different sizes, types, configurations, etc.), can operate under a greater variety of conditions (e.g., in complex urban environments), carries more state-of-the-art technologies (robotics, computer vision, etc), and is more susceptible to environmental conditions (wind, weather, local climate, etc). It would be problematic to apply a unified set of regulatory rules on the entire system. Extensive studies are required to investigate the safety requirements for a diverse set of representative scenarios, such that tailored and more effective decisions can be made. Also due to the diversity of UASs, every existing study in its risk modeling is a result of specific relevant underlying assumptions (on the aircraft properties, operating conditions, environmental conditions, etc). Consequently, each model in the literature has a fairly limited range of applicability, and that critical attention must be taken when generalizing or transferring such results. The second dimension of challenge resides in the high levels of uncertainty in most sub-models. Treatment of uncertainty is highly critical when estimating the impact location of an Unmanned Aerial Vehicle (UAV) crash and predicting the level of harm/damage a crash can bring to people/property. Moderate uncertainty also exists in other sub-models such as the failure model and the stress model. When assessing UAS safety risk in a complex urban environment, such assessment must be conducted at high granularity such that the spatio-temporal aspect of uncertainty is considered. Overall, both diversity and uncertainty have a significant impact on risk modeling and the resulting regulations. Therefore, considering the rate of development in UAS operations, accurate modeling of UAS safety risk will be an active research area for many years to come.

\subsection{Mitigation of UAS Safety Risk}


Mitigation of safety risk is among the main objectives of UAS system design. It can be achieved through improvement in either or both engineering design and operations management. Engineering design includes both hardware and software of the aerial vehicle (and other infrastructure in the system). The present work emphasizes on the operations management aspect, i.e., the planning and optimization of UAS operations to reduce the system's TPR. The operations management workflow is analogous to that of prediction-driven optimization, where the prediction comes as a result of the modeling of UAS safety risk described in the last subsection. In most cases, risk-aware UAS operations planning is based on risk maps -- a `heat map' that characterizes the spatial distribution of safety risk in an urban or suburban area. The operations planning problem then becomes a trajectory planning problem which either completely avoid high risk areas that are above certain thresholds or conduct trade-off to minimize the total risk an operation poses to people and property in the area. 

\begin{table}[h!]
\centering
\caption{A summary of literature on the use of risk map for operations planning}
\begin{tabular}{l|l|l}
\hline
\textbf{Paper}                                    & \textbf{Dimensions} & \textbf{Basic Models Used}                              \\ \hline
\cite{lum2011assessing}          & \multirow{9}{*}{2D}  & Failure, Impact Location, Exposure               \\ 
\cite{bertrand2017ground}        &                   & Failure, Impact Location, Stress, Harm, Exposure \\ 
\cite{cour-harbo2017quantifying} &                   & Failure, Impact Location, Stress, Harm, Exposure \\ 
\cite{levasseur2019accurate}     &                   & Impact Location                                  \\ 
\cite{lin2020failure}            &                   & Failure, Impact location, Stress, Harm, Exposure \\ 
\cite{hu2020risk}                &                   & Failure, Stress, Harm, Exposure                  \\ 
\cite{primatesta2020ground}      &                   & Failure, Impact Location, Stress, Harm, Exposure \\ 
\cite{kim2022risk}               &                   & Failure, Impact location, Stress, Harm, Exposure \\ 
\cite{berard2022risk}            &                   & Failure, Impact location, Harm                   \\  \hline
\cite{pang2022uav}               & \multirow{2}{*}{3D}                  & Failure, Stress, Harm, Exposure   \\ 
\cite{zhang2023uav}              &                   & Failure, Exposure                                \\ \hline
\end{tabular}
\label{tbl:riskmap}
\end{table}

In most research efforts in the literature, the creation of a risk map for UAS operations planning involves the integration of multiple basic models for UAS safety risk. Table~\ref{tbl:riskmap} is a list of 12 of the most relevant papers in the literature which build risk maps for UAS operations, in chronological order. These works utilized between 2 and 5 basic models to construct risk maps for different areas and/or purposes. We can observe that the failure model, impact location model, and exposure model are the most frequently chosen building blocks, whilst none of these works included the recovery model in their approach. Another important aspect to consider is the number of dimensions in the risk map. The majority of works in Table~\ref{tbl:riskmap} build 2D risk maps, therefore the corresponding trajectory planning problems are in the 2D domain. Two most recent works~\citep{pang2022uav,zhang2023uav} have started to construct 3D risk maps that consider 3D UAS trajectory planning problems. A risk map can enable efficient ways for flight trajectory planning and optimization. Optimization methods such as the Dijkstra algorithm, A$^*$, Ant Colony Optimization (ACO) and their variants can be applied to generate trajectories with limited or minimal safety risk.

\subsection{Research Gap}

The existing studies in the modeling and mitigation of UAS safety risk have made considerable contributions to the advancement of UAS and AAM. In an attempt to further the state-of-the-art UAS operations planning practices, we identify a research gap at the intersection of three limitations in the current literature. The first limitation is that very few existing approaches have comprehensively integrated the basic models of UAS safety risks. Although some works in Table~\ref{tbl:riskmap} have considered as many as five basic models in their risk maps, several basic models include over-simplified assumptions, such as uniform distribution in impact location and certainty (probability = 1) in serious injury after an impact. The lack of basic models and the use of oversimplified assumptions can have a significant impact on a model's precision. The second notable limitation is that, most relevant works in the literature can only create 2D risk maps for UAS operations planning. In a complex urban environment, however, UAS trajectory planning must be informed by a high-fidelity 3D risk map. To date, 3D risk mapping remains as a widely unexplored research area. Third, and most important to our research, none of the existing works can provide explicit guidance to risk-aware urban airspace management. Converting risk constraints into space constraints will have many benefits, such as insights for regulatory planning and consistency with other societal impact constraints. We use a concept called 3D ``virtual terrain'' to construct 3D `no-fly' zones for better airspace management. Among the existing literature, the closest concept to the 3D virtual terrain is the one described in \citep{zhang2023uav}, where the authors developed a 3D collision risk heatmap for an airport terminal area. In contrast, our work focuses on the TPR in complex urban environments and builds a risk map of a different nature. To address this research gap, we propose an integrated risk-based approach for generating 3D virtual risk terrain for UAS operations planning in complex urban environments. This approach includes all basic models of UAS safety risks, develops risk terrain in the 3D domain, and offers explicit guidance on urban airspace management. 


\section{The Proposed Approach}\label{sec:models}

\subsection{Overview}

\begin{figure}[h!]
	\centering
        \includegraphics[width=0.99\textwidth]{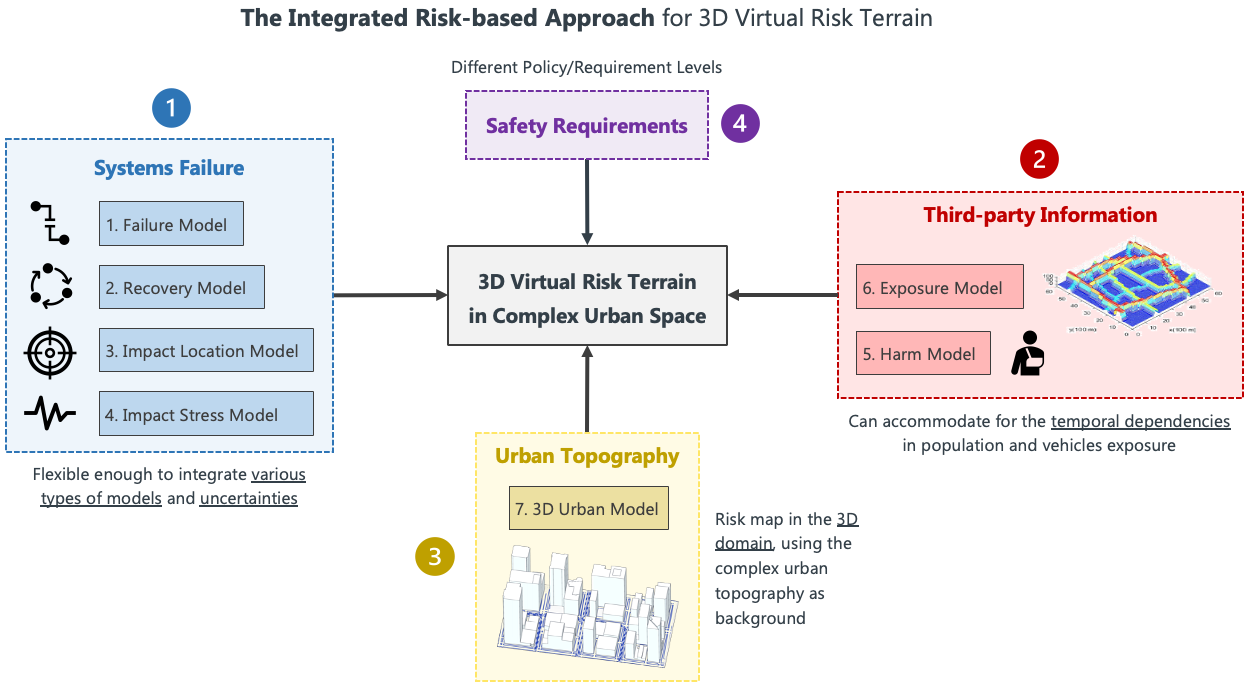}
	\caption{Flowchart of the overall integrated risk-based approach for generating 3D virtual risk terrain.}
	\label{fig:Approach}
\end{figure}

In section we introduce details of the proposed risk-based approach for formulating 3D virtual risk terrain for airspace management and UAS trajectory planning. Figure~\ref{fig:Approach} displays the main modules and sub-models in our approach. The overall approach consists of four primary modules and seven sub-models. The systems failure module includes the failure model, recovery model, impact location model, and impact stress model to characterize the ground risk of UAS operations. The analytical frameworks in the systems failure module are flexible to accommodate the variabilities and uncertainties in the domain. The third-party information module focuses on the TPR posed to people and properties on the ground. It considers the harm model and the exposure model which can accommodate for the temporal dependencies in population and vehicle exposure in an urban environment. This provides the flexibility necessary to investigate the safety terrains and requirements at different times of a day and at different locations in a city. The urban topography module employs a 3D urban model as the physical background of the virtual risk terrain. Eventually, the physical and virtual terrains will collectively determine a `no-fly' zone in the urban airspace for UAS trajectory planning. Lastly, the safety requirements will determine the position of a 3D virtual surface, above which the UAS operations can maintain a satisfactory risk level. The final risk terrain result provides inputs to the airworthiness requirements of the system, i.e., given a safety requirement (risk standard), what failure and recovery reliability levels (in sub-models 1 and 2) are required for the UAS to operate at a certain altitude above the ground in complex urban environments.  

In this risk-based framework, the first five sub-models are conditional probability models; the exposure model can be obtained through analyzing mobility and traffic data in the city; the 3D urban model can be obtained from public sources such as OpenStreetMap~\citep{hakley2008openstreet}, NASA, and the United States Geological Survey (USGS). Because a probability model is defined by sample space, events, and probabilities associated with each event, we first formally define the sample spaces that are involved in this flight risk analysis. 
\begin{itemize}
    \item \textbf{Aircraft type}: $\mathcal{A} = \{A_1, ..., A_n\}$. The type of aircraft has a discrete sample space. At macro level, there are several general categories of UAV, such as fixed-wing, multi-rotor, and helicopter. Within each category, there are UAVs with various detailed configurations, weights, parameters, etc. Each aircraft type in $\mathcal{A}$ can have different features and behaviors related to safety analysis.
    \item \textbf{Operating conditions}: $\mathcal{O} = \{O_1, ..., O_n\}$. Typical operating conditions of a UAV include takeoff, landing, hovering, and level flight at various speeds. These conditions compose a discrete sample space. 
    \item \textbf{Environmental conditions}: $\mathcal{E} = \{E_1, ..., E_n\}$. The environmental conditions can broadly include wind, temperature, other weather conditions, and urban topography. Therefore, the sample space of $\mathcal{E}$ is inherently multivariate and continuous. However, in reliability and safety related analysis, it is more realistic to discretize $\mathcal{E}$ into a finite number of representative scenarios. 
    \item \textbf{Failure mode}: $\mathcal{F} = \{F_1, ..., F_n\}$. This is also a discrete sample space. Some general types of failure mode include Loss of Control (LOC), Unpremeditated Descent Scenario (UDS), Controlled Flight into Terrain (CFIT), and Dropped or Jettisoned Components (DOJC). Each general failure type can be divided into multiple detailed scenarios. For example, LOC includes partial and complete losses; DOJC can happen on different types and/or numbers of components on the aerial vehicle. 
    \item \textbf{Recovery outcome}: $\mathcal{R} = \{R_1, R_2\}$. The recovery model in the framework is takes into account state-of-the-art aerial vehicle technologies in robotics, control, and detection that can help a vehicle avoid catastrophic consequences when certain failure types occur. In this work, the recovery outcome indicates whether the aerial vehicle can recover (by its own system or with the help of a remote pilot) and land safely. $R_1$ refers to successful recovery; $R_2$ refers to unsuccessful recovery, which means that the aerial vehicle will enter into a crash trajectory.
    \item \textbf{Contingency capabilities}: $\mathcal{C} = \{C_1, ..., C_n\}$. This sample space indicates what contingency capacities are onboard the aircraft. Typical examples include parachute, air bag, and emergency landing functions. An event $C_i \in \mathcal{C}$ indicates the availability of each option. For example, given a total of six possible contingency capabilities, a vector $C_i = [1, 0, 1, 0, 0, 0]$ indicates that the first and third options are onboard. 
    \item \textbf{Kinetic energy}: $\mathcal{K}$. Kinetic energy is the most frequently used property to indicate the incident stress of a falling object. It is a continuous property, yet can be discretized to facilitate the analysis. 
    \item \textbf{Harm level}: $\mathcal{H} = \{H_1, H_2, H_3, H_4, H_5, H_6\}$. The six harm levels on people or properties are Minor, Moderate, Serious, Severe, Critical, Unsurvivable, according to the Abbreviated Injury Scale (AIS). 
    \item \textbf{Initial location}: $(x_0, y_0)$ indicates the 2-D coordinates of aircraft when the failure occurs. 
    \item \textbf{Initial altitude}: $h_0$ is the Above Ground Level (AGL) altitude of aircraft when the failure occurs. 
    \item \textbf{Failure time}: $\mathcal{T}$. Time is a crucial factor in the determination of third-party risk, as the presence and density of people and vehicles vary at different times of the day. 
\end{itemize}

For simplicity, we further denote the `current level' in each discrete sample space as $\tilde{A} = a \in \mathcal{A}$, $\tilde{O} = o \in \mathcal{O}$, $\tilde{E} = e \in \mathcal{E}$, $\tilde{F} = f \in \mathcal{F}$, $\tilde{R} = r \in \mathcal{R}$, $\tilde{C} = c \in \mathcal{C}$, $\tilde{K} = k \in \mathcal{K}$, $\tilde{H} = h \in \mathcal{H}$, and $\tilde{T} = t \in \mathcal{T}$ respectively. Now, we define the form of each sub-model and relate them to the final flight risk metric that will be utilized to construct the 3D virtual flight risk terrain in an urban space. Below are the formal conditional probability forms of the first six sub-models.
\begin{enumerate}
    \item \textbf{Failure model}: an important indicator of a UAV's reliability, the uncertainty in the occurrence of a failure mode $\tilde{F}$ is dependent mainly on the aircraft type/configuration $\tilde{A}$, the operating condition $\tilde{O}$, and the environmental condition $\tilde{E}$, given by
    \begin{equation}
    P_F \left(f | a, o, e\right) = \text{Pr} \left\{\tilde{F} = f | \tilde{A} = a, \tilde{O} = o, \tilde{E} = e\right\}
    \end{equation}
    \item \textbf{Recovery model}: the uncertainty in the ability of a UAV to recover from the failure mode and avoid catastrophic outcomes such as ballistic descent $\tilde{R}$ is dependent on the aircraft type/configuration $\tilde{A}$, the failure mode $\tilde{F}$, the contingency capability $\tilde{C}$, and the initial altitude $h_0$. Note that sub-models 3 to 6 are considered only if the recovery is unsuccessful, i.e., $\tilde{R} = R_2$. Here we further denote $\bold{p}_0 = (x_0, y_0, h_0)$ for simplicity, the recovery model is given by
    \begin{equation}\label{eqn:Rmodel}
    P_R \left(r | a, f, c, \bold{p}_0\right) = \text{Pr} \left\{\tilde{R} = r | \tilde{A} = a, \tilde{F} = f, \tilde{C} = c, \bold{p}_0\right\}
    \end{equation}
    \item \textbf{Impact Location model}: the spatial uncertainty in the ground location of a UAV’s ground impact once an unrecoverable failure occurs $(x, y, 0)$ is influenced by many factors, including the initial location and altitude $(x_0, y_0, h_0)$, the aircraft type/configuration $\tilde{A}$, the operating condition $\tilde{O}$, the environmental condition $\tilde{E}$, the failure mode $\tilde{F}$, and the recovery outcome $\tilde{R}$. The probability density function on the ground impact location is given by $f_G \left((x, y, 0) | (x_0, y_0, h_0), \tilde{A}, \tilde{O}, \tilde{E}, \tilde{F}, \tilde{R}\right)$. We further denote $\bold{p}$ as a small area around $(x, y, 0)$, then the probability that the UAV falls into $\bold{p}$ is given by
    \begin{equation}
    P_G \left(\bold{p} | \bold{p}_0, a, o, e, f, r\right) = \text{Pr} \left\{\bold{p} | \bold{p}_0, \tilde{A} = a, \tilde{O} = o, \tilde{E} = e, \tilde{F} = f, \tilde{R} = r\right\}
    \end{equation}
    \item \textbf{Impact stress model}: like most relevant works in the literature, we use kinetic energy as the metric to measure a UAV's stress characteristic. The uncertainty in impact stress level $\tilde{K}$ depends on the impact location $\bold{p}$, the initial failure location $\bold{p}_0$, the aircraft type/configuration $\tilde{A}$, the operating condition $\tilde{O}$, the failure mode $\tilde{F}$, and the contingency capability $\tilde{C}$, given by
    \begin{equation}
    P_S \left(k | \bold{p}, \bold{p}_0, a, o, f, c\right) = \text{Pr} \left\{\tilde{K} = k | \bold{p}, \bold{p}_0, \tilde{A} = a, \tilde{O} = o, \tilde{F} = f, \tilde{C}=c\right\}
    \end{equation}
    \item \textbf{Harm model}: the uncertainty in an EoV's harm level $\tilde{H}$ is influenced by the aircraft type/configuration $\tilde{A}$, the contingency capability $\tilde{C}$, and the kinetic energy level $\tilde{K}$, given by
    \begin{equation}
    P_H \left(h | a, c, k\right) = \text{Pr} \left\{\tilde{H} = h | \tilde{A} = a, \tilde{C} = c, \tilde{K} = k\right\}
    \end{equation}
    \item \textbf{Exposure model}: characterizes density of a specific EoV at time $t$ and location $\bold{p}$ in the domain. The model is given by
    \begin{equation}
    E \left(\bold{p}, t\right) = E \left(\bold{p}, \tilde{T} = t\right)
    \end{equation}
\end{enumerate}

We define the general model of the individual risk $R^i$ at ground location $\bold{p}$ as the probability that the UAV failure at location $\bold{p}_0$ can cause a certain harm level $h$ to the EoV at ground location $\bold{p}$ and time $t$. With all six sub-models, we have
\begin{subequations}\label{eqn:totalprob}
\begin{align}
    & \quad~ R^i_{~\mathbb{I}|\mathbb{A}}\left(h \odot \bold{p}, t | \bold{p}_0\right) \label{eqn:totalprob1}\\
    &= P\left((k, h) \odot \bold{p}, t, f, r| \bold{p}_0, a, o, e, c\right) \label{eqn:totalprob2}\\
    &= \underbrace{E \left(\bold{p}, t\right)}_{\begin{subarray}{c}\text{Exposure}\\
    \text{Model}\end{subarray}} \underbrace{P_H \left(h | a, c, k\right)}_{\begin{subarray}{c}\text{Harm}\\
    \text{Model}\end{subarray}} \underbrace{P_S \left(k | \bold{p}, \bold{p}_0, a, o, f, c\right)}_{\begin{subarray}{c}\text{Impact Stress}\\
    \text{Model}\end{subarray}} \underbrace{P_G \left(\bold{p} | \bold{p}_0, a, o, e, f, r\right)}_{\begin{subarray}{c}\text{Impact Location}\\
    \text{Model}\end{subarray}} \underbrace{P_R \left(r | a, f, c, \bold{p}_0\right)}_{\begin{subarray}{c}\text{Recovery}\\
    \text{Model}\end{subarray}} \underbrace{P_F \left(f | a, o, e\right)}_{\begin{subarray}{c}\text{Failure}\\
    \text{Model}\end{subarray}} \label{eqn:totalprob3}
\end{align}
\end{subequations}
where $\mathbb{I}$ denotes the incident conditions $(f, r)$, and $\mathbb{A}$ denotes representative aircraft and operation conditions $(a, o, e, c)$. From Equations~\eqref{eqn:totalprob1} to \eqref{eqn:totalprob2}, between $(k, h)$ we can further eliminate $k$ because only $h$ is the true result of interest. In the next step, we extend the individual risk terrain to the cumulative risk terrain -- a virtual risk terrain that contains flight risk information for all locations of interest on the ground. 

Let $\mathcal{G}$ denote the finite set of all locations of interest on the ground. The individual and cumulative risk terrains differ in their perspectives on the primary object. The individual risk terrain is a property of a ground location $\bold{p} \in \mathcal{G}$ and depicts the risk map around $\bold{p}$. In contrast, the cumulative risk terrain focuses on each location $\bold{p}_0$ in the air. The cumulative risk at location $\bold{p}_0$ above the ground is the maximum possible individual risk it applies to all ground locations in $\mathcal{G}$, given by
\begin{equation}
    R^c_{~\mathbb{I}|\mathbb{A}, h}\left(\bold{p}_0, t\right) = \max_{\bold{p} \in \mathcal{G}} \left[R^i_{~\mathbb{I}|\mathbb{A}}\left(h \odot \bold{p}, t | \bold{p}_0\right)\right]
\end{equation}

Let $\mathcal{L}$ denote the finite set of all possible locations in the air where the UAV could fly. Points in $\mathcal{L}$ could be waypoints in classic aircraft trajectory planning, or simply the discretization of the continuous urban airspace. The collection of $R^c_{~\mathbb{I}|\mathbb{A}, h}\left(\bold{p}_0, t\right)$ from all locations $\bold{p}_0 \in \mathcal{L}$ results in the virtual flight risk terrain in the 3D urban space, the final outcome of this work. 

Upon the computation of the complete virtual flight risk terrain in a complex urban space, the process can be decomposed into two steps for better efficiency. In the first step, we compute the ``individual probability terrain'' around any single location $\bold{p}$. In Equation~\eqref{eqn:totalprob3}, we further denote
\begin{equation}
    P_{~\mathbb{I}|\mathbb{A}}\left(h \odot \bold{p} | \bold{p}_0\right) = P_H \left(h | a, c, k\right) P_S \left(k | \bold{p}, \bold{p}_0, a, o, f, c\right) P_G \left(\bold{p} | \bold{p}_0, a, o, e, f, r\right) P_R \left(r | a, f, c\right) P_F \left(f | a, o, e\right)
\end{equation}
such that Equation~\eqref{eqn:totalprob} can be further written as
\begin{equation}\label{eqn:totalprobsim}
    R^i_{~\mathbb{I}|\mathbb{A}}\left(h \odot \bold{p}, t | \bold{p}_0\right) = E \left(\bold{p}, t\right) P_{~\mathbb{I}|\mathbb{A}}\left(h \odot \bold{p} | \bold{p}_0\right) 
\end{equation}

Between the two items in the RHS of Equation~\eqref{eqn:totalprobsim}, $E \left(\bold{p}, t\right)$ is the exposure at point $\bold{p}$, while $P_{~\mathbb{I}|\mathbb{A}}\left(h \odot \bold{p} | \bold{p}_0\right)$ is the individual probability terrain above $\bold{p}$. In practice, we compute $P_{~\mathbb{I}|\mathbb{A}}\left(h \odot \bold{p} | \bold{p}_0\right)$ for $\bold{p}_0 \in \mathcal{L}'$, where $\mathcal{L}'$ is a limited space above $\bold{p}$. In the 3D space, $\mathcal{L}'$ can be a cubic space above $\bold{p}$. It is important to note that, with fixed conditions $\mathbb{I}$ and $\mathbb{A}$, $P_{~\mathbb{I}|\mathbb{A}}\left(h \odot \bold{p} | \bold{p}_0\right)$ represents a relative probability pattern that is identical at all $\bold{p} \in \mathcal{G}$ on the ground. To compute the individual risk terrain, $P_{~\mathbb{I}|\mathbb{A}}\left(h \odot \bold{p} | \bold{p}_0\right)$ is multiplied by a constant $E \left(\bold{p}, t\right)$ at each $\bold{p} \in \mathcal{G}$. Therefore, to obtain the cumulative risk terrain, we can first pre-compute $P_{~\mathbb{I}|\mathbb{A}}\left(h \odot \bold{p} | \bold{p}_0\right)$ in a cubic space $\mathcal{L}'$ around a single ground location, then multiply this ``building unit'' with every point in the ground exposure model $E \left(\bold{p}, t\right)$, an action similar to convolution.

\subsection{Failure Model}

As the first sub-model in the framework, the failure model characterizes the probability and/or uncertainty in the occurrence of specific failure modes. The failure model is dependent on the aircraft type and configuration, operating conditions, and environmental conditions. According to~\citep{clothier2018modelling}, failure modes of UASs can be broadly classified into four categories:
\begin{enumerate}
    \item \textsl{Unpremeditated Descent Scenario (UDS)}: a failure (or combination of failures), which results in the inability of the aerial vehicle to maintain a safe altitude above the surface or distance from objects and structures.
    \item \textsl{Loss of Control (LOC)}: a failure (or combination of failures), which results in the loss of control of the aerial vehicle and may lead to impact at high velocity.
    \item \textsl{Controlled Flight into Terrain (CFIT)}: when an airworthy aerial vehicle is flown, under the control of a qualified remote pilot or certified autopilot system, unintentionally into terrain (water, structures, or obstacles).
    \item \textsl{Dropped or Jettisoned Components (DOJC)}: failures that result in a component of the aerial vehicle (including its payload or stores) being dropped or jettisoned from the aerial vehicle.
\end{enumerate}

Each failure mode can be attributed to a list of failures. For example, UDS can be caused by the propulsion system failure~\citep{burke2011system}; components involved in DOJC can be propellers, camera, and packages. The type of failure mode is a crucial factor in the safety analysis of an UAS as it largely influences the subsequent three sub-models -- recovery model, impact location model, and impact stress model. For instance, operations with UDS have better controllability on the impact location over those having LOC or CFIT~\citep{washington2017review}. Overall, the failure models can be developed using (1) historical data (from failure, accident, incident), (2) expert opinions, and (3) reliability info on the components, subsystems, and systems of the UAS. Due to the limited historical data on UAS failures, functional and structural decomposition approaches, as well as the elicitation of expert opinions have been employed to assess the failure rate of a vehicle or system. 

Although dependencies exist between different failure modes, most models in the literature considered a single failure mode in their analyses. The majority of models also assumed constant failure rates, although uncertainty exists because of a lack of data and knowledge on UAS failure. Going forward, with improved accessibility to UAS reliability data, a system-informed data-driven approach can advance the assessment of UAS system failure. Among the existing failure models, \citep{clothier2007casualty} assumed 10$^{-5}$ per flight hour for unrecoverable flight critical event. \citep{ford2010assessment} assumed 10$^{-5}$ per flight hour for catastrophic failure (with uncontrolled flight termination) and 10$^{-4}$ per flight hour for hazardous failure. \citep{stevenson2015estimated} assumed the Mean Time Between Failures (MTBF) to be 10$^{5}$ for sub-urban and 10$^{6}$ for urban areas. \citep{petritoli2018reliability} conducted a more detailed reliability evaluation on UAV and produced a Failure In Time (FIT) rate table for commercial drone system and components. Table~\ref{tbl:failure} is a summary of their estimations on UAV reliability.

\begin{table}[h!]
\centering
\caption{Reliability information of a commercial drone (from~\citep{petritoli2018reliability})}
\begin{tabular}{cccc}
\hline
\textbf{System Description} & \textbf{System FIT (Failure/10$^{6}$ hrs)} & \textbf{MTBF (hours)} & \textbf{Incidence (\%)} \\ \hline
Ground control system       & 2.00                            & 500,000.0             & 6.62                    \\ 
Mainframe                   & 2.77                            & 360,984.8             & 9.16                    \\ 
Power plant                 & 9.94                            & 100,603.6             & 32.88                   \\ 
Navigation system           & 9.41                            & 106,269.9             & 31.13                   \\ 
Electronic system           & 5.01                            & 199,600.8             & 16.57                   \\ 
Payload                     & 1.10                            & 909,090.9             & 3.64                    \\ \hline
TOTAL                       & 30.23                           &                       &                         \\ \hline
\end{tabular}
\label{tbl:failure}
\end{table}

In the case study described in Section~\ref{sec:case}, we will create 3D virtual risk terrains for catastrophic failures such as LOC, because of improved accessibility to models and evidence in the literature. Since most relevant works in the literature have adopted 10$^{-5}$ to 10$^{-6}$ per flight hour for catastrophic system or component failures, and considering the existence of uncertainty in this process, we will conduct a failure rate sensitivity analysis  by using a range of failure rates and then compare the results.

\subsection{Recovery Model}

The recovery model characterizes the uncertainty in the UAV's ability to recover to a nominal or degraded operational state immediately after the failure occurs. In general, there are two types of recovery from UAV failure or incident. The first type of recovery refers to the UAV's ability to execute an emergency landing given the occurrence of a failure. This requires the UAV system (with or without a remote pilot) to at least maintain partial function such that a crash or other catastrophic consequences can be avoided. Under certain failure modes, it is possible for the UAV to land safely without causing damage to any EoV on the ground. When this type of recovery is attained, there is zero risk to EoVs on the ground and the remaining sub-models are not invoked. The second type of recovery refers to the mitigation of ground impact via contingency equipment such as air bags. Such capabilities are helpful in mitigating damage when an uncontrollable descent is unavoidable, which can happen under certain failure models such as LOC. The recovery model in this framework considers the first type of recovery, while the effect of ground impact mitigation equipment is considered later in the harm model.

\begin{figure}[h!]
	\centering
        \includegraphics[width=0.425\textwidth]{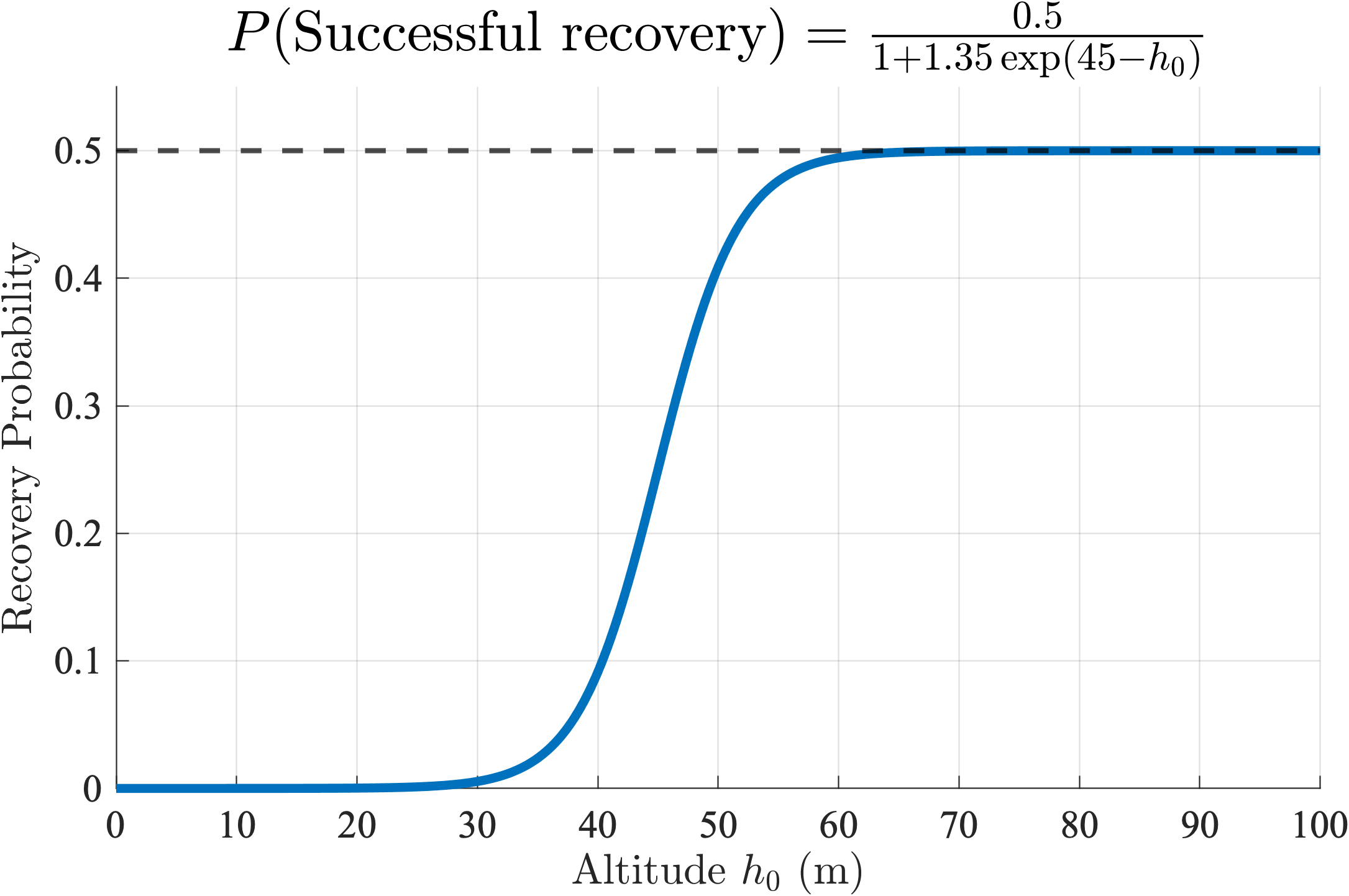}
	\caption{The probability of successful recovery vs. altitude when the multi-rotor UAV is deployed with parachute under LOC}
	\label{fig:Parachuterec}
\end{figure}

The recovery model is built upon the existence and reliability of a failure recovery system. It is dependent on factors such as failure type, contingency capability, and the rate of success of the recovery system. When determining the amount of risk reduction that would be achieved through the failure recovery system, very few works in the literature have provided quantitative assessment. One significance of the recovery model in this framework is the space it provides to incorporate state-of-the-art control and robotics technologies. New developments in both hardware and software could facilitate the first type of recovery and reduce the probability of a crash during particular failure types or incidents. For example, at the start of UDS, the recovery capability in the control system could restart the engine to avoid further descent. When a component failure (e.g., rotor, blade) occurs, advanced control algorithms could facilitate the maintenance of an acceptable flight altitude, resulting in a safe landing. When a loss of communication happens, some UAVs could rely on the sensing and perception capabilities to make decisions that protect public safety. However, when a critical failure occurs, such as LOC or massive component failure, the probability of success for the first type of recovery is close to zero, i.e., $P_R(\text{the recovery is unsuccessful}) = 1$. In the case study described in this paper, we are building risk terrains for the LOC failure mode, which is the worst case scenario of UAV failure for public safety. During LOC, the UAV cannot recover to safe landing. Use of contingency equipment, such as the correct deployment of a parachute, can increase the probability of successfully mitigating ground impact. In the case study we assume the following recovery model for LOC. When the UAV is not equipped with parachute:
\begin{equation}
    P_R \left(\text{Unsuccessful recovery} | \text{Multi-rotor UAV}, \text{LOC}, \text{Without parachute}, \bold{p}_0\right) = 1
\end{equation}
which follows the model format in Equation~\eqref{eqn:Rmodel}. The recovery model with parachute takes into account the AGL altitude $h_0$ of the UAV operation. The deployment of a safety parachute system on UAV has been considered by many studies and designs in the literature~\citep{hasan2019parachute,al2018design,panta2018dynamics}. Ballistic parachute deployment requires less deployment time and is therefore well suited to multi-rotor UAVs in the urban environment. Each parachute system has a minimum deployment altitude, such that the chance of a successful recovery (in terms of not damaging EoVs on the ground) increases with higher altitude. At the time of this writing, the minimum deployment altitude for relevant parachute systems ranges from 20 m to 50 m. Assuming that under LOC, a UAV with a parachute system has a maximum probability of 50\% to avoid a crash, the sigmoid function in Figure~\ref{fig:Parachuterec} models the probability of successful recovery versus altitude. Consequently, the probability of unsuccessful recovery is given by
\begin{equation}
    P_R \left(\text{Unsuccessful recovery} | \text{Multi-rotor UAV}, \text{LOC}, \text{With parachute}, \bold{p}_0\right) = 1-\frac{0.5}{1+1.35\exp(45-h_0)}
\end{equation}

The conditional probabilities in the recovery model can be further updated when more information on the failure recovery system's effectiveness or Subject-Matter Expert (SME) opinions are available.

\subsection{Impact Location Model}

The impact location model is a critical component in the integrated risk-based approach and one that dominates the spatial pattern of the 3D virtual risk terrain. Formally, an impact location model in UAS risk analysis characterizes the spatial-temporal uncertainty in the location and area of a UAV's ground impact once a failure occurs~\citep{washington2017review}. In our approach, the impact location model is invoked under the assumption that the result of the recovery model is unsuccessful -- the system (with or without human interference) is unable to recover to a nominal or degraded operational state such that the UAV can safely land. The potential impact locations due to system failure are dependent on the crash trajectory of the UAV, which can be influenced by the following five factors:
\begin{itemize}
    \item \textsl{Type of UAV}: a significant influencer on the impact distribution. For example, a fixed-wing UAV without power has the capability to glide a certain distance before impacting with the ground; a multi-rotor UAV under various failure modes is likely to start a free fall, leading to a smaller impact area.
    \item \textsl{Failure mode}: an indicator of the remaining controllability the UAV system or the remote pilot can maintain. The simple ballistic trajectory model could be appropriate for DOJC or LOC; more complex dynamic models are required for LOC or UDS~\citep{washington2017review}. 
    \item \textsl{Initial conditions at failure}: the initial altitude and locations are key factors in determining the impact range; the initial velocity determines the skewness or movement of the impact distribution.
    \item \textsl{Environmental conditions}: prevailing wind condition is an influencer on the trajectory; urban topography defines the physical constraints in the 3D space.
    \item \textsl{Contingency capabilities}: such as parachutes, air bags, and other damage control functions, will also influence the spatial and temporal patterns of the impact distribution.
\end{itemize}

Because (1) the crash trajectory can be jointly determined by multiple factors in this list, and (2) of the considerable variabilities and uncertainties in these factors, high-fidelity modeling of the ground impact location of UAV under off-nominal conditions has been a challenging task. In the meantime, the importance of the problem has made it an active research area in the UAV and system safety communities. Due to the lack of experimental and real-world data to apply statistical approach for characterizing the impact model, most of the existing works in the literature utilize a combination of modeling and simulation approach and analytical (probabilistic) approach for estimations. The simulation approach predicts the crash trajectory of UAV using aerodynamics models, flight dynamics models, and laws of physics. The probabilistic approach accounts for the epistemic and aleatory uncertainties in the process. A range of specific impact location models are inevitably required to capture the diversity and uncertainty in the impact location for different combinations of factors above. Some works~\citep{dalamagkidis2008evaluating,foster2017highfidelity} modeled a single point impact with a low level of uncertainty. Because of the high degree of uncertainty associated with the potential impact locations under the occurrence of a failure, we focus on the impact location models in the literature which use a probability distribution to characterize the impact location of a UAV. 

The current literature includes impact location models for both fixed-wing and multi-rotor configurations. As previously mentioned, we use multi-rotor UAV in this work because its VTOL capability has a unique advantage to operate in complex urban environments. Most earlier works have been limited to ballistic descent, which is applicable only to catastrophic failure conditions. Some works~\citep{wu2012development,ancel2017risk,kim2022risk} assumed uniform distribution of impact location inside a certain crash area for simplicity, which cannot reflect the higher impact probabilities at locations surrounding the crash trajectory. Among some of the representative works in recent years, \citep{aalmoes2015conceptual} used bivariate normal distribution to model the potential impact area of general types of system failure. \citep{haartsen2016simulation} used flight simulation to determine the potential impact locations for quadcopter UAVs under various failure types and flight conditions and found that the impact areas are elliptical shaped, depending on the altitude and initial velocity. \citep{courharbo2020ground} developed an analytical solution to estimate the 2D ground impact probability distribution for the ballistic descent of a UAV, while considering uncertainty sources from aircraft and wind parameters. \citep{lin2020failure} simulated Newton's laws of motion and Galileo's free fall to assess a UAV's crash probability density (CPD) under loss of power. \citep{man2022crash} used simulation to investigate the crash trajectory area for different failure modes of a Quadrotor UAV, such as when only one or two motors/blades fail with or without control system. They found that even when crashing to the ground is unavoidable, the control system helps reduce the size of the potential impact area. 

Although variabilities exist in failure modes, initial flight conditions, and environmental conditions, from a probabilistic modeling perspective, we think that the impact location models can generally be classified as either a Gaussian model or a Non-Gaussian model. Below is a summary of the reasoning behind each of these two probabilistic models.
\begin{itemize}
    \item \underline{Gaussian model}: a Gaussian impact location model is useful in both individual and cumulative ground impact estimations. On the individual case, when the UAV is hovering~\citep{man2022crash}, during takeoff and landing, or when the initial velocity and wind play a weak role, it is most likely that its ground impact location is immediately below the initial failure location, since a multi-rotor UAV's crash trajectory is dominated by spin and autorotation while falling to the ground~\citep{lin2020failure}. On the cumulative case, researchers~\citep{lum2011assessing,haartsen2016simulation} also found that when simulating a large number of failure types and flight conditions, the vast majority of crashes occur in close proximity to the initial failure location. 
    \item \underline{Non-Gaussian model}: a non-Gaussian impact location model mainly applies to individual failure cases when the initial velocity and/or wind condition play a considerable role in the crash trajectory~\citep{courharbo2020ground,primatesta2020ground}. In these individual cases, it is unlikely that the ground impact location is immediately below the initial failure location. Instead, the UAV would travel along a declining horizontal trajectory such that the impact area is at a distance from the initial failure event. Then, a non-Gaussian probabilistic model is required to characterize the probable impact locations. 
\end{itemize}

In this work, we propose two probabilistic frameworks to accommodate the Gaussian and non-Gaussian impact location models. Specifically, we add two new considerations in these frameworks to better serve the generation of 3D virtual risk terrain for UAS trajectory planning. The first consideration is the development of 3D impact location models. Each of the impact location models mentioned in this section are 2D models, i.e., in a probabilistic sense, the density function of the ground impact location $f(x,y)$ is always obtained for a specific initial altitude such that it is only a function of the 2D initial location $(x_0,y_0)$. Here, we add altitude as the third dimension and propose model forms that are functions of 3D initial location $(x_0,y_0,h_0)$, where $h_0$ is the initial altitude. The 3D impact location models can better capture the variation in an impact area as altitude changes. The second consideration is that, for each initial failure location $(x_0,y_0,h_0)$, we aim to obtain a impact location model for all flight directions (360 degrees). This is because the resulting 3D virtual risk terrain used for flight trajectory planning should be independent from flight direction, i.e., a UAV can pursue any flight direction at a ``safe'' location in the air. The following two subsections introduce details of the two probabilistic frameworks: the 3D Gaussian impact model, and the Rayleigh impact model for the non-Gaussian case. Both parametric models are flexible to accommodate variations of different UAV types and flight conditions. The model parameters can be obtained by fitting the model to real-world experiment or simulation data.

\subsubsection{Gaussian Impact Model}

\begin{figure}[h!]
	\centering
        \includegraphics[width=0.75\textwidth]{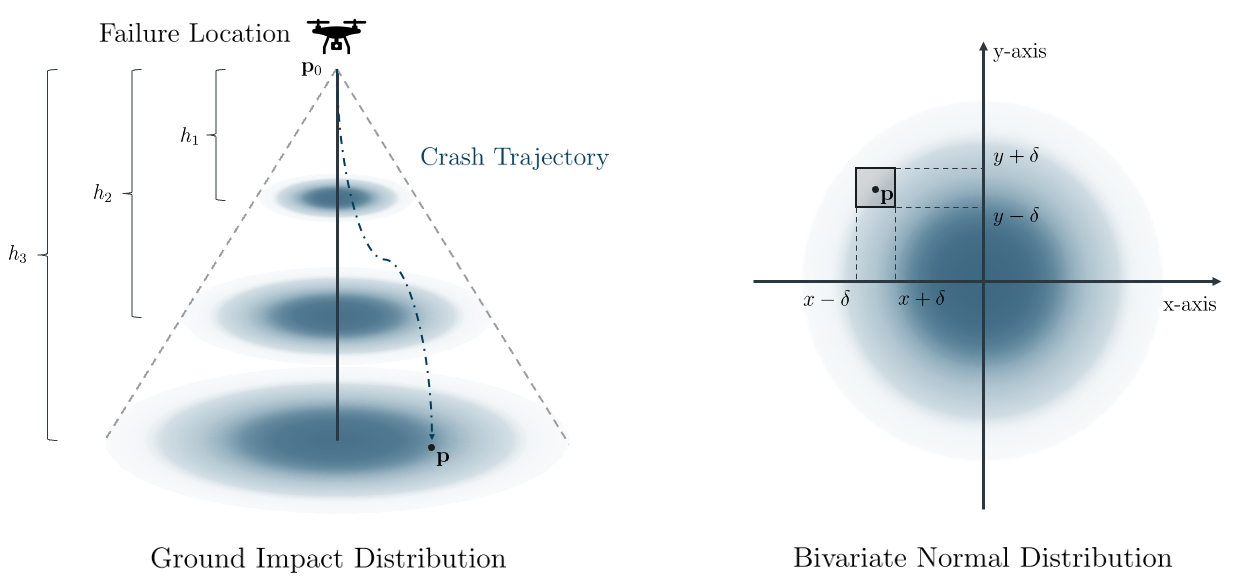}
	\caption{Illustrative Gaussian impact model: change of impact distribution with altitude (left), and calculation of crash probability into a specific area through integration (right).}
	\label{fig:GaussianImpactModel}
\end{figure}

The 3D Gaussian impact model captures cases where the majority of crashes occur in close proximity to the ground location that is immediately below the initial failure location. Because the ground impact area generally becomes larger with increasing initial altitude, a visualization of the 3D Gaussian impact model is similar to a 3D cone, as displayed in the left plot of Figure~\ref{fig:GaussianImpactModel}. Below we define the Gaussian impact model. 

Suppose that the UAV's failure happens at location $(\bold{p}_0, h_0)$, where $\bold{p}_0 = (x_0, y_0)$ and $h_0$ is the AGL altitude in the 3D space, and that the UAV itself is unable to recover, we use the following modified model form of the multivariate Gaussian distribution to represent the impact location density on the ground. Let $\bold{p} = (x,y)$, we have
\begin{equation}\label{model:impactgaussian}
 \displaystyle f_G (\bold{p}|\bold{p}_0, \bold{\Sigma}) = \frac{1}{\sqrt{(2 \pi)^2|\bold{\Sigma}|}} \exp \left(-\frac{1}{2} \left( \bold{p} - \bold{p}_0\right)^\top \bold{\Sigma}^{-1} \left( \bold{p} - \bold{p}_0\right) \right)
\end{equation}
where $\bold{\Sigma} = f(h_0) \bold{\Sigma}_0$ and $f(h_0)$ has the form $f(h_0) = \alpha h_0^2$, and $\alpha$ is a scaling constant. We further assume that the impact model has spherical symmetry. Therefore, $\bold{\Sigma} = \bold{I}$, and $\bold{\Sigma} = \alpha h_0^2 \bold{I}$. Equation~\eqref{model:impactgaussian} therefore has the following simplified model form
\begin{equation}\label{model:impactgaussian2}
 \displaystyle f_G (\bold{p}|\bold{p}_0, h_0) = \frac{1}{\sqrt{(2 \pi)^2 \alpha^2 h_0^4}} \exp \left(-\frac{1}{2\alpha h_0^2} \left( \bold{p} - \bold{p}_0\right)^\top \left( \bold{p} - \bold{p}_0\right) \right)
\end{equation}

Now, it can be seen that the only unknown parameter in Equation~\eqref{model:impactgaussian2} is the scaling constant $\alpha$, which depends on properties such as vehicle type and flight conditions. In this work, we estimate $\alpha$ using results and data from the existing literature. With the bivariate Probability Density Function (PDF) in Equation~\eqref{model:impactgaussian2}, the next step is to calculate the probability that the UAV crashes into a specific area on the ground. This can be computed by integrating the bivariate PDF over the specific area, as shown in the right plot of Figure~\ref{fig:GaussianImpactModel}. In this work, we define the specific area of interest as a square with an area of 4 square meters in the proximity of point $\bold{p}$. Let $\delta$ be the distance from $\bold{p}$ to the middle of the square's side length ($\delta = 1$ when area is 4), we can express the probability as follows
\begin{equation}\label{model:impactgaussian3}
 P_G^g(\bold{p}|\bold{p}_0, h_0) = \frac{1}{\sqrt{(2 \pi)^2 \alpha^2 h_0^4}} \int_{x-\delta}^{x+\delta} \int_{y-\delta}^{y+\delta} \exp \left(-\frac{1}{2\alpha h_0^2} \left( \bold{p} - \bold{p}_0\right)^\top \left( \bold{p} - \bold{p}_0\right) \right) dy dx
\end{equation}
where the value can be obtained through numerical integration methods and tools. The Gaussian model in Equation~\eqref{model:impactgaussian3} computes the probability that the UAV will crash into the small square area around ground point $\bold{p}$ if the initial failure location is $(\bold{p}_0, h_0)$.

\subsubsection{Rayleigh Impact Model}

\begin{figure}[h!]
	\centering
        \includegraphics[width=0.65\textwidth]{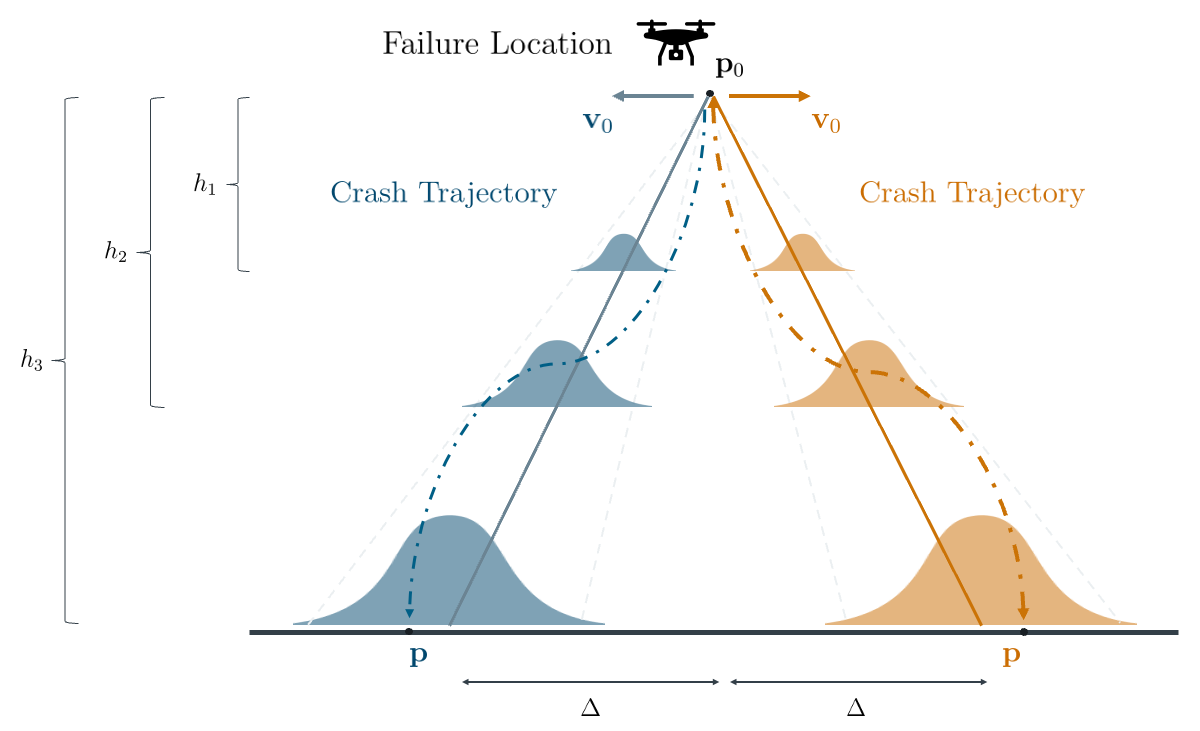}
	\caption{Illustrative Rayleigh impact model: change of impact distribution with altitude and flying direction.}
	\label{fig:RayleighImpactModel}
\end{figure}

The 3D Rayleigh impact model captures cases where the impact area is at a horizontal distance from the initial failure event. When different flight directions are considered, the nature of this model is a multi-modal spatial impact distribution. Figure~\ref{fig:RayleighImpactModel} provides a simple illustration of the Rayleigh impact model on a 2D plane. Suppose that the initial failure location is at $\bold{p}_0$ in the air, the mean location of the impact location density depends on the direction of the initial velocity. Assuming level flight, the direction of the initial velocity can be either left or right. Therefore, the impact area can have two possibilities; centered around a location that has a certain horizontal distance relative to $\bold{p}_0$, either to the left or to the right. This results in a bi-modal impact distribution. When this simple illustration is extended to the complete 3D case where we consider all 360 degrees in the horizontal plane as possible flight directions, the Rayleigh impact model for each initial altitude is analogous to a 2D ring on the ground. In this concept, the most likely impact locations reside at the circumference that is distance $\Delta$ from the origin immediately below $\bold{p}_0$. Below we define the Rayleigh impact model.

The Rayleigh impact model has the following analytical form. Given the failure location $(\bold{p}_0 = (x_0, y_0), h_0)$, the Rayleigh probability density function for the crash location is circularly symmetric around the origin $(\bold{p}_0 = (x_0, y_0), 0)$ that is directly below the failure location. For any point on the ground $\bold{p} = (x,y)$, we have
\begin{equation}\label{model:impactrayleigh}
 \displaystyle f_R (\bold{p}|\bold{p}_0, \sigma, \Delta) = \frac{1}{2 \pi \sigma^2} \exp \left(-\frac{1}{2\sigma^2} \left(\| \bold{p} - \bold{p}_0 \|_2 - \Delta \right)^2\right) 
\end{equation}
where $\Delta = \beta h_0$ is the displacement distance from the origin, and $\sigma = l(h_0) = \gamma h_0$ represents the location of the mode that is $\Delta+\sigma$ from the origin. Therefore, Equation~\eqref{model:impactrayleigh} can be further written as
\begin{equation}\label{model:impactrayleigh2}
 \displaystyle f_R (\bold{p}|\bold{p}_0, h_0) = \frac{1}{2 \pi \gamma^2 h_0^2} \exp \left(-\frac{1}{2\gamma^2 h_0^2} \left(\| \bold{p} - \bold{p}_0 \|_2 - \beta h_0 \right)^2\right) 
\end{equation}

For the Rayleigh impact model, two unknown parameters $\beta$ and $\gamma$ need to be obtained through fitting statistical models to data from computer simulations or real-world flight tests for vehicle types and flight conditions. Like the Gaussian impact model, in this work we obtain these two parameters from the crash distributions of representative aerial vehicles in the literature. Under the Rayleigh model, the probability that the UA crashes into a specific area on the ground is computed through the integration
\begin{equation}\label{model:impactrayleigh3}
  P_G^r(\bold{p}|\bold{p}_0, h_0) = \frac{1}{2 \pi \gamma^2 h_0^2} \int_{x-\delta}^{x+\delta} \int_{y-\delta}^{y+\delta} \exp \left(-\frac{1}{2\gamma^2 h_0^2} \left(\| \bold{p} - \bold{p}_0 \|_2 - \beta h_0 \right)^2\right)  dy dx
\end{equation}

\begin{figure}[h!]
	\centering
        \includegraphics[width=0.16\textwidth]{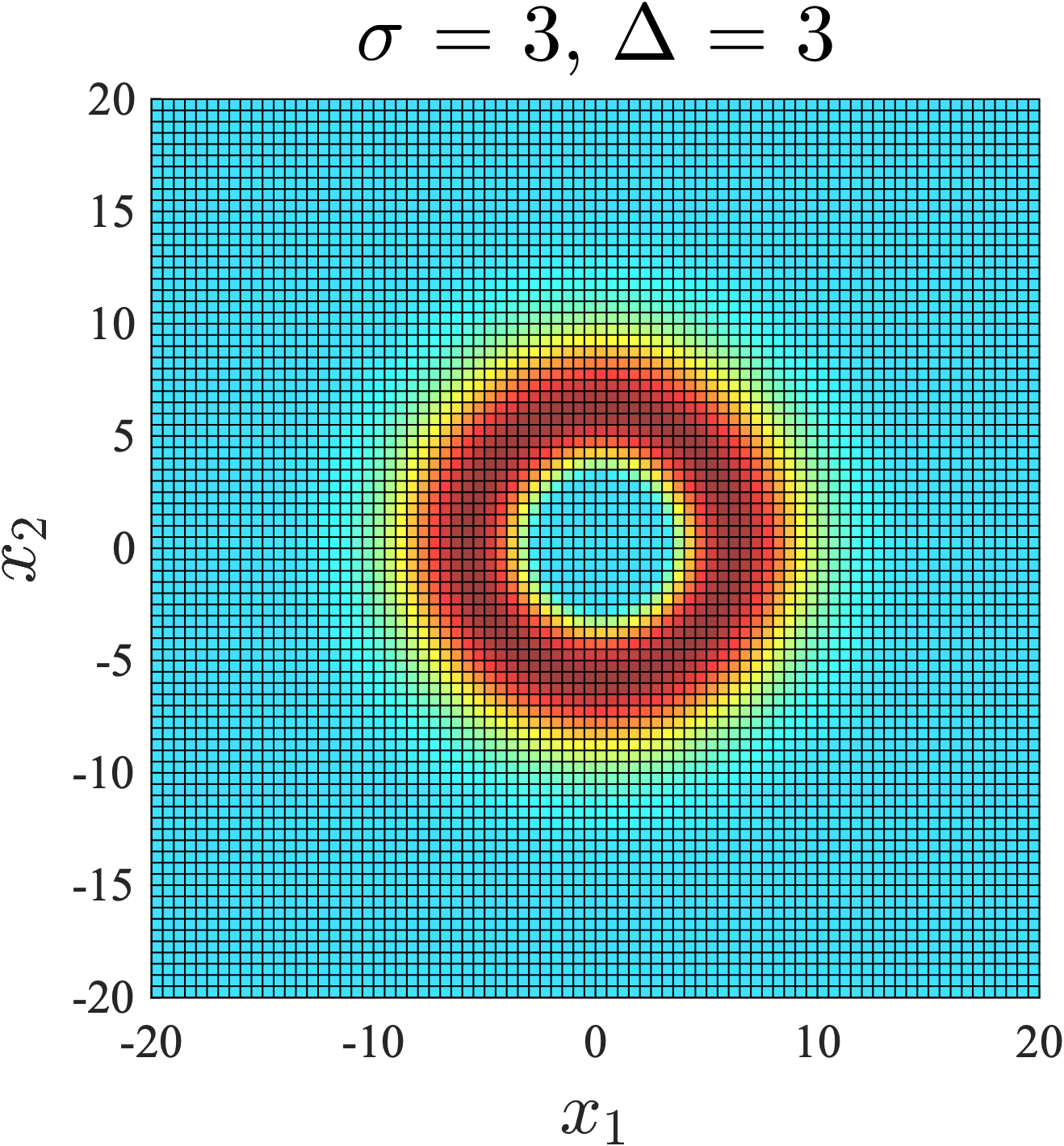}
        \includegraphics[width=0.16\textwidth]{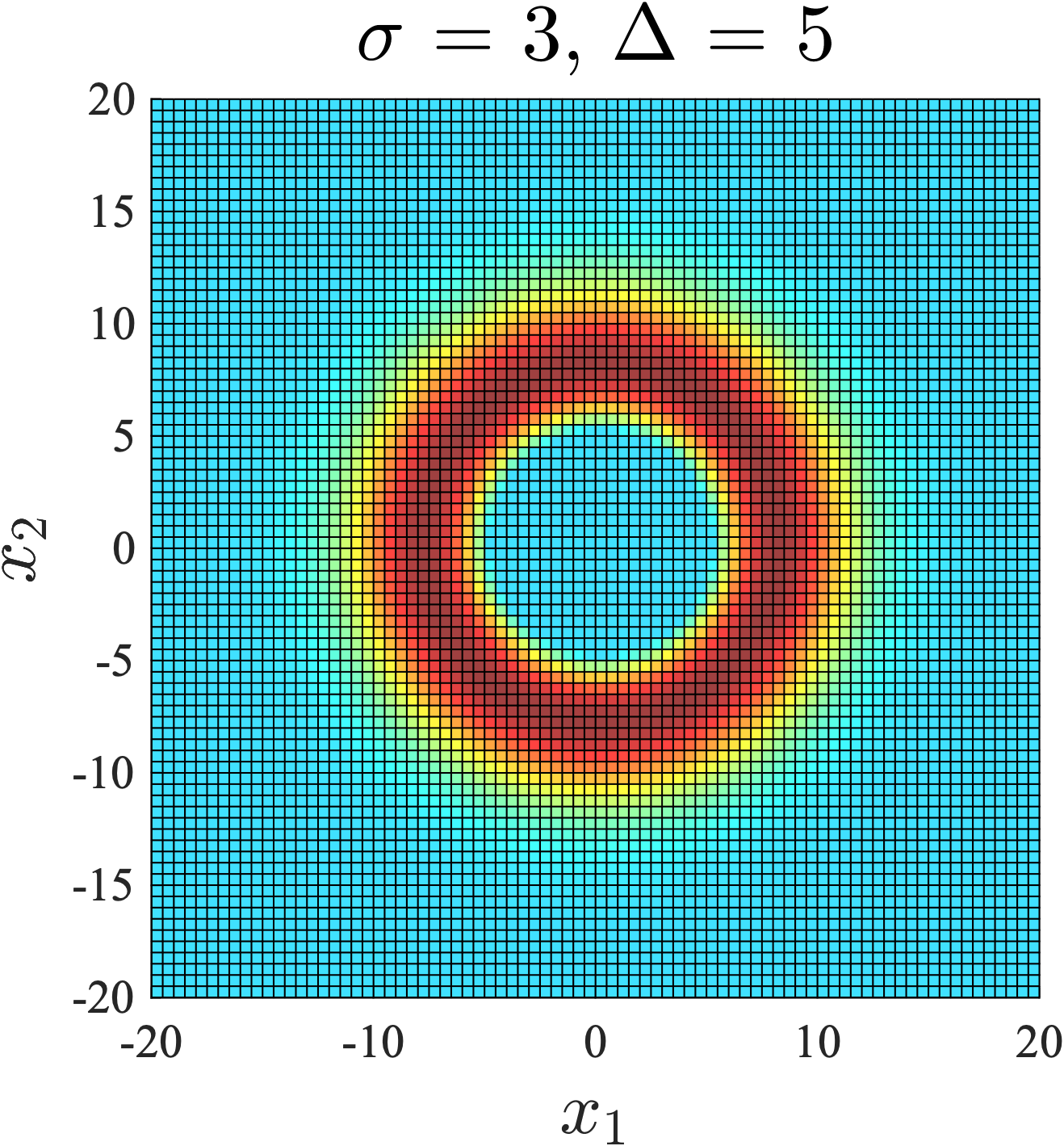}
        \includegraphics[width=0.16\textwidth]{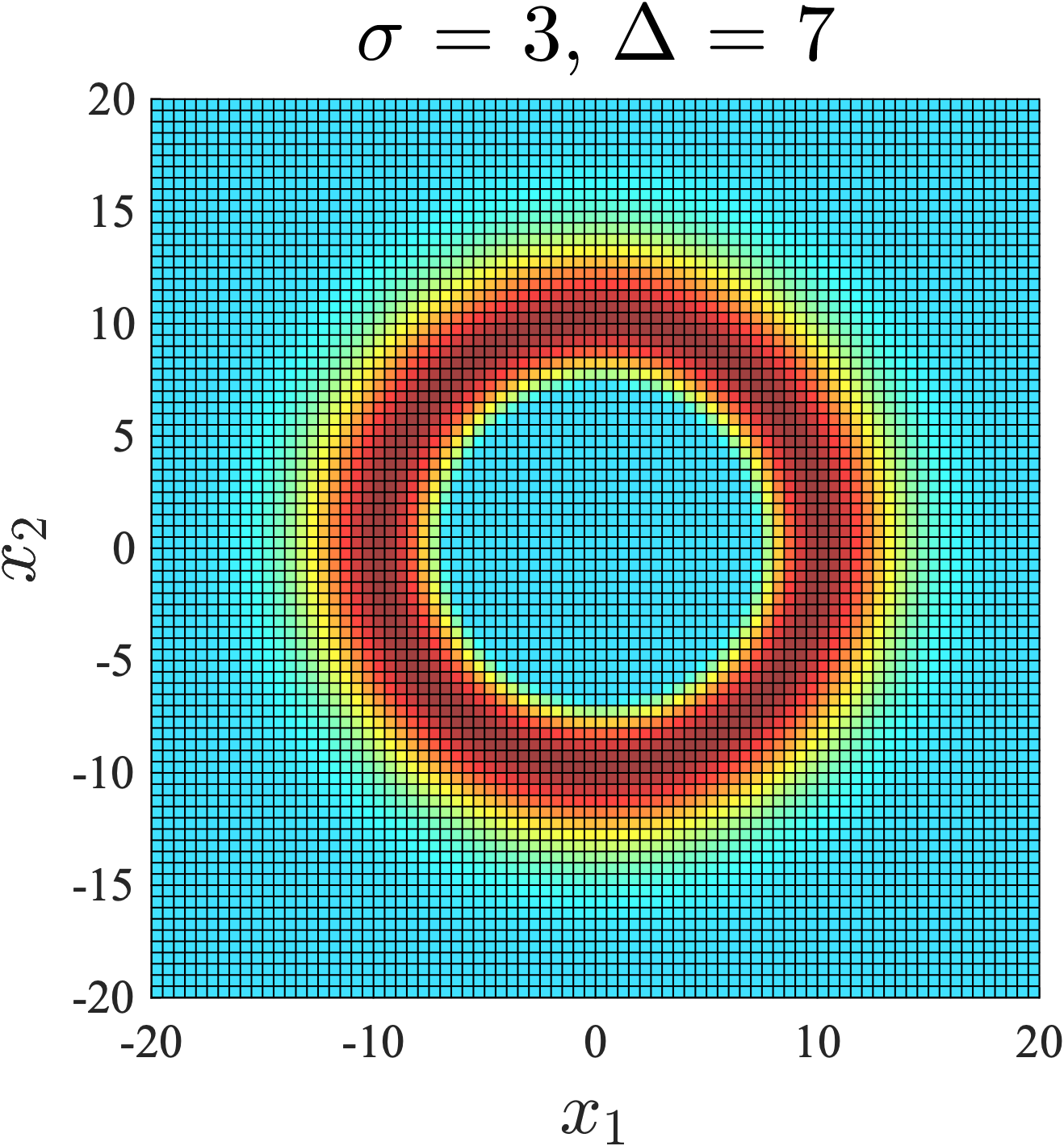}
        \includegraphics[width=0.16\textwidth]{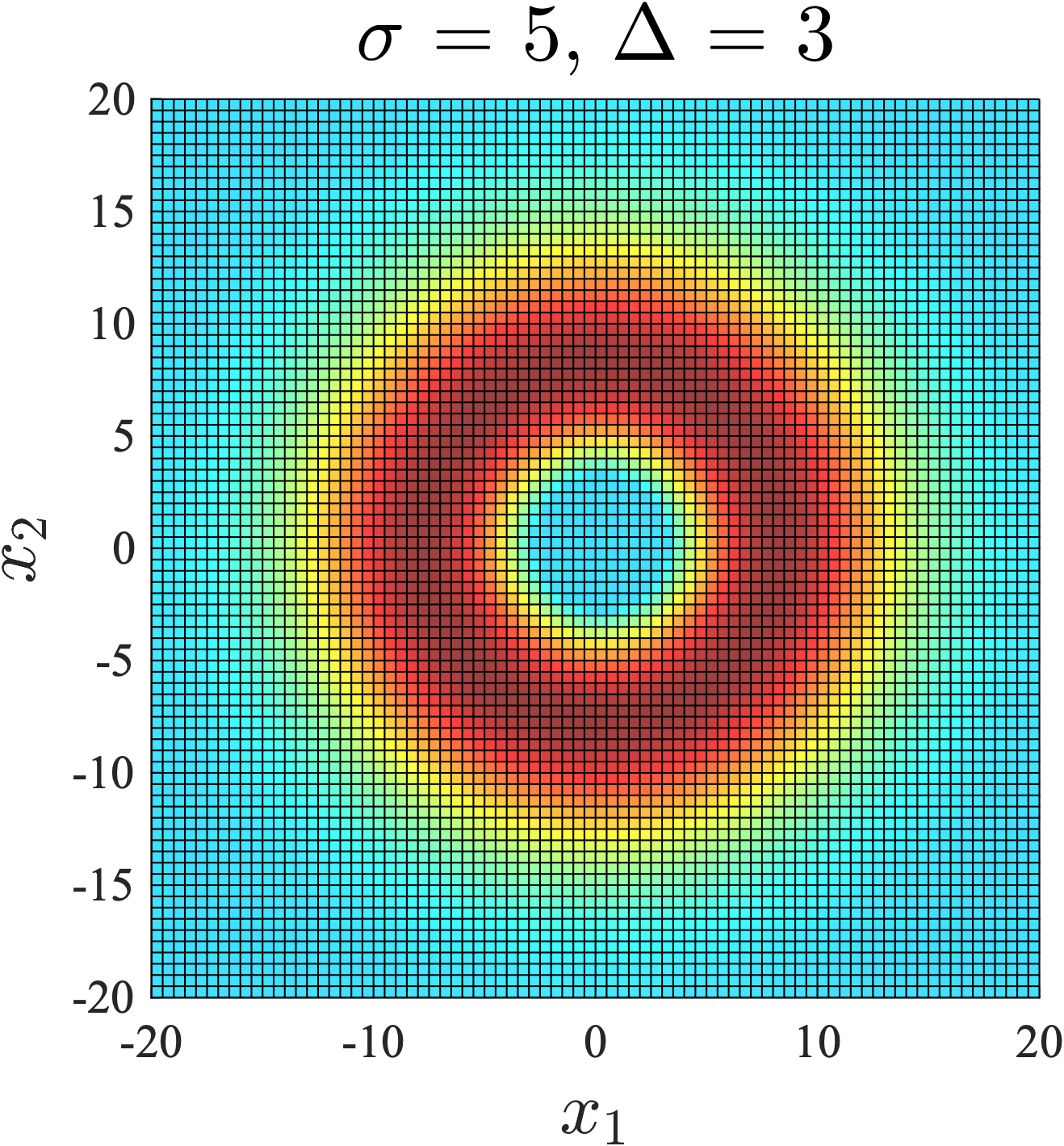}
        \includegraphics[width=0.16\textwidth]{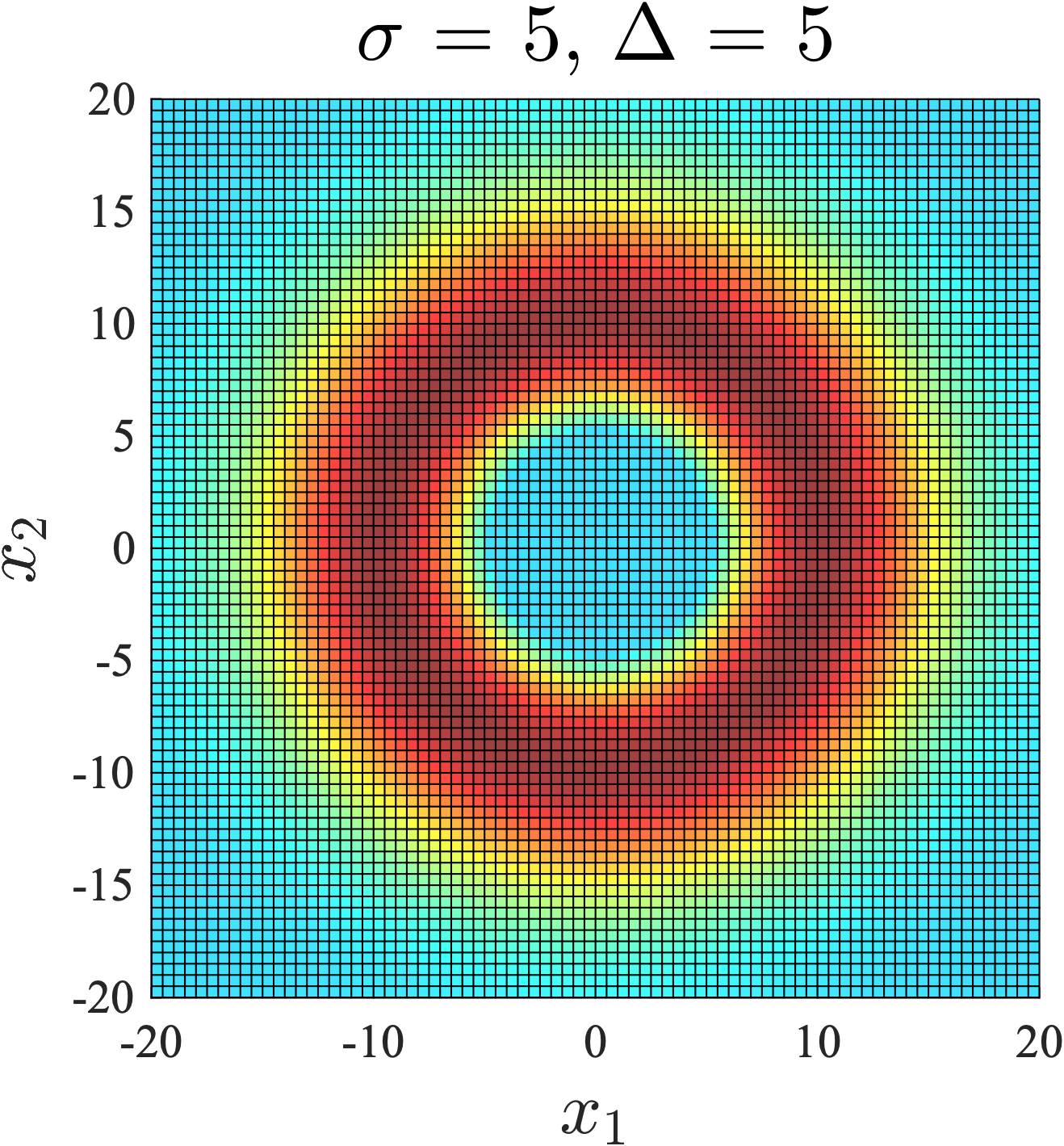}
        \includegraphics[width=0.16\textwidth]{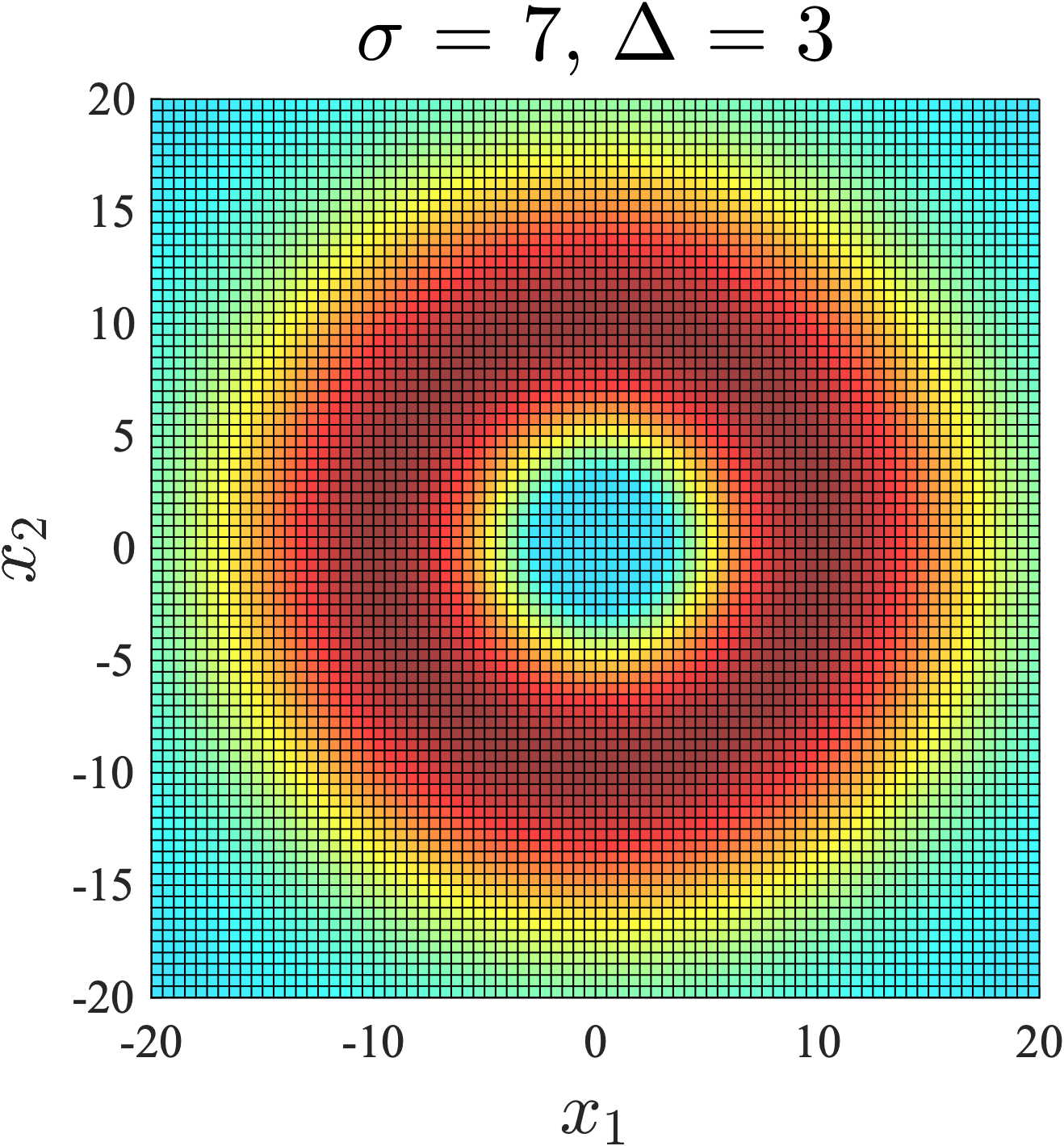}\\
        \includegraphics[width=0.16\textwidth]{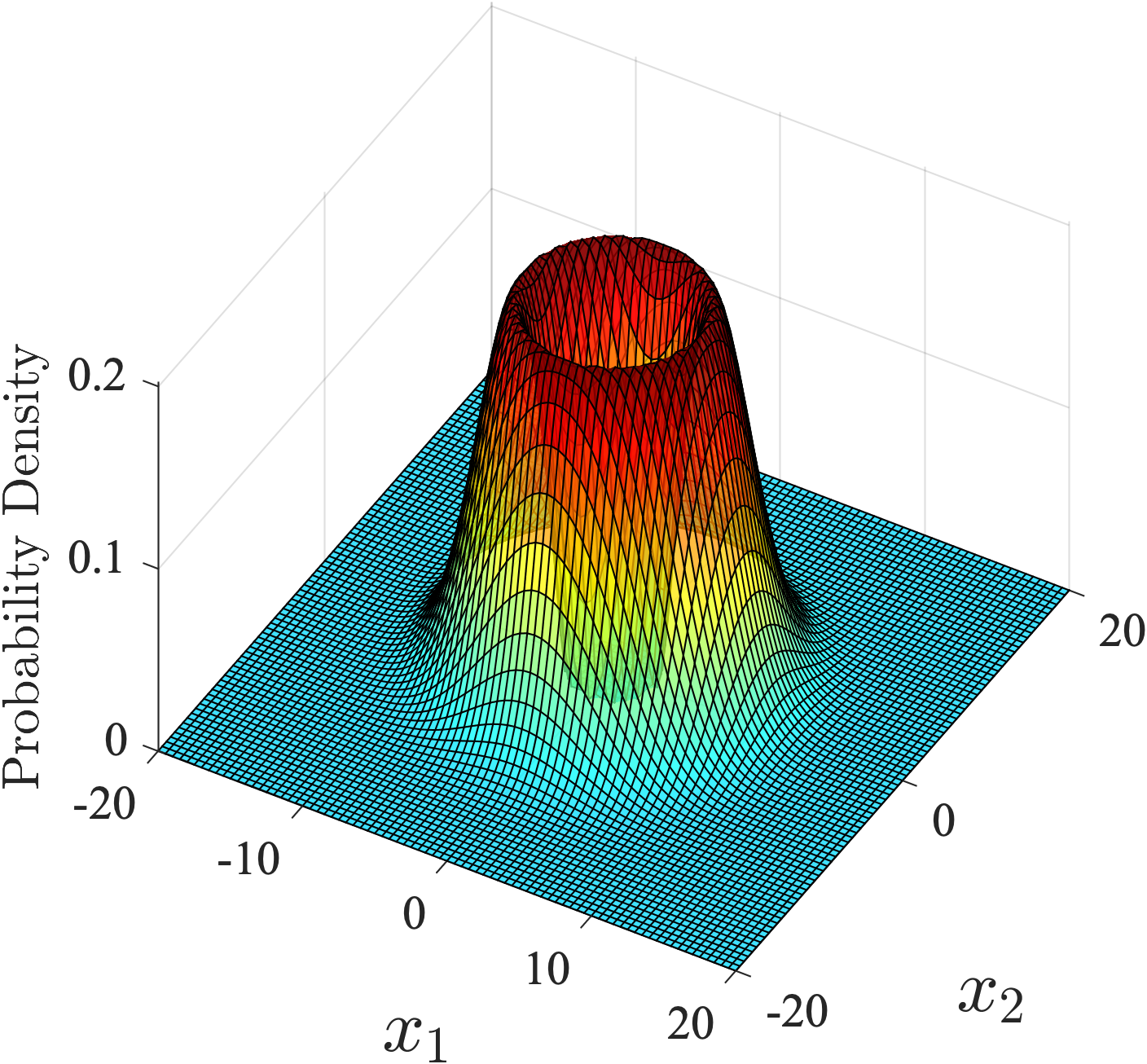}
        \includegraphics[width=0.16\textwidth]{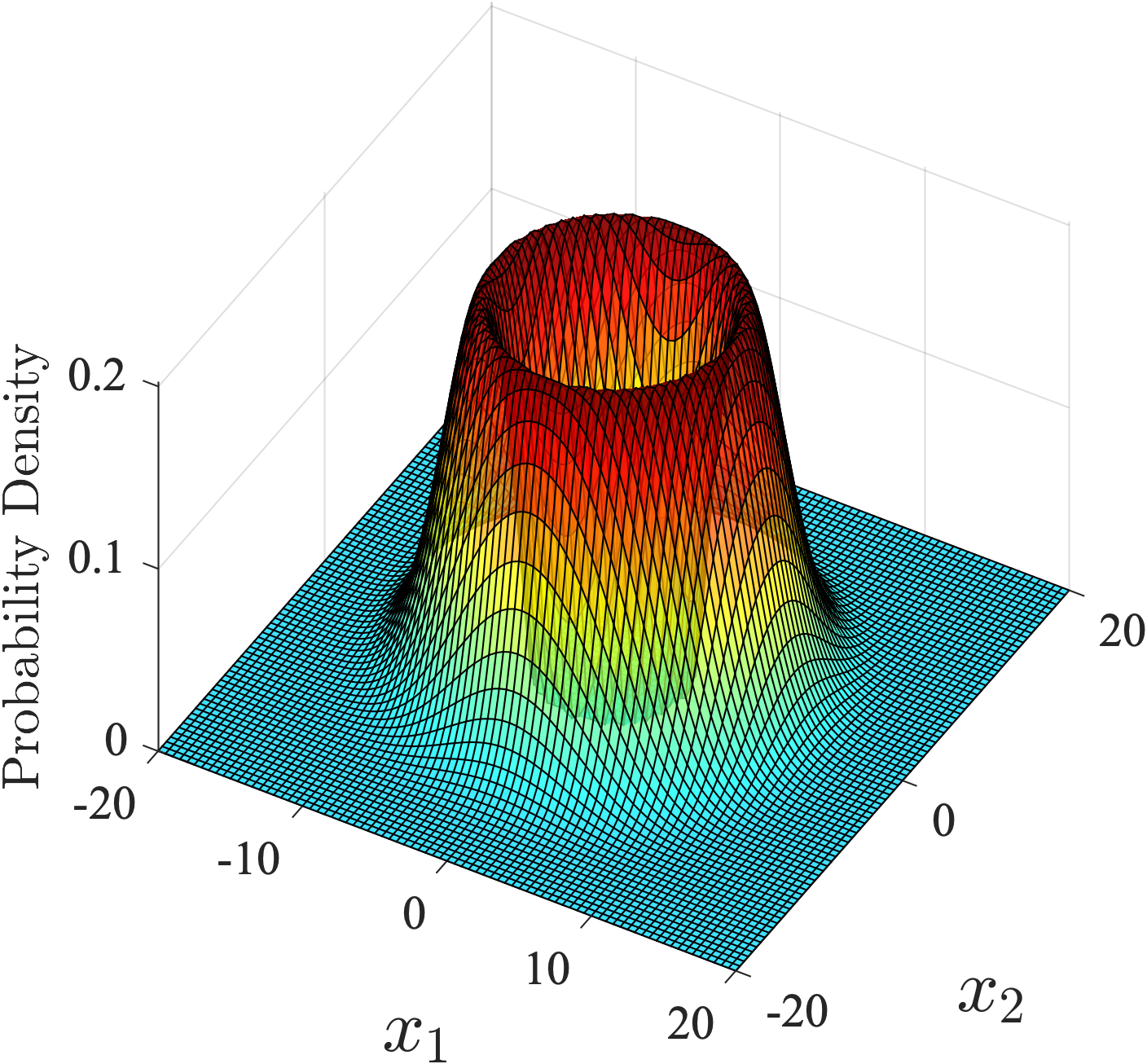}
        \includegraphics[width=0.16\textwidth]{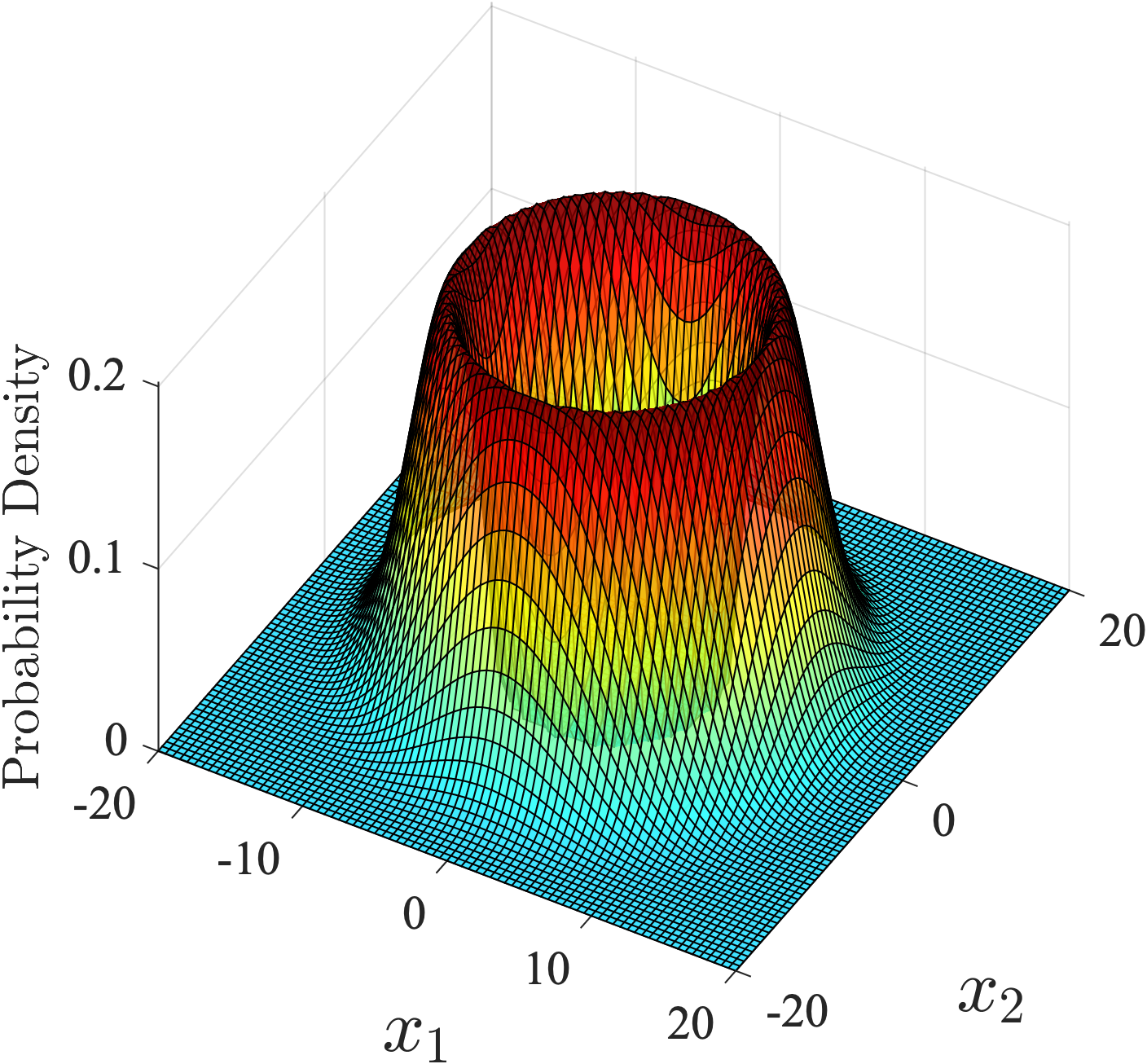}
        \includegraphics[width=0.16\textwidth]{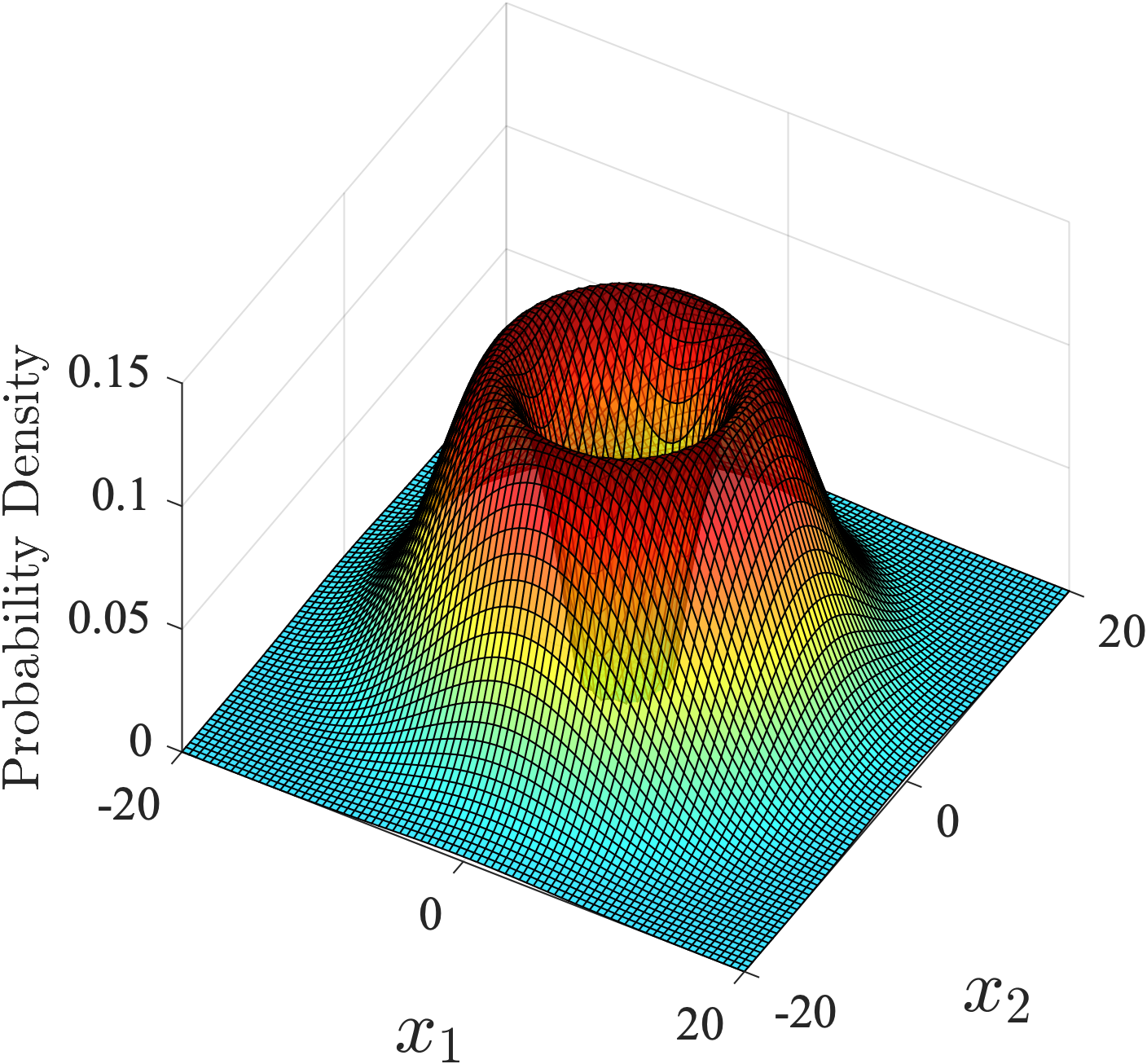}
        \includegraphics[width=0.16\textwidth]{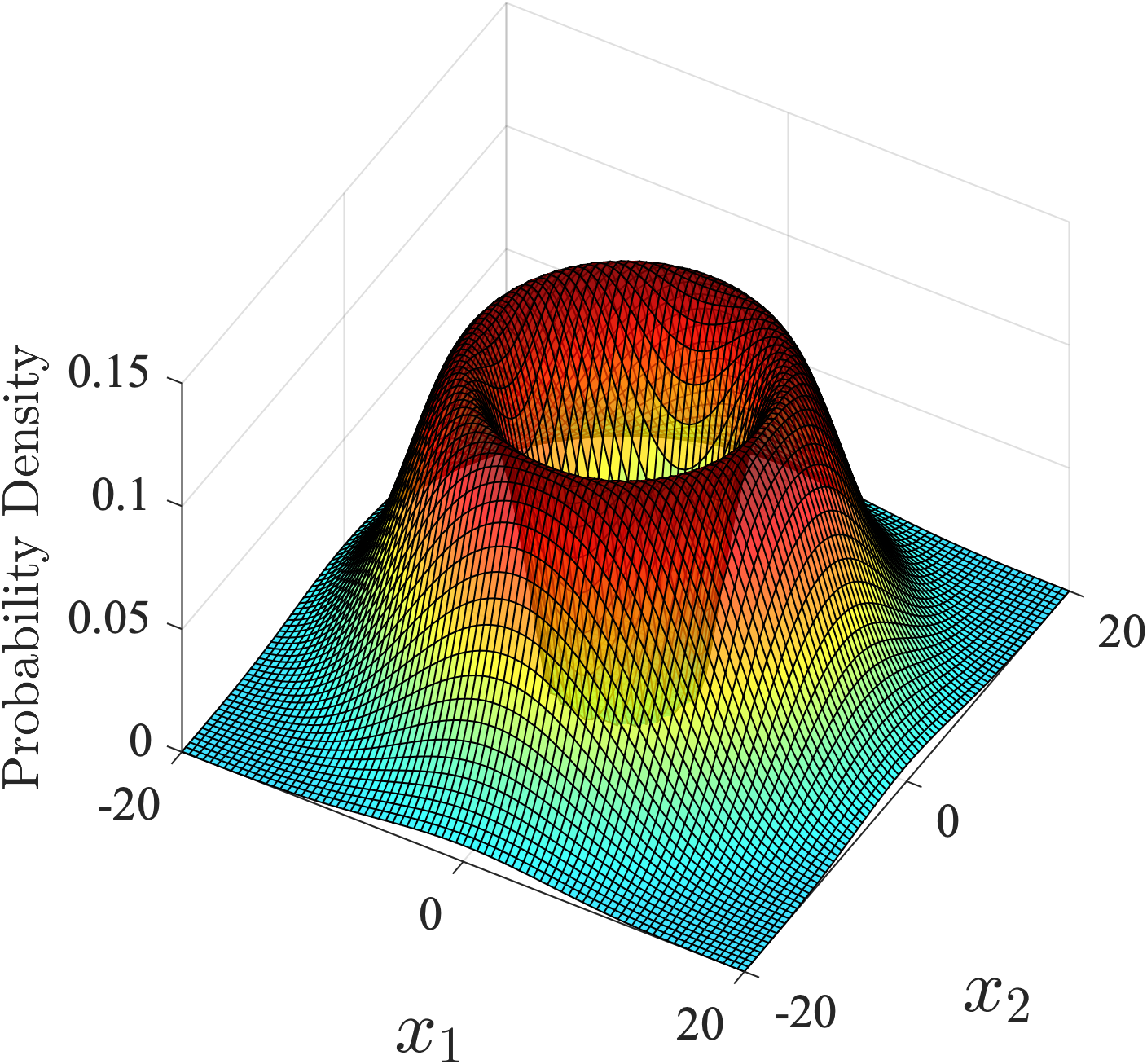}
        \includegraphics[width=0.16\textwidth]{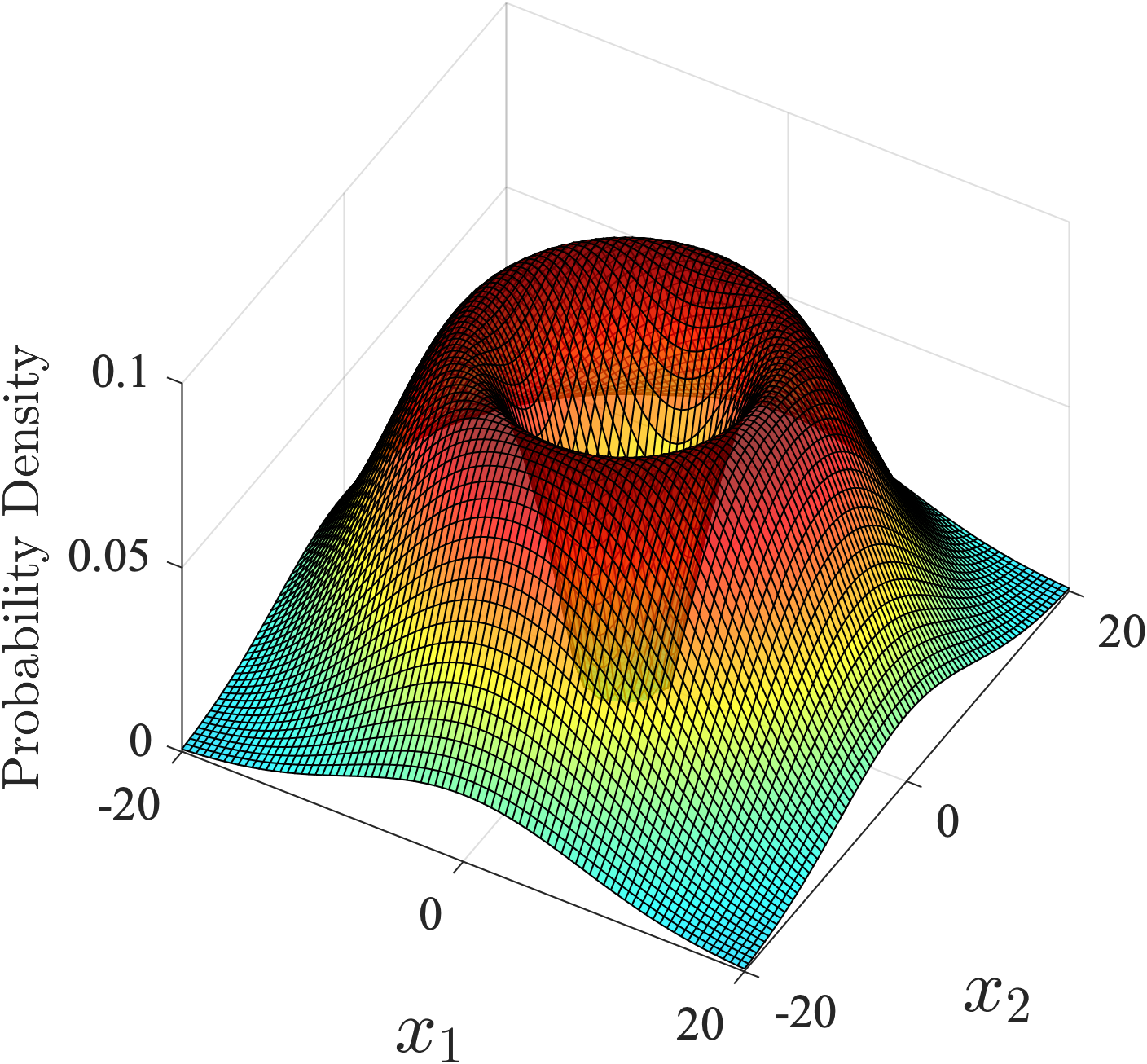}
	\caption{Illustrative examples of of the circular bivariate Rayleigh distribution at different parameter settings}
	\label{fig:RayleighExamples}
\end{figure}

Figure~\ref{fig:RayleighExamples} displays a set of the Rayleigh impact models, where the parameters $\sigma$ and $\Delta$ in Equation~\eqref{model:impactrayleigh} vary. When $\bold{p}_0$ is at (0, 0), the circular bivariate Rayleigh distribution at different parameter settings is an illustration of how this framework can accommodate non-Gaussian impact location densities with different horizontal displacements and variances.

\subsubsection{The Individual Risk Terrain} 

Both the Gaussian and Rayleigh impact models are parametric probabilistic models. The Gaussian impact model in Equation~\eqref{model:impactgaussian2} has one parameter $\alpha$; the Rayleigh impact model in Equation~\eqref{model:impactrayleigh2} has two parameters $\beta$ and $\gamma$. These parameters can be obtained by analyzing real-world experiment data or simulation data. A qualified dataset must have: (1) information regarding the mode and dispersion of the potential impact locations at a certain altitude, and (2) results from multiple altitudes. In the case study, we use a medium sized multi-rotor delivery UAV that weighs 25 kg and normally flies at a speed of 10 m/s. Although the current regulation requires that such UAVs only fly up to a maximum of 400 feet (122 m), we will also explore the flight risk above this altitude to gain better insights on urban airspace management for safety purposes. To pursue close estimations to such conditions, we use simulation results from \citep{lin2020failure} and \citep{courharbo2020ground} to obtain the parameter(s) for the Gaussian and Rayleigh impact models, respectively. Both works investigated, at multiple altitudes, the 2D ground impact distributions of multi-rotor UAV under a major in-flight incident, such as (near) complete loss of lift. The three model parameters used in the case study are: $\alpha = 0.0244$, $\beta = 0.2790$, $\gamma = 0.0918$.

\begin{figure}[h!]
	\centering
        \includegraphics[width=0.295\textwidth]{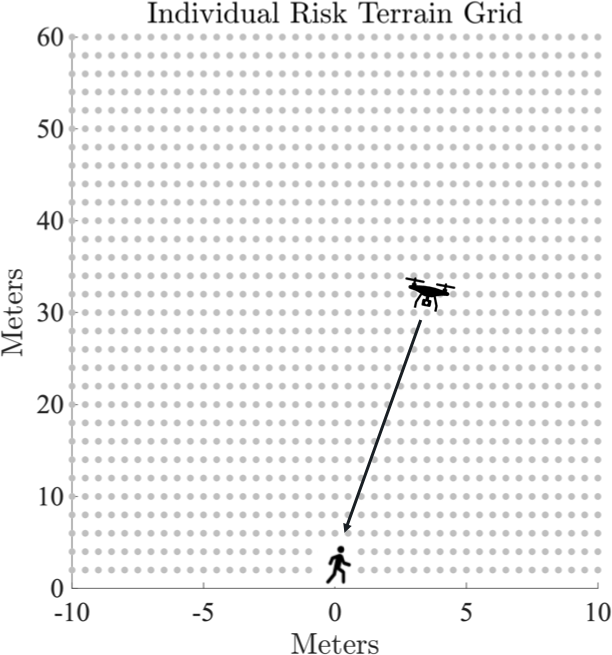}
        \hspace*{0.35cm}
        \includegraphics[width=0.3\textwidth]{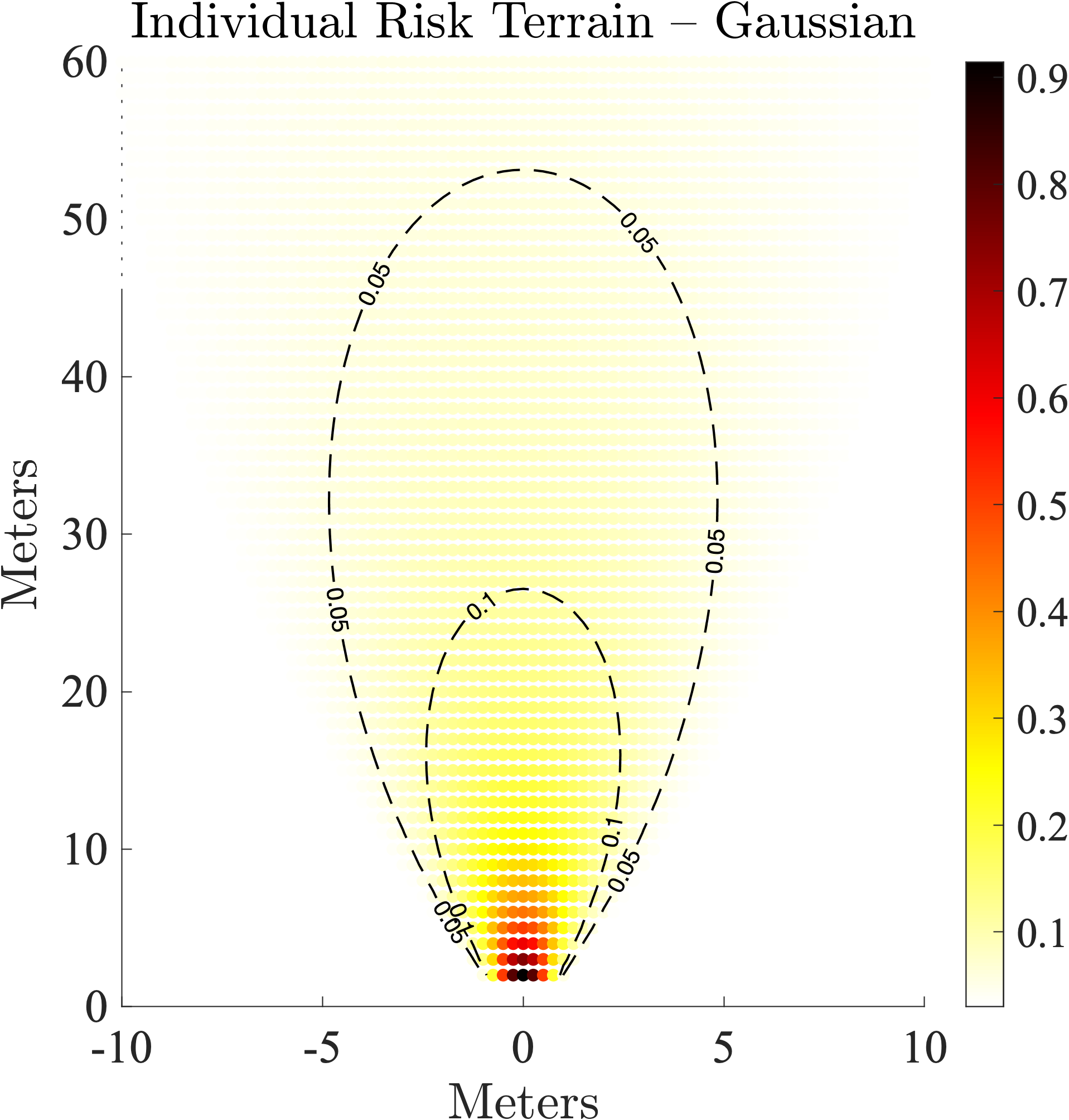}
        \hspace*{0.35cm}
        \includegraphics[width=0.3\textwidth]{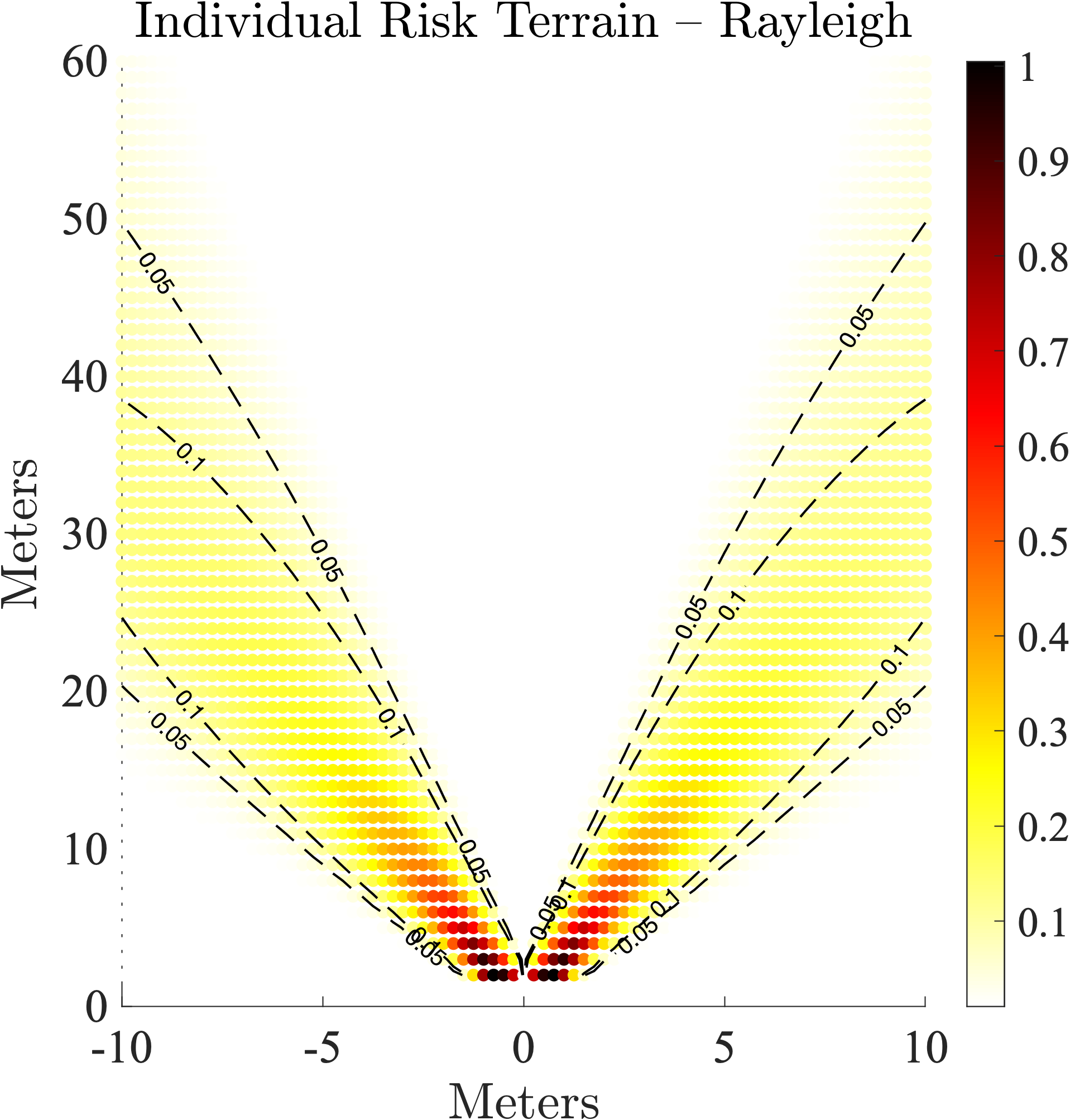}
	\caption{Example of individual risk terrains in the 2D space: the risk terrain grid (left), risk terrain under Gaussian impact model (middle), and risk terrain under Rayleigh impact model (right)}
	\label{fig:IRisk}
\end{figure}

With the impact location models, we can build some illustrative visualizations of the individual risk terrains to demonstrate the spatial variation of ground impact risk. Figure~\ref{fig:IRisk} displays a group of 2D illustrations of the individual risk terrain. The left plot of Figure~\ref{fig:IRisk} depicts a grid in the 2D space with dimensions 20 meters x 60 meters. Suppose that a pedestrian is at $\bold{p} = (0, 0)$, while a quadrotor UAV can fly in the space above the pedestrian. In this example a UAV impacts at a point on the ground if it falls within 1 meter of the point (for $\bold{p}$, this interval is $[-1, 1]$). For each location $\bold{p}_0$ on the grid, with an impact location model, the UAV's risk to $\bold{p}$ is defined as
\begin{equation}
    R^i = P(\text{Impact location is~} \bold{p}|\text{Unrecoverable failure happens at~} \bold{p}_0)
\end{equation}

The individual risk terrain for $\bold{p}$ consists of such probability values at all points on the grid. The middle plot of Figure~\ref{fig:IRisk} visualizes the individual risk terrain under the Gaussian impact location model, where the black dashed lines indicate contours at levels of 0.05 and 0.1. Under the Gaussian impact location model, locations with the highest risks are the grid points that are immediately above and not too distant from (0, 0). Along the vertical line that goes through (0, 0) -- the ``centerline'', the risk is below 0.1 when the altitude is higher than 25 meters; the risk further decreases to below 0.05 when the altitude is over 50 meters. The right plot of Figure~\ref{fig:IRisk} visualizes the individual risk terrain under the Rayleigh impact location model, which shows a different pattern. Under the Rayleigh impact location model, points on the ``centerline'' generally have very low risks. The high-risk regions and the contours become oblique, and the two symmetric regions capture risk terrains for both flight directions -- left and right.

\subsection{Impact Stress Model}

The impact stress model describes the probability or uncertainty in the impact's consequential harmful conditions (stresses) at a given location and time~\citep{washington2017review}. Common metrics for measuring stress include kinetic energy (KE), momentum, and energy density of the UAV. The levels of impact stress can then be related to the harm levels through the harm model. The impact stress model is mainly characterized by type of UAV, initial failure location and ground location, operating conditions (e.g., initial velocity), and contingency equipment (e.g., air bags). More complex stress models also consider secondary effects such as debris scattering, explosions, and the release of hazardous materials. Most stress models in the literature used the KE associated with the UAV's impact as the primary impact stress metric. Thus, the primary harm mechanism considered is trauma caused through blunt force impact. Sources of uncertainty in the impact stress model include mass, speed, and orientation at the point of impact. Because of limited data availability to develop the impact stress model, physics-based models have been used to determine the amount of KE potentially transferred on impact. Below we follow a procedure used by some recent literatures~\citep{pang2022uav,koh2018weight} to calculate the impact KE.

The falling of UAV from the initial failure location is impacted by two main types of forces: (1) the gravitational force $F_g = mg$, where $m$ is the mass of the UAV and $g = 9.8$m/s$^2$ and (2) the (vertical) drag force $F_d$, which can be determined by
\begin{equation}
    F_d = \frac{1}{2} \rho v_{\perp}^2 S C_D 
\end{equation}
where $\rho = 1.225$kg/m$^3$ at sea level is the air density, $v_{\perp}$ is the vertical velocity of the falling UAV, $S$ is the cross-sectional area of the UAV in the direction of falling, and $C_D$ is the drag coefficient. Then, the acceleration of the falling UAV can be calculated as
\begin{equation}
    a = \frac{F_g-F_d}{m} = g - \frac{\rho v_{\perp}^2 S C_D }{2m}
\end{equation}

When the initial vertical velocity is zero, the final impact velocity (also called terminal velocity) of the UAV from altitude $h$ AGL can be obtained as
\begin{equation}
    u = \int_{0}^T \left(g - \frac{\rho v_{\perp}^2 S C_D }{m}\right) dt = \sqrt{\frac{2mg}{\rho S C_D}\left(1-\exp\left(- \frac{\rho S C_D h}{m}\right)\right)}
\end{equation}

Finally, the falling UAV's KE at the ground impact location can be obtained as
\begin{equation}
    K_g = \frac{1}{2} m u^2 = \frac{m^2g}{\rho S C_D}\left(1-\exp\left(- \frac{\rho S C_D h}{m}\right)\right)
\end{equation}

In addition to the impact stress model, some literature~\citep{washington2017review} also mentioned a similar term called ``incident stress model'', which describes the uncertainty in the magnitude of stress that is actually transferred to an EoV when a particular attenuating or amplifying factor is present. The attenuating factors include various types of shelters (e.g., building structures), vehicles, and personal protective equipment (e.g., helmets). The probability of people being protected by these attenuating factors varies with the space and time that the overflight occurs. Many impact models in the literature do not consider this dimension and assume that 100\% of the falling UAV's KE is transferred to an EoV. In this work, we focus our consideration on two types of EoVs in an urban environment -- pedestrians and vehicles. For pedestrians, we assume that no attenuating factor is present in daily situations. This framework can further accommodate the incident stress model and relevant considerations when more knowledge becomes available.

\subsection{Harm Model}

The harm model describes the uncertainty in the level of harm/damage caused to an EoV by an impact stress. More specifically, it relates an impact stress of a certain magnitude to the type and severity of the unwanted outcome. Two key factors involved in a harm model are EoV type and harm mechanism. Common EoV types include people, animals, vehicles, other properties, and the environment. An individual EoV's characteristics, such as a person's height, weight, and age, can also affect the EoV's physical response to an impact stress. Common harm mechanisms include penetration and laceration for small multi-rotor UAVs, crushing and blunt force trauma for larger UAVs, as well as blast and burns~\citep{washington2017review}. Compared to physical harms, psychological harms have not been adequately considered in the literature. The failure of a UAV could cause harm to an EoV through one or more harm mechanisms. The most commonly studied harm mechanism in UAS safety analysis is blunt force trauma~\citep{shelley2016model}, which can cause serious head injuries such as skull fracture. Some works~\citep{weibel2006safety,authority2013human} have also investigated certain aspects of cutting and penetration. It is more challenging to assess the combined effects of multiple harm mechanisms, which could happen when there is no dominating harm mechanism for some specific UAV configurations. It is also common in previous harm response investigations~\citep{authority2013human,feinstein1968personnel} to assume a specific demographic EoV model, such as an adult male with average physical fitness. 

Harm models in the literature are built upon expert judgment, historical data, and impactor studies. Models based on inputs from SMEs output fixed probabilities of fatality, such as 100\%~\citep{weibel2006safety,clothier2007casualty} or 50\%~\citep{lum2011assessing}, for certain types of UAV or all UAV strikes. The use of models informed by historical data~\citep{melnyk2014risk,authority2013human} from accidents and incidents should take into account information such as the specific UAV type and configuration. Models based on experimental or simulation data~\citep{ball2012crash,dalamagkidis2008evaluating} can provide information on specific harm mechanisms, such as blunt force trauma, on specific body parts such as the head~\citep{raymond2009tolerance}. Overall, modeling harm caused by a falling object is a highly complex problem~\citep{melnyk2014risk} with multiple sources of uncertainty. Every model in the literature has both its assumptions and limitations. 

We next discuss two categories of harm models in the literature -- fatality models and casualty models. Some harm models directly relate impact energy to the probability of fatality. \citep{shelley2016model} modeled the probability of fatality using the following logistic curve:
\begin{equation}
    P_1(\text{fatality}|E_i) = \frac{1}{1+\exp\left(-k(E_i-E_0)\right)} 
\end{equation}
where $E_0$ is the impact energy associated with a 50\% probability of a fatality, $E_i$ is the impact energy, and $k$ is a constant. In a more recent work, \citep{primatesta2020ground} further included the sheltering factor and suggested the computation of the fatality rate using 
\begin{equation}
    P_2(\text{fatality}|E_i) = \frac{1-k}{\displaystyle 1-2k+\sqrt{\frac{\alpha}{\beta}} \left(\frac{\beta}{E_i}\right)^{\frac{1}{4 C_S}}}
\end{equation}
where $k = \min(1,(\beta/E_i)^{1/4 C_S})$, $C_S \in (0,1]$ is the sheltering coefficient, $\alpha$ is the impact energy needed to cause 50\% probability of fatality with $C_S = 0.5$, and $\beta$ is the impact energy required to cause fatality as $C_S$ approaches zero. In addition to the probability of fatality, works~\citep{ball2012crash,burke2011system,dalamagkidis2009integrating,melnyk2014risk} also suggested energy level cut-offs for the nonlethal impact KE, which are mostly in the range of 70 to 90 J. 

On the other hand, studies on non-fatal injuries of varying severity~\citep{arterburn2017ground,barr2017preliminary}, also called casualty models, are useful resources for assessing the harm of a falling UAV. Different criteria that relate the impact KE to the severity of various types of injuries have been established. In this work we highlight and use two most relevant harm criteria: Abbreviated Injury Scale (AIS) and Blunt Criterion (BC). The AIS~\citep{greenspan1985abbreviated,gennarelli2008abbreviated} is the most widely-used criterion to assess the severity of individual injury based on medical diagnosis. This global, anatomical-based coding system, initially proposed to classify injuries sustained in vehicle accidents by the Association for the Advancement of Automotive Medicine (AAAM) and later on widely adopted by other industries, defines the severity of injuries throughout the body. Table~\ref{tbl:ais} includes details of six injury classifications, where a higher AIS level indicates an increased threat to life. An advantage of the AIS system, as a mature assessment tool, is that the AIS score can be calculated or converted from a wide range of impact stress metrics, such as KE, forces, and acceleration. In the safety analysis of UAS systems and beyond, most literature~\citep{arterburn2017ground,authority2013human,magister2010small} used AIS level 3 as a reference level, i.e., any injury greater than AIS level 3 is considered substantial. 

\begin{table}[h!]
\centering
\caption{Levels and details in the Abbreviated Injury Scale (AIS)}
\begin{tabular}{l|l|l|l}
\hline
\textbf{AIS Code} & \textbf{Injury} & \textbf{Example}                 & \textbf{Probability of Death} \\ \hline
1                 & Minor           & Superficial Laceration (Skin cut) & 0\%                           \\ 
2                 & Moderate        & Minor Skull Fracture              & 1--2\%                        \\ 
3                 & Serious         & Major Skull Fracture              & 8--10\%                       \\ 
4                 & Severe          & Severe Life-Endangering Fracture  & 5--50\%                       \\ 
5                 & Critical        & Ruptured Liver with Tissue Loss   & 5--50\%                       \\ 
6                 & Unsurvivable    & Death                             & 100\%                         \\ \hline
\end{tabular}
\label{tbl:ais}
\end{table}

The BC correlates the KE deforming the body on impact with the body's ability to tolerate the energy on impact~\citep{sturdivan2004analysis,bir2004design,magister2010small}, and has been extensively used to predict the level of injury due to blunt impacts. A model to compute the magnitude of BC is given by
\begin{equation}\label{eqn:bc}
    BC = \ln \left(\frac{E_i}{TDM^{\frac{1}{3}}}\right) = \ln \left(\frac{E_i}{kDM^{\frac{2}{3}}}\right)
\end{equation}
where $M$ (kg) is the mass of the struck body, $T$ (cm) is the combined thickness of the soft tissue, $D$ (cm) is the  UA characteristic diameter (impact diameter), and $k$ is the coefficient for determining the body wall thickness $T$, with $k=0.593$ for female and $k=0.711$ for male. As a stronger criterion than KE, the BC is recognized by previous works~\citep{magister2010small} as a suitable UAS design and airworthiness criterion for minimizing ground injuries in unsheltered populated areas. The final puzzle piece in the harm model is a relationship between BC and AIS. Researchers~\citep{bir2004design} studied injury data from ballistic impacts and developed a logistic regression model which predicts the probability of AIS 2--3 injuries using BC. The logistic regression model is given by
\begin{equation}\label{eqn:pais}
    P(AIS = 3) = \frac{1}{1+\exp\left(17.76- 38.50 BC\right)}
\end{equation}

By combining Equations~\eqref{eqn:bc} and \eqref{eqn:pais}, we obtain an estimation of the probability that an individual will sustain an AIS level 3 injury given the impact KE of the UAV. This casualty model is used as the human injury model in this risk-based framework. 

Another EoV we consider in this risk model is ground vehicles. There exist two types of treatments when considering the impact of UAV failure on ground vehicles. In one treatment, the focus is on the probability that people sitting in the a vehicle will get injured; vehicles are considered as people with a sheltering factor. In the second treatment, the focus is on the damage to a vehicle and possible secondary hazards, such as vehicle accidents. Research efforts on the evaluation of UAVs' impacts on ground vehicles are still at an early stage. Some existing studies have utilized methods such as Finite Element Method (FEM)~\citep{che2022preliminary} and collision tests~\citep{zhang2021collison,lee2019impact} to investigate the damage resulting from a UAV collision on vehicles and glass panels. The extent of damage can be influenced by many factors such as the UAV type, UAV weight, impact angle, size and material of the impact location, and temperature. Therefore, a mature and generalized model is still lacking. Because our modeling scope consider small to medium sized multi-rotor UAVs, a great amount of their impact energy can be absorbed by the windshield or metal structures of the vehicle, such that penetration rarely happens~\citep{che2022preliminary}. As a result, the probability that the UAV impact will cause direct (and serious) harm to people in the car is very low.

However, if a UAV crashes into a car windshield, the damage could reduce driver visibility and lead to serious traffic accident. This consideration is utilized as the criterion to assess the level of damage on ground vehicles. Using the Impact Effect Assessment (IEA) proposed by EASA in their `Drone Collision' Task Force~\citep{easa2016}, standards for three drone collision damage levels -- High, Medium, and Low are displayed in Table~\ref{tbl:iea}. We refer to the IEA standards and use medium level damage (similar to AIS level 3 in the human case) as the threshold for serious damage on ground vehicles. 

\begin{table}[h!] 
\centering
\caption{Impact Effect Assessment (IEA) at Component Level}
\begin{tabular}{l|l|l|l}
\hline
\textbf{Component/Effects}  & \textbf{High}                                                                                        & \textbf{Medium}                                                                                 & \textbf{Low}                                                                                                          \\ \hline
General Components & \begin{tabular}[c]{@{}l@{}}Penetration, major\\ deformation, part\\ detachment\end{tabular} & \begin{tabular}[c]{@{}l@{}}No penetration\\ but limited\\ deformation\end{tabular}     & Only dents or scratches                                      \\ \hline
Windshield         & \begin{tabular}[c]{@{}l@{}}Penetration or total\\ loss of\\ visibility\end{tabular}         & \begin{tabular}[c]{@{}l@{}}No Penetration,\\ partial loss\\ of visibility\end{tabular} & \begin{tabular}[c]{@{}l@{}}No or limited damage,\\ Nonsignificant loss\\ of external visibility\end{tabular} \\ \hline
\end{tabular}
\label{tbl:iea}
\end{table}

\begin{figure}[h!]
	\centering
        \includegraphics[width=0.425\textwidth]{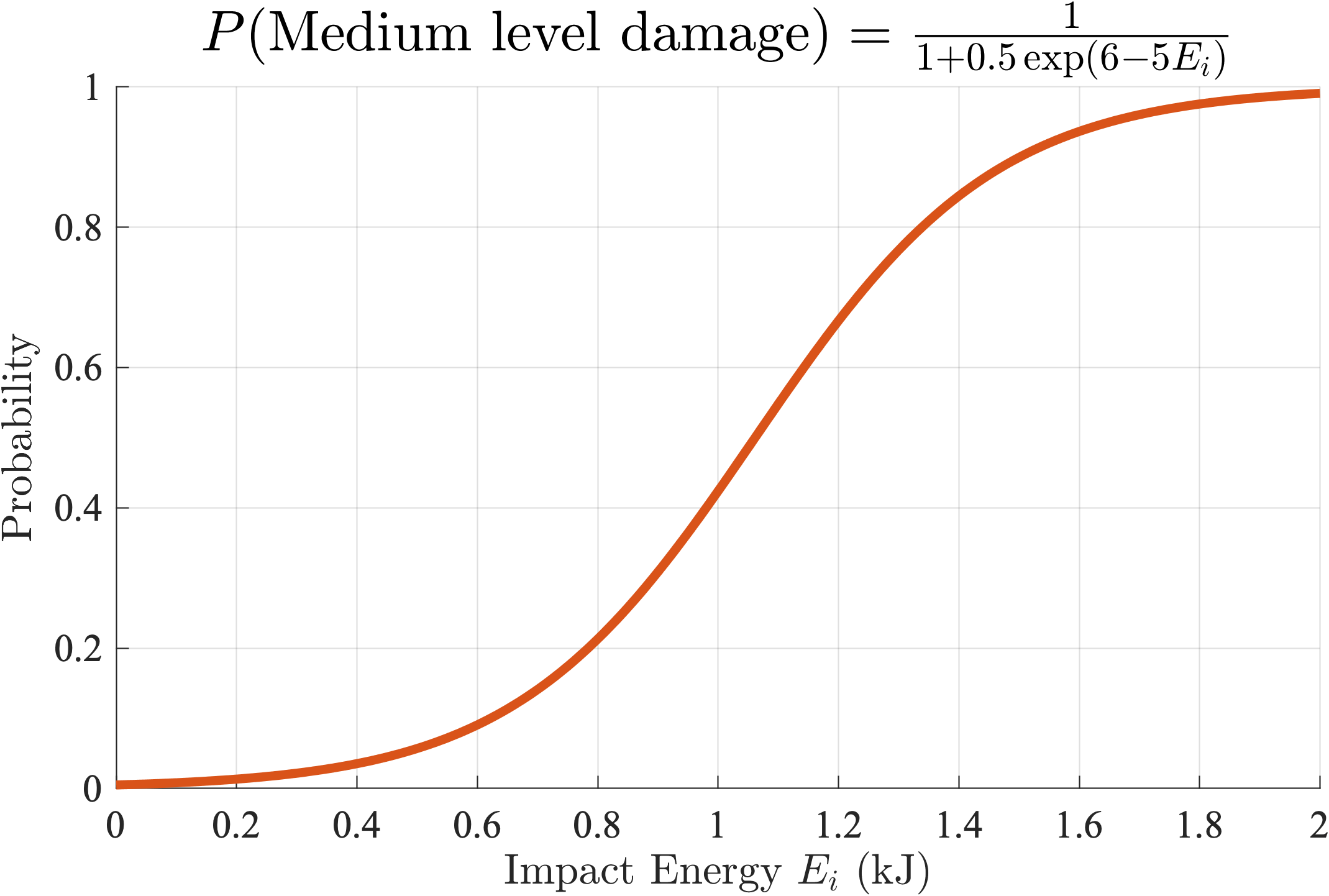}
	\caption{The probability of medium level damage to the car windshield vs. impact energy}
	\label{fig:Mdamage}
\end{figure}

By referring to results and model forms from \citep{che2022preliminary} and \citep{lee2019impact}, we use a sigmoid function to model the probability that a UAV crash will cause medium level damage to a car windshield, assuming that the vehicle speed is 50 km/h (31 mi/h) in an urban setting and that the impact angle is 90 degrees. The ground vehicle damage model is given by
\begin{equation}
    P(\text{Medium level damage}) = \frac{1}{1+0.5\exp\left(6-5 E_i\right)}
\end{equation}
where $E_i$ is the impact energy (KE in kJ) of the UAV. This ground vehicle damage model is based on the impact energy of the UAV, while some works~ (e.g., \citep{lee2019impact}) also suggested the use of BC adjusted to car windshield. Figure~\ref{fig:Mdamage} displays the shape of this function within 0 to 2 kJ. For example, a multi-rotor UAV with a weight of 2 kg and an impact velocity 40 m/s results in an impact KE of 1.6 kJ, which has a probability of 0.937 to cause medium level damage to a car windshield under this representative condition. 

\subsection{3D Urban Model}

This sub-model aims to develop a 3D CAD model for a complex real-world urban environment. Here, the preparation of a 3D urban model involves two steps. The first step is to extract urban information in the form of geospatial data. In this work, we obtain the general building and terrain information of a representative city from an online platform CADMAPPER. CADMAPPER is a tool commonly utilized by architects, urban planners, and designers to create 3D CAD model of a city's terrain, buildings, roads, etc. CADMAPPER can transform data from public sources such as OpenStreetMap~\citep{hakley2008openstreet}, NASA, and USGS into organized CAD files. In the second step, the urban information from CADMAPPER is further processed by the software Autodesk 3ds Max. This step further modifies the model format and refines the 3D urban model for better usage and visualization purposes.

\begin{figure}[h!]
     \centering
     \begin{subfigure}[b]{0.475\textwidth}
         \centering
         \includegraphics[width=\textwidth]{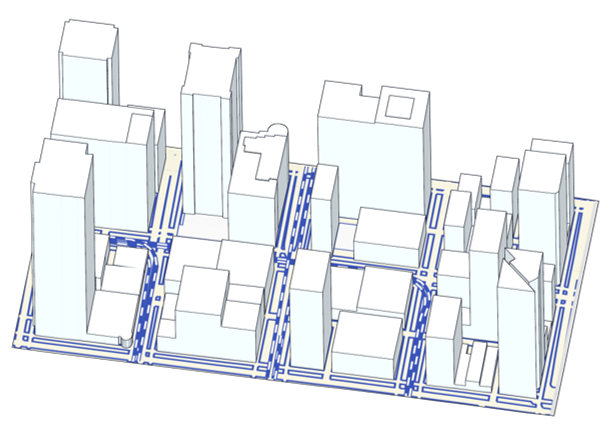}
         \caption{3D axonometric view from CADMAPPER}
         \label{fig:UrbanModel1}
     \end{subfigure}
     \hspace{0.8cm}
     \begin{subfigure}[b]{0.425\textwidth}
         \centering
         \includegraphics[width=\textwidth]{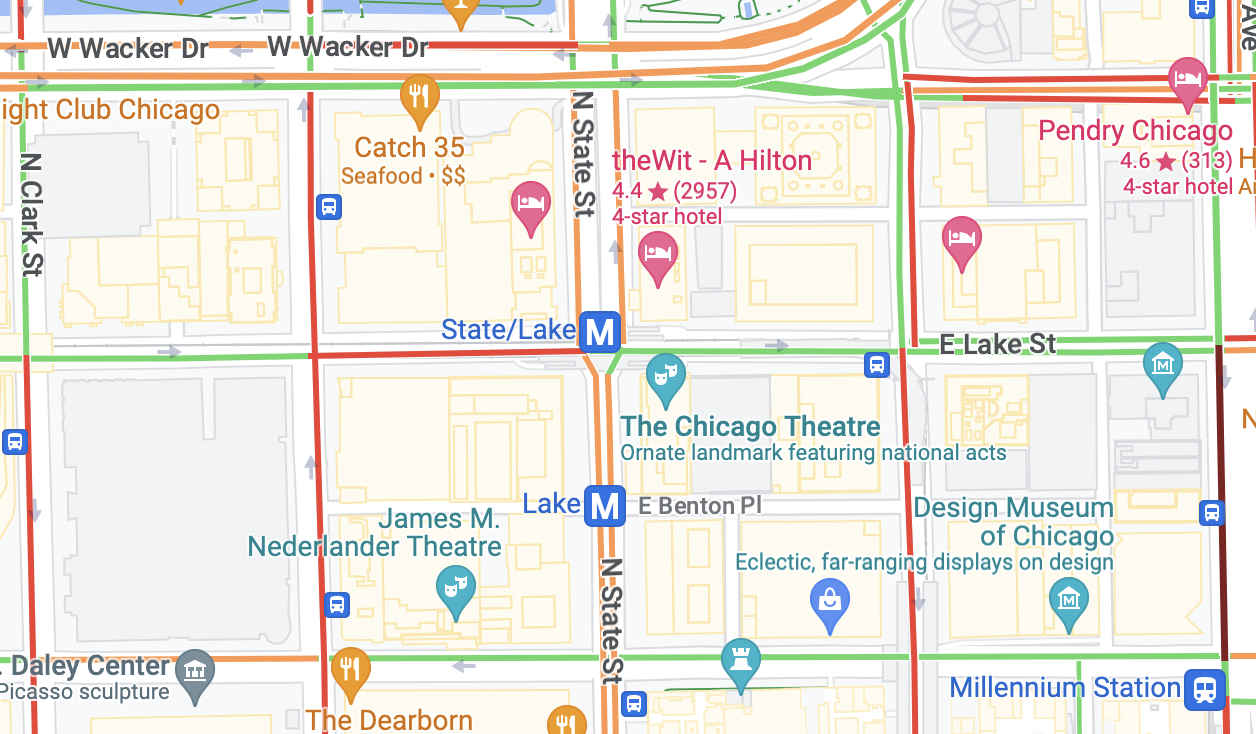}
         \vspace*{0.2mm}
         \caption{2D traffic view from Google Maps}
     \end{subfigure}
        \caption{Views of the representative 3D urban model used in this study}
        \label{fig:UrbanModel}
\end{figure}

Although the proposed approach also applies to suburban and rural environments, in this study we choose a complex urban scene for demonstration. Our case study focuses on a downtown urban model with tall and high-density buildings. Figure~\ref{fig:UrbanModel} displays the 3D urban model used in this study, which is from the Chicago downtown area. The left plot of Figure~\ref{fig:UrbanModel} shows the 3D axonometric view from CADMAPPER; the right plot of Figure~\ref{fig:UrbanModel} shows the 2D traffic view of the same region from Google Maps. The model is part of the vibrant and famous Chicago Loop area, a neighborhood that is comprised of high-rises and a combination of various facilities commonly found in a city. The dimensions of the model are (approximately) 500 m $\times$ 250 m $\times$ 250 m. Although our method is scalable to larger urban models (as demonstrated in Section~\ref{sec:case}), this model size is the most appropriate for data visualization and comparison. On the distribution of buildings, this model has more high-rises on the periphery and less height and density in the centre, which is beneficial for the effective visualizations of the 3D virtual risk terrains.

\subsection{Exposure Model}

The exposure model estimates the probability of the presence of an EoV at time $t$ and location $\bold{p}$ in the evaluation domain. The EoV involved in a UAS operation has three classifications~\citep{clothier2018modelling}. First parties are individuals and assets that are directly involved in the operation; second parties are ones that are not directly involved in the operation but gain direct benefits from its usage; third parties are ones that are neither involved in, nor derive any direct benefit from the operation of the UAS. This study (and exposure model) focuses on third parties, which mainly includes pedestrians and vehicles on the ground. The exposure model takes into account the population density ($P$), the vehicle density ($V$), and the temporal effect ($T$). A general mathematical representation of the exposure model is given by:
\begin{equation}
    E \left(\bold{p}, t\right) = P(\bold{p}) T_p(t) + V(\bold{p}) T_v(t)
\end{equation}
where:
\begin{itemize}
    \item $E \left(\bold{p}, t\right)$ is the exposure at location $\bold{p}$ and time $t$.
    \item $P(\bold{p})$ is the base population (or crowd, pedestrian) density at location $\bold{p}$.
    \item $T_p(t)$ is a temporal factor for population density at time $t$, derived from time series data.
    \item $V(\bold{p})$ is the base vehicle density at location $\bold{p}$.
    \item $T_v(t)$ is a temporal factor for vehicle density at time $t$, derived from time series data.
\end{itemize}

The temporal factor represents the relative density at a specific time point compared to the base density. For example, if the temporal factor for population density at 2 pm is 1.5, it means that the population density at 2 pm is 50\% higher than the base population density. In our case study, we consider that people and vehicles have separate spaces (sidewalks and roads) in an urban environment. While most related works in the literature assumed uniform exposure models within a specific geographic area in a city, in our work we integrate a comprehensive exposure model to capture the spatial and temporal variations in the density of pedestrians and vehicles in a complex urban environment. 

Because our exposure model needs to operate at a very fine-scale granular level (e.g., at the street level distribution of people and vehicles), the development of an accurate spatial and temporal model requires finer data and models on the complex dynamics and fluctuations of mobility in urban environments. Fortunately, in this big data era, data-driven approaches can provide powerful means of deriving high-fidelity exposure models. Urban transportation researchers have utilized data sources such as mobile phone data, traffic data, and techniques such as statistical models, deep learning, and computer vision to obtain accurate estimations of people and vehicle densities in a city. In this work, we are particularly interested in modeling and comparing the 3D virtual risk terrains at the following three representative times of a weekday:
\begin{itemize}
    \item \textsl{Midday (12 pm)}: This time represents the middle of the day when many people are taking their lunch breaks. Moderate densities of people and vehicles are expected on the ground.
    \item \textsl{Evening rush hour (5 pm)}: This time is selected to represent the peak evening commuting hours, when people are leaving their workplaces or schools to return home. Just as with the morning commute, public transportation hubs and office-dense areas are expected to have high densities of people and vehicles on the ground. 
    \item \textsl{Night time (10 pm)}: This time is chosen to represent the late evening activities. At this time most people stay indoors and the ground densities of people and vehicles are at the lowest point of the day. We don't consider midnight and early in the morning because there is hardly any business activity during those hours.
\end{itemize}

For the Chicago urban area in Figure~\ref{fig:UrbanModel}, there is currently no publicly available data with the required spatiotemporal granularity. Precise pedestrian/crowd density data is hardly publicly available, because of the high cost of data collection and processing, and concerns about privacy and public security. Traffic data, on the other hand, is more readily available, although the granularity of many publicly available datasets is still inadequate for our analysis. Here, we utilize a combination of publicly available Chicago transportation datasets and the latest research outcomes on urban analytics to estimate the exposure model for this Chicago neighborhood case. We conduct separate analyses for the pedestrian density and the vehicle density.

The estimation of pedestrian/crowd density in urban areas has been advanced through mobile phone data analytics~\citep{huo2021short,weppner2011collaborative,weppner2013bluetooth,fu2021spatial} and deep learning~\citep{fu2015fast,ding2020crowd,jiang2021deepcrowd,zhu2020crowd}. We first use the existing data and results to estimate $P(\bold{p})$, the base pedestrian density on the sidewalks of the Chicago downtown area. In this work, the base pedestrian density is set as the density at the peak hour of the day (5 pm). Using the computer vision literature~\citep{fu2015fast,weppner2011collaborative,weppner2013bluetooth}, we first establish the following five classes of pedestrian density: very low ($<$ 0.05 people/m$^2$), low (0.05--0.1 people/m$^2$), moderate (0.1--0.2 people/m$^2$), high (0.2--0.3 people/m$^2$), and very high ($>$ 0.3 people/m$^2$). The very high pedestrian density ($>$ 0.3 people/m$^2$, up to 2 people/m$^2$) usually occurs during crowd gathering activities such as a square rally. The high pedestrian density can also occur as a daily routine on the business streets of a densely populated city, such as some megacities in Asia, during rush hours. For the Chicago downtown area (and many similar Central Business District (CBD) areas in North America), the average pedestrian density during rush hours is at the moderate level (0.1--0.2 people/m$^2$), excluding special events. Hence, we use 0.15 people/m$^2$ as the estimation of pedestrian density on the sidewalks of the Chicago downtown area during evening rush hour. For the estimation of $T_p(t)$, the temporal factor for population density at time $t$, it can be derived from several previous works in the literature~\citep{huo2021short,fu2021spatial,jiang2021deepcrowd}. These works reported time series data which uncover the trends of pedestrian density at different times of the day. Since we set the pedestrian density at 5 pm as the base, we have $T_p(5~\text{pm}) = 1$. For the temporal factor at midday and night time, using data from \citep{huo2021short} and \citep{jiang2021deepcrowd}, we conclude that $T_p(12~\text{pm}) = 0.5$ and $T_p(10~\text{pm}) = 0.1$ are valid projections.

\begin{figure}[h!]
	\centering
        \includegraphics[width=0.925\textwidth]{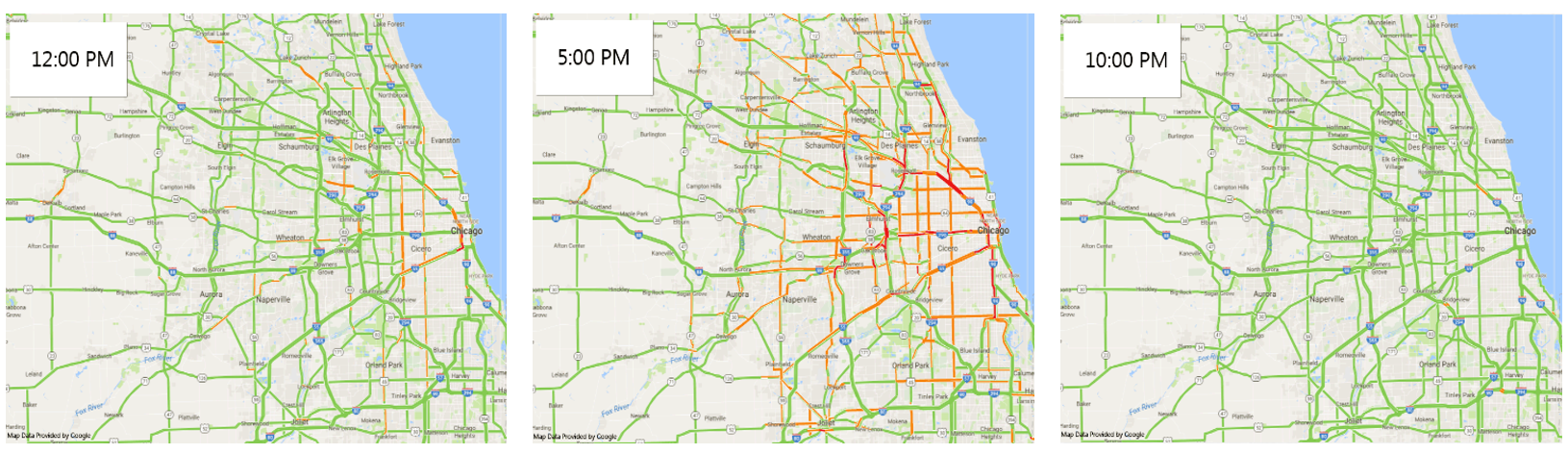}
	\caption{Visualization of Chicago traffic patterns by time of the day: 12 pm (left), 5 pm (middle), 10 pm (right) (Sources: data from U.S. DOT HPMS Public Release, visualization from Illinois Vehicle Auto Insurance)}
	\label{fig:ChicagoTraffic}
\end{figure}

The estimation of traffic density is enabled by similar data types and techniques. Compared to pedestrian density, traffic density data is more publicly available. For example, the U.S. Department of Transportation (DOT) Highway Performance Monitoring System (HPMS) database contains the Chicago area traffic data. For example, Figure~\ref{fig:ChicagoTraffic} is a set of visualizations which display the Chicago area traffic patterns at 12 pm, 5 pm, and 10 pm, respectively. However, at the current stage, the publicly available traffic data does not adequately provide the desired traffic density information. We therefore estimate the detailed traffic density in our selected area using a combination of publicly available traffic data and urban analytics results in the literature. On the vehicle density $V(\bold{p})$, a commonly used measure in the literature is the number of vehicles per unit length (e.g., 100 m, 1000 m)~\citep{zeroual2019road,raj2016application,sakai2019traffic} or the number of vehicles per unit length per lane~\citep{lim2022spatiotemporal}. Based on vehicles/100 m/lane (veh/100m/lane), we again establish five classes of vehicle density: very low ($<$ 2 veh/100m/lane), low (2--5 veh/100m/lane), moderate (5--10 veh/100m/lane), high (10--15 veh/100m/lane), and very high ($>$ 15 veh/100m/lane). For the downtown area of Chicago, 10 veh/100m/lane is a valid estimation of vehicle density at the peak hour (5 pm). With the urban car lane width 10 ft (3 m) and average windshield projection area 14 ft$^2$ (1.28 m$^2$), we have the base vehicle density $V(\bold{p}) = 10 \cdot 1.28~\text{m}^2/100~\text{m} \cdot 3~\text{m} = 0.04$ counts/m$^2$. For the temporal factor, we use time series data in the literature~\citep{po2019sensors,raj2016application,zeroual2019road} to estabilish the following estimates: $T_v(5~\text{pm}) = 1$, $T_v(12~\text{pm}) = 0.6$ and $T_v(10~\text{pm}) = 0.2$.

\section{Case Study}\label{sec:case}

\subsection{Study Set-up}

In this section we conduct a comprehensive case study and generate prototypes of the 3D virtual risk terrains for the Chicago downtown example. Table~\ref{tbl:case} is a summary of the settings and parameters used in the case study. Overall, we are simulating cargo delivery operations in a complex urban environment, enabled by a medium sized multi-rotor UAV. The failure type is LOC, a catastrophic failure that is unrecoverable and has little to no  controllability on the impact location. Therefore, the resulting 3D virtual risk terrains are conservative and represent the `worst case scenarios' on recovery and impact location. Both Gaussian and Rayleigh impact location models are considered, and their results are compared. On the type of EoV, we model the ground risk of both pedestrians and vehicles. Average human body characteristics are used, although a more conservative study could model more vulnerable populations. On the reference human injury level, our investigation centers around AIS level 3, which is more conservative because most relevant works in the literature have used the probability of fatality in the harm model. Likewise, we use Medium level damage under IEA as the threshold for vehicle windshield damage, which represents a probable risk for causing a traffic accident. 

\begin{table}[htbp]
\centering
\caption{Factors and settings of the case study}
\label{tbl:case}
\begin{tabular}{l|l}
\hline
\textbf{Factors} & \textbf{Details}                                        \\ \hline
UAV type         &  Multi-rotor configuration        \\ 
UAV weight $m$      &  25 kg\\
UAV flying speed &  10 m/s\\
UAV drag coefficient $C_D$    &   1.8 (estimated via method in~\citep{hattenberger2023evaluation})\\
UAV cross-sectional area size $S$    &   0.2 m$^2$\\
UAV characteristic diameter $D$    &  50 cm\\
UAV failure type    &  Loss of Control (LOC) \\
UAV contingency equipment    &  With and without parachute \\ 
Impact location models    &  Gaussian model, Rayleigh model \\ \hline
EoVs and their locations    &  Pedestrians and vehicles on the ground\\
Mass of the struck body $M$    &  70 kg\\
Human body wall thickness coefficient $k$    &  0.652\\ 
Reference human injury level    &  AIS Level 3\\ 
Reference vehicle damage level    &  Medium -- no penetration, partial loss of visibility\\ \hline
Air density $\rho$    & 1.225 kg/m$^3$\\
Standard gravity $g$    &  9.8 m/s$^2$\\
Weather (used in the noise part)   &  Standard day weather \\ \hline
Size of unit area around $\bold{p}$    & 4 m$^2$ (2 m $\times$ 2 m)\\
Density of $\bold{p}$ on the ground    & One every 2 meters\\
Size of individual risk terrain around $\bold{p}$    & 40 m $\times$ 40 m $\times$ 200 m\\
Size of the urban model    & 500 m $\times$ 250 m $\times$ 250 m\\
Times of the day    &  12 pm, 5 pm, 10 pm\\
\hline
\end{tabular}
\end{table}

\begin{figure}[h!]
     \centering
     \begin{subfigure}[b]{0.322\textwidth}
         \centering
         \includegraphics[width=\textwidth]{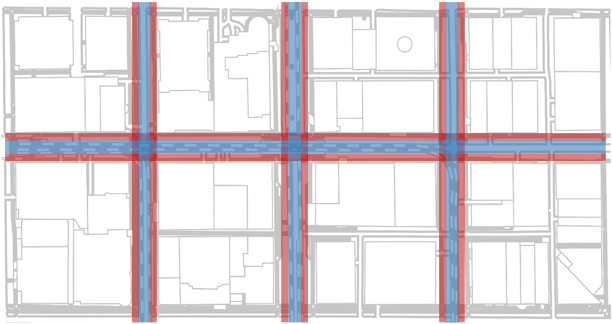}
         \caption{12 pm}
     \end{subfigure}
     \hspace{0.01cm}
     \begin{subfigure}[b]{0.322\textwidth}
         \centering
         \includegraphics[width=\textwidth]{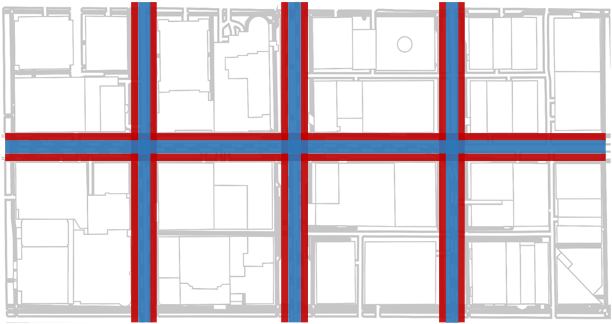}
         \caption{5 pm}
     \end{subfigure}
     \hspace{0.01cm}
     \begin{subfigure}[b]{0.322\textwidth}
         \centering
         \includegraphics[width=\textwidth]{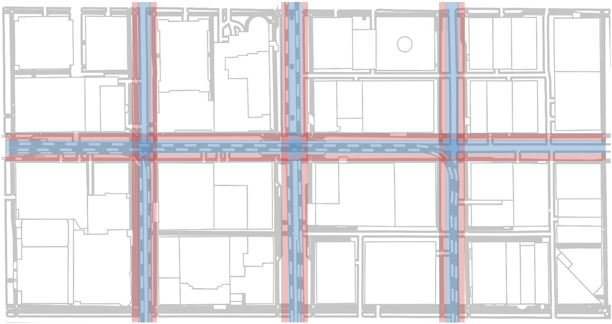}
         \caption{10 pm}
     \end{subfigure}
        \caption{Spatial distribution and density of pedestrians (red) and vehicles (blue) on the ground, at 12 pm (left), 5 pm (middle), and 10 pm (right)}
        \label{fig:densities}
\end{figure}

For a specific location $\bold{p}$ on the ground, we use an area of 4 m$^2$ (2 m $\times$ 2 m) around $\bold{p}$ as its impact range. Within the available ground space of the urban model in Figure~\ref{fig:UrbanModel1}, we place a grid of $\bold{p}$ with a density of one point every 2 meters. Therefore, the evaluation range is continuous on the ground. For the range of individual risk terrain around a certain $\bold{p}$, we consider a cube with size 40 m $\times$ 40 m $\times$ 200 m, which, according to our analysis, is a large enough volume to cover locations in the air that have notable risks to $\bold{p}$. Although the current regulations have set a maximum altitude 400 ft (122 m) for the operations of similar UAS in an urban space, we compute the risk terrain for up to 200 m for a more sufficient exploration. Each ground location $\bold{p}$ represents either pedestrian or vehicle, but not both. Figure~\ref{fig:densities} is a visualization of the exposure model that shows the distributions of pedestrians and vehicles in the urban model at the three representative times of the day. The background of each subfigure in Figure~\ref{fig:densities} is the overhead view of the urban model in Figure~\ref{fig:UrbanModel1}. In Figure~\ref{fig:densities}, the pedestrians (red) and vehicles (blue) appear on the sidewalks and roads (motor vehicle lanes) of the urban model, respectively. The intensity of color is an indicator of the density of an EoV at different times of the day. 

\subsection{Visualizations of Virtual Risk Terrains}

\begin{figure}[h!]
	\centering
        \includegraphics[width=0.325\textwidth]{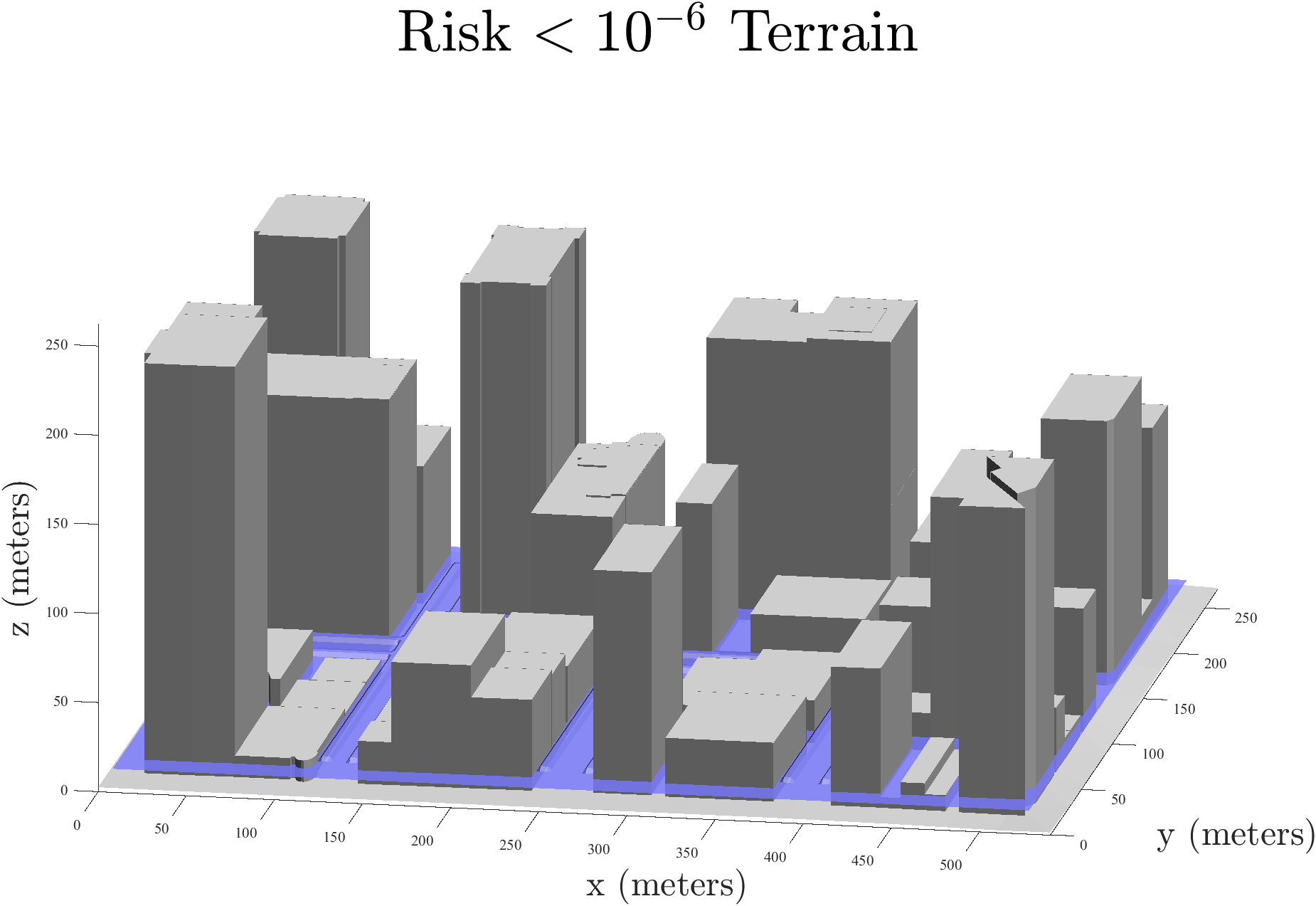}
        \includegraphics[width=0.325\textwidth]{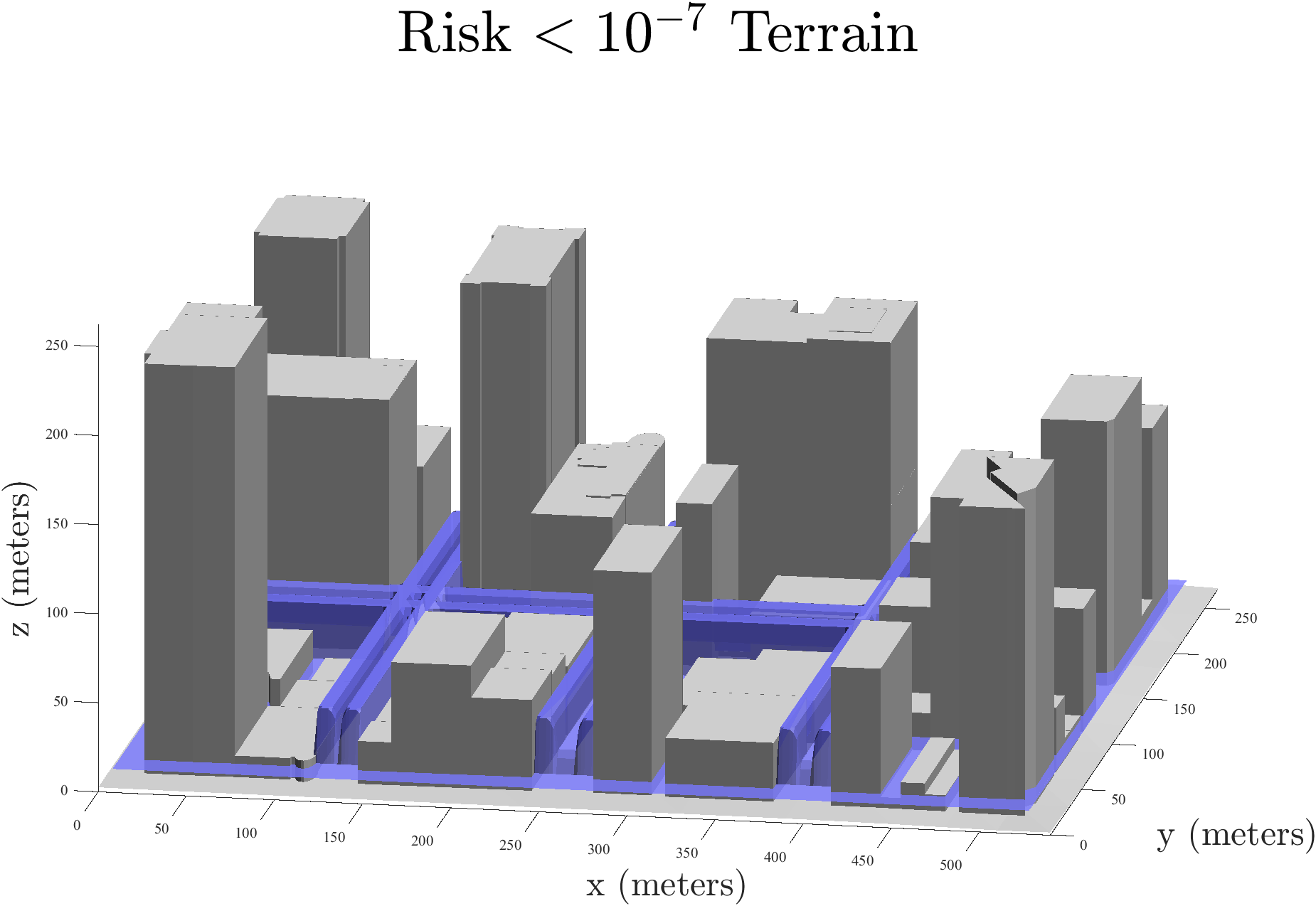}
        \includegraphics[width=0.325\textwidth]{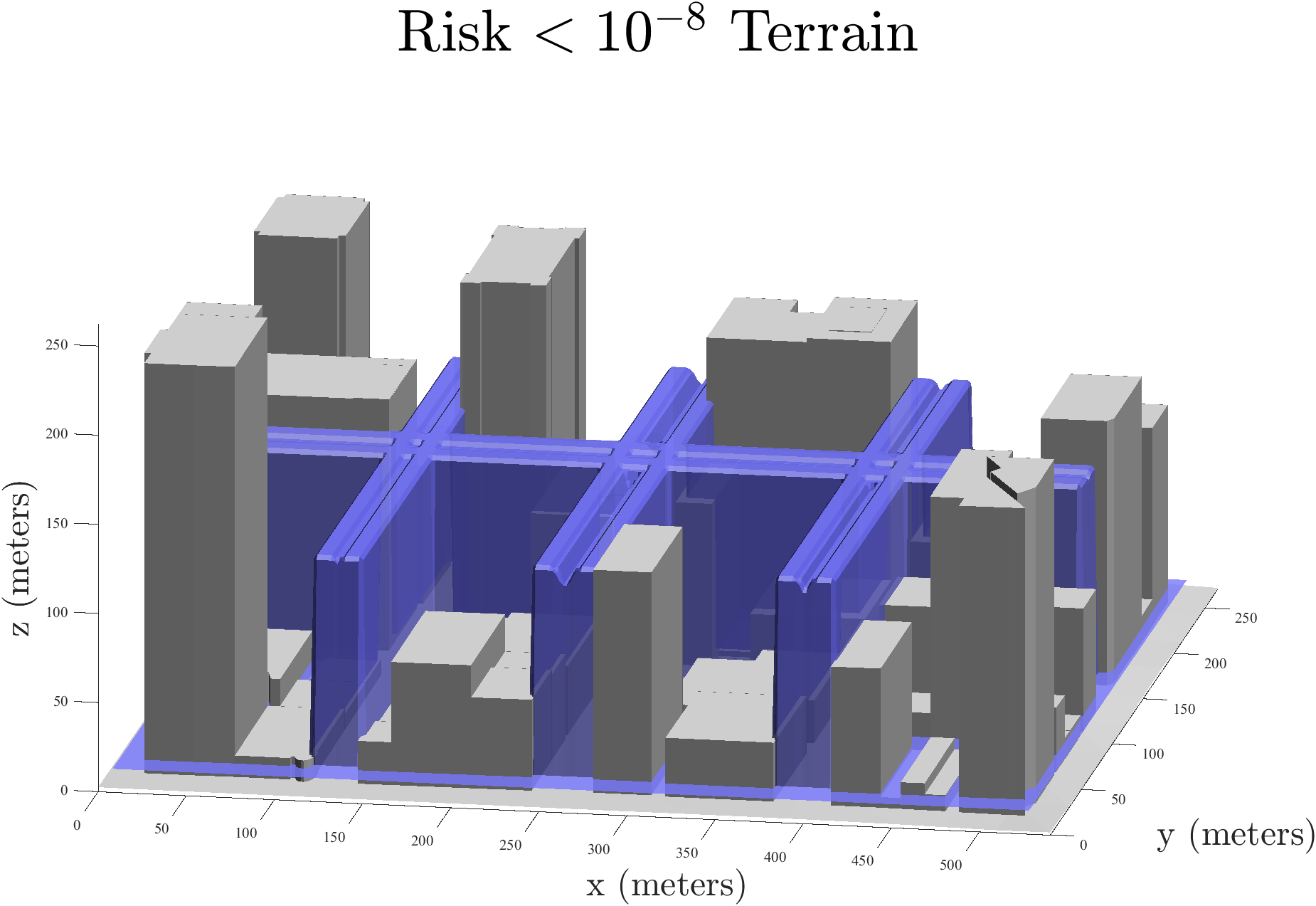}
	\caption{Virtual risk terrains at different risk levels: 10$^{-6}$ (left), 10$^{-7}$ (middle), and 10$^{-8}$ (right), under the failure rate 10$^{-5}$ and evening rush hour (5 pm).}
	\label{fig:terrains1}
\end{figure}

In a computational environment, we implement our integrated approach in Figure~\ref{fig:Approach} with every sub-model described in Section~\ref{sec:models} and apply the computations to the Chicago urban model. This subsection first displays visualizations of the 3D virtual risk under different parameter settings. Specifically, we are interested in investigating how the characteristics of the 3D virtual risk terrain change in relation to the risk requirement, UAS reliability level, time of the day, and impact location model. In the first group of visualizations, Figure~\ref{fig:terrains1} shows the 3D virtual risk terrains at three different risk levels: 10$^{-6}$, 10$^{-7}$, and 10$^{-8}$, assuming maximum risk settings in both failure model (10$^{-5}$/flight hour) and exposure model (evening rush hour). The interpretation of a virtual risk terrain is straightforward: to avoid a certain level of safety risk to crowd and vehicles on the ground, the UAS operation must avoid and fly above the corresponding virtual risk terrain. From the perspective of airspace management, the 3D virtual risk terrain clearly defines the ``no-fly'' zones, the space within the 3D surface, where the ground risk exceeds a certain threshold. In Figure~\ref{fig:terrains1}, we can observe that the height of the virtual risk terrain increases as the risk requirement becomes more stringent -- from 10$^{-6}$/flight hour to 10$^{-8}$/flight hour. This indicates that UAS must maintain a larger clearance distance (and altitude) from crowds and vehicles on the ground under more stringent safety regulations. In other words, with all other conditions held the same, more stringent safety policies will result in less space for UAS to operate in an urban environment. 

\begin{figure}[h!]
	\centering
        \includegraphics[width=0.325\textwidth]{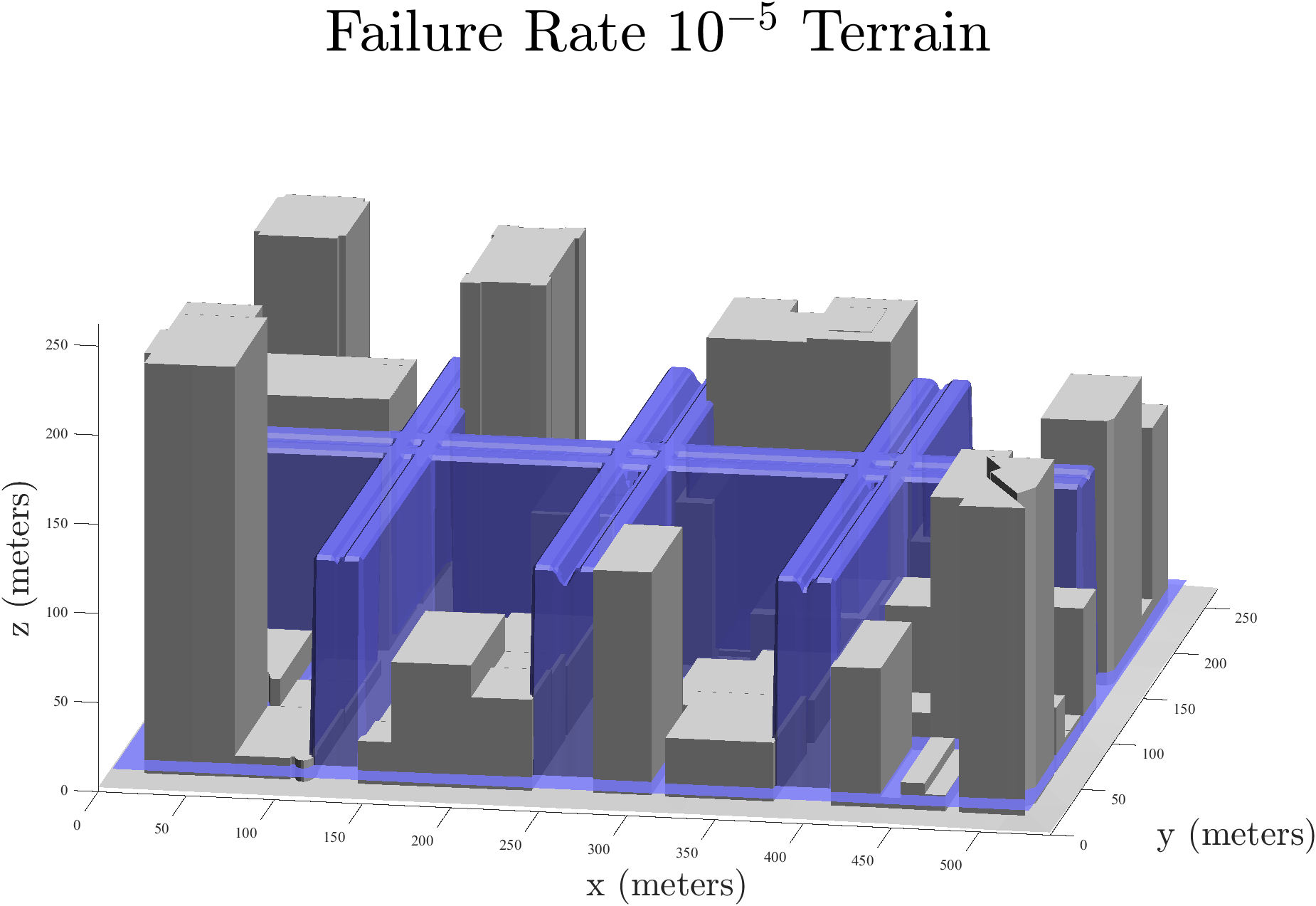}
        \includegraphics[width=0.325\textwidth]{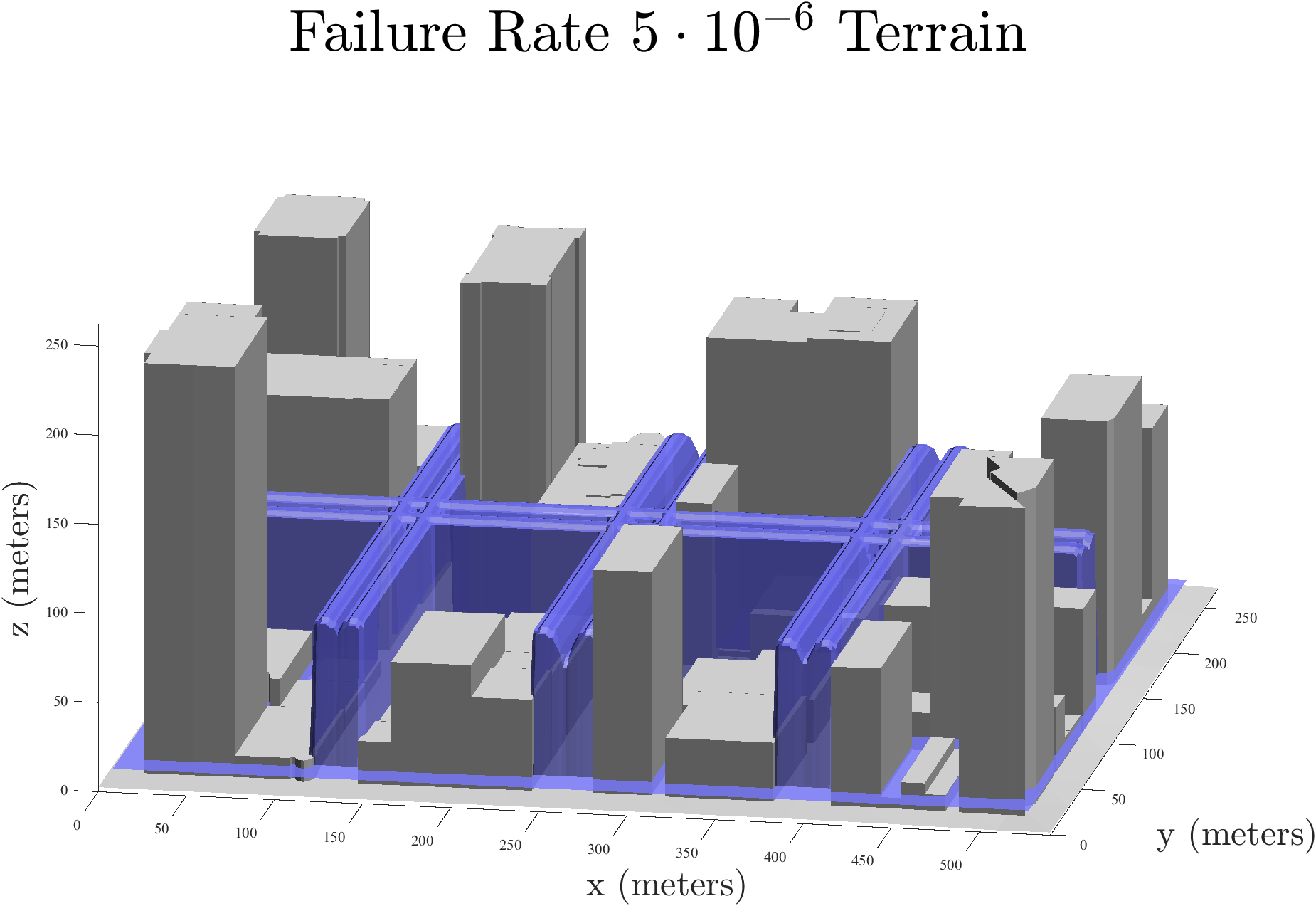}
        \includegraphics[width=0.325\textwidth]{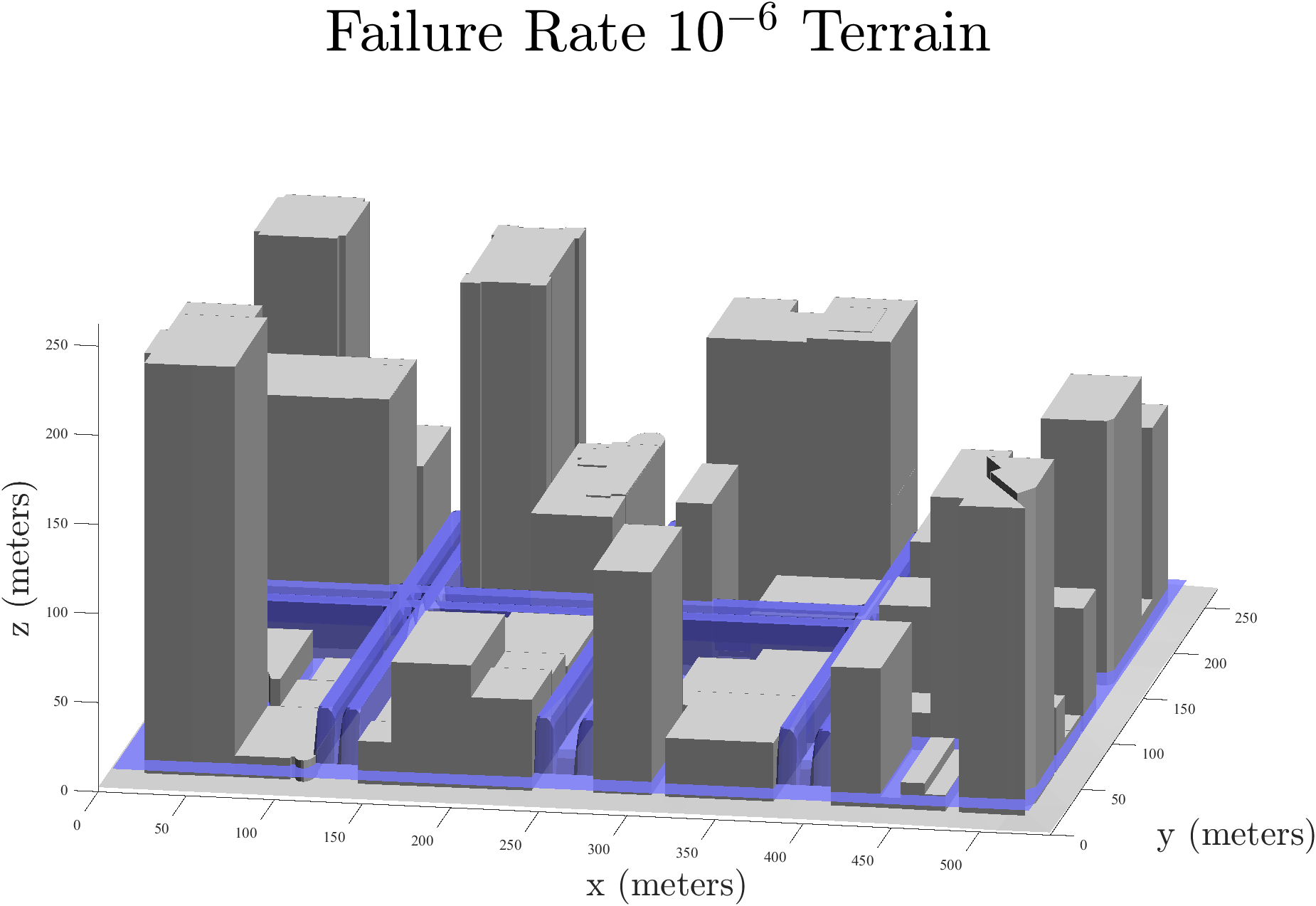}
	\caption{Virtual risk terrains at different failure rates: 10$^{-5}$ (left), $5 \cdot 10^{-6}$ (middle), and 10$^{-6}$ (right), under the risk level 10$^{-8}$ and evening rush hour (5 pm).}
	\label{fig:terrains2}
\end{figure}

\begin{figure}[h!]
	\centering
        \includegraphics[width=0.325\textwidth]{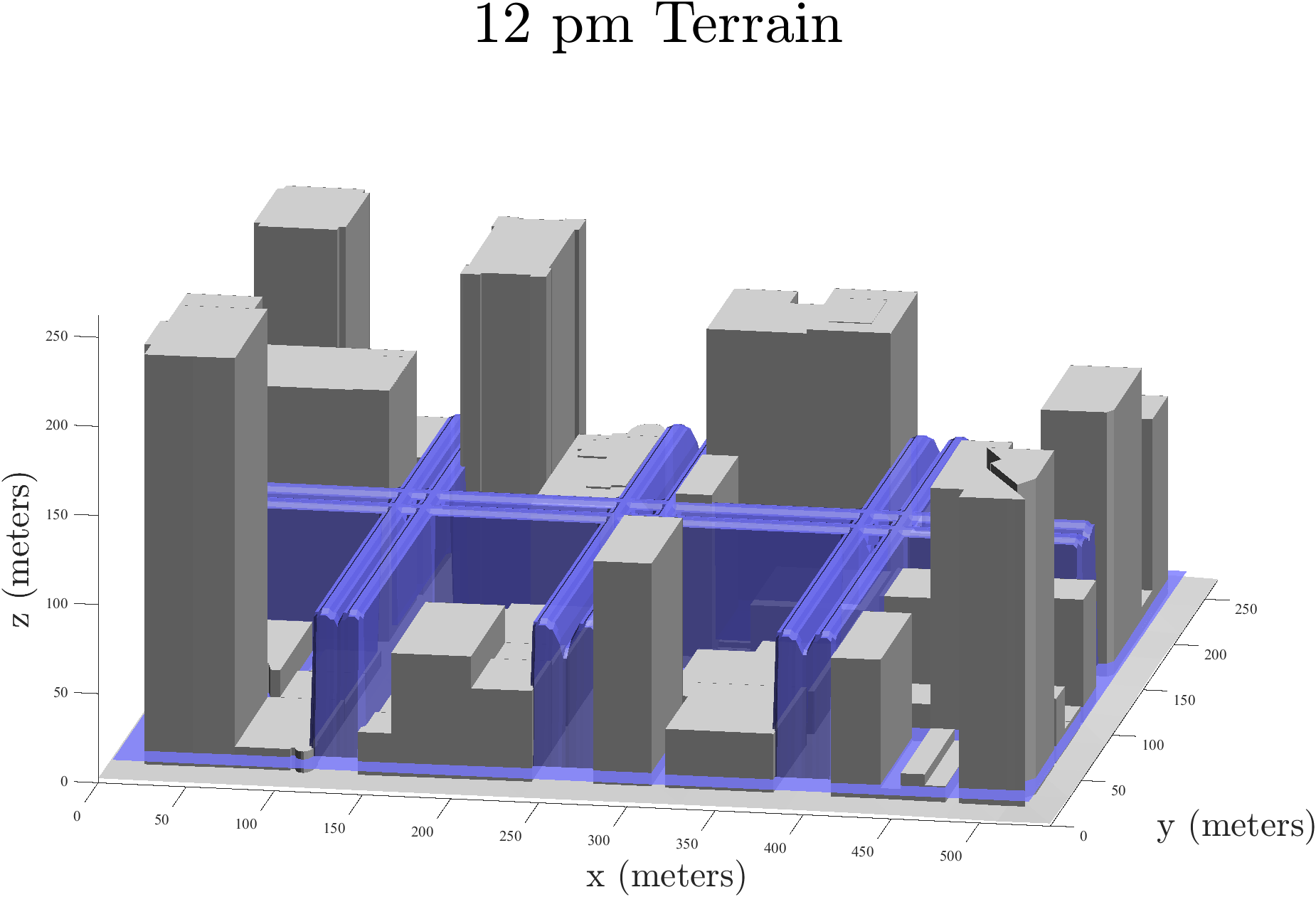}
        \includegraphics[width=0.325\textwidth]{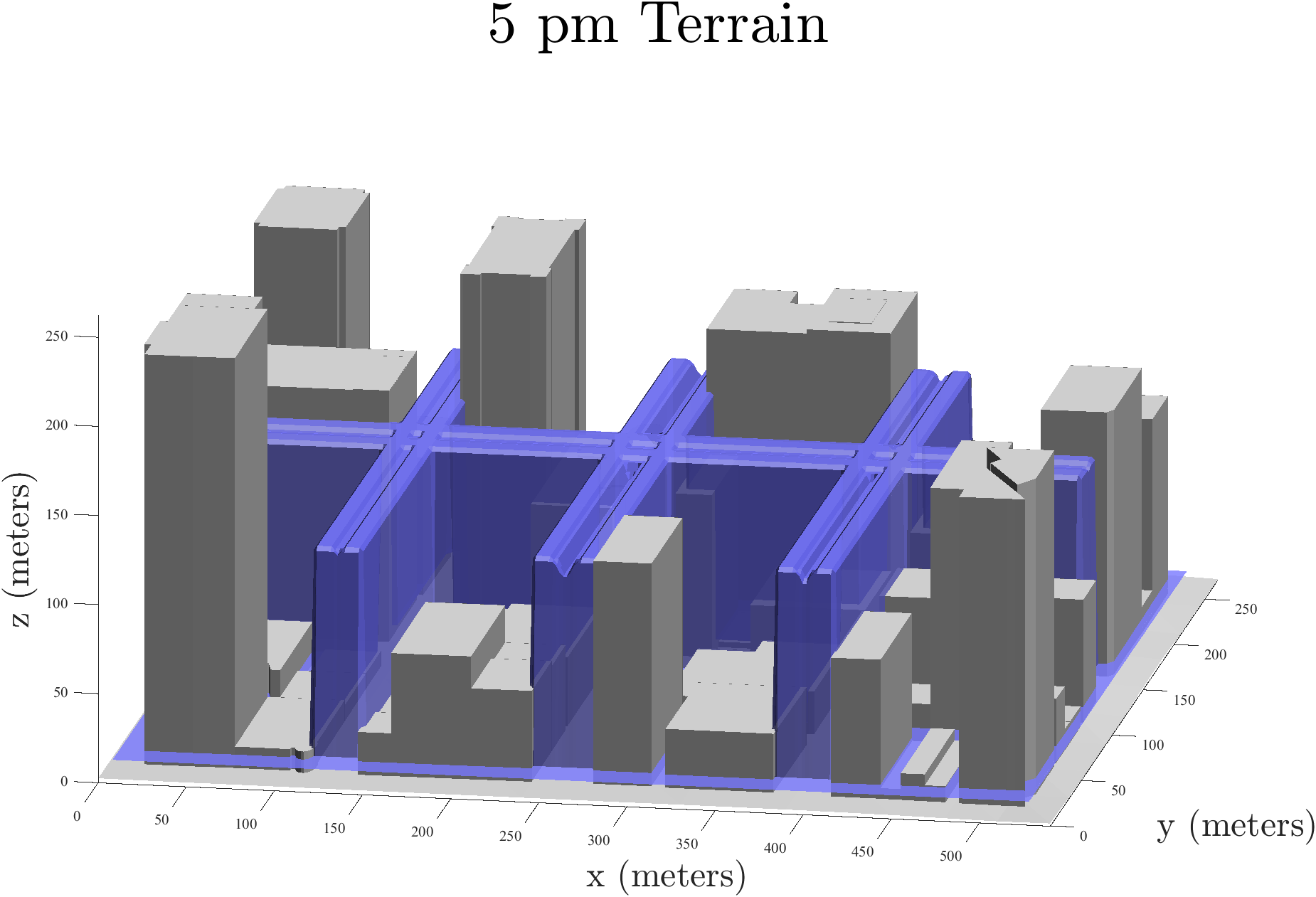}
        \includegraphics[width=0.325\textwidth]{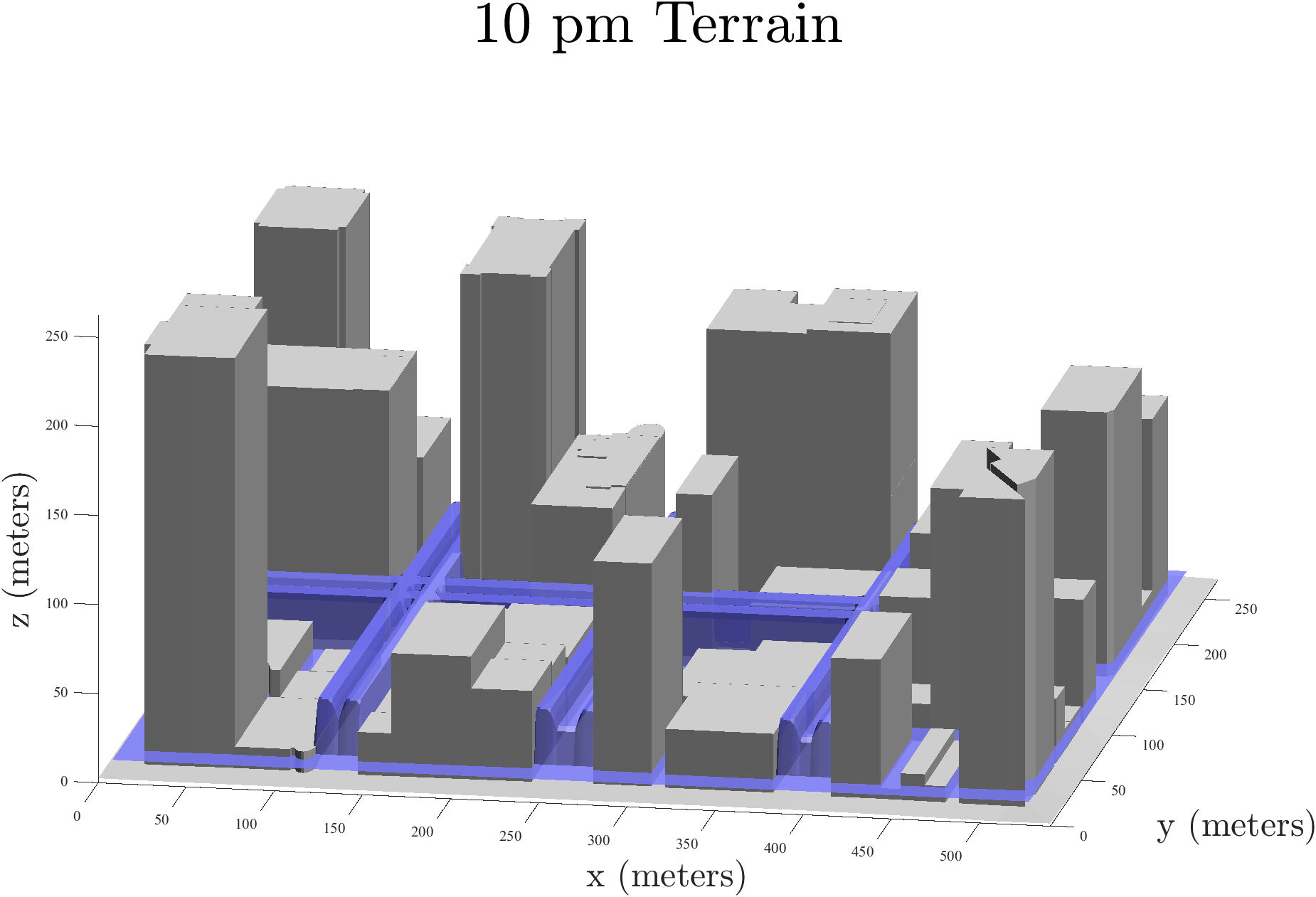}
	\caption{Virtual risk terrains at different times of the day: 12 pm (left), 5 pm (middle), and 10 pm (right), under the risk level 10$^{-8}$ and failure rate 10$^{-5}$.}
	\label{fig:terrains3}
\end{figure}

We next demonstrate how improved UAS reliability can affect the virtual risk terrain. In Figure~\ref{fig:terrains2}, while assuming the maximum risk settings for risk requirement (10$^{-8}$/flight hour) and exposure model (evening rush hour), we generate 3D virtual risk terrains for three different UAS failure rates: 10$^{-5}$/flight hour, $5 \cdot 10^{-6}$/flight hour, and 10$^{-6}$/flight hour. More airspace becomes available as the system failure rate decreases. Therefore, UAS can operate closer to crowds and vehicles on the ground when the system itself becomes more reliable. This feature allows policy makers to work backwards to derive the required UAS reliability level for a specific airspace design scheme. For example, one can answer the question: if UAS is allowed to fly at a minimum of 40 meters above pedestrians during evening rush hour, what would be the required reliability of the system? In the same format, Figure~\ref{fig:terrains3} explores how the virtual risk terrain varies with time of the day. Under the maximum risk settings in risk requirement (10$^{-8}$/flight hour) and failure model (10$^{-5}$/flight hour), the UAS can operate in a broader urban airspace and at a lower altitude when there are fewer people and vehicles on the ground. For better illustrative purposes, we choose to compare three settings for each of the risk requirements, UAS reliability levels, and times of the day. Similar studies can be conducted considering wider ranges of parameters and at finer granularity levels. Although many other factors can also play a critical role, risk requirement, UAS reliability level, and time of the day are potentially the three most significant factors in this decision making problem.

\begin{figure}[h!]
	\centering
        \includegraphics[width=0.325\textwidth]{Figures/Risk-6.png}
        \includegraphics[width=0.325\textwidth]{Figures/Risk-7.png}
        \includegraphics[width=0.325\textwidth]{Figures/Risk-8.png}\\
        \vspace{0.2cm}
        \includegraphics[width=0.325\textwidth]{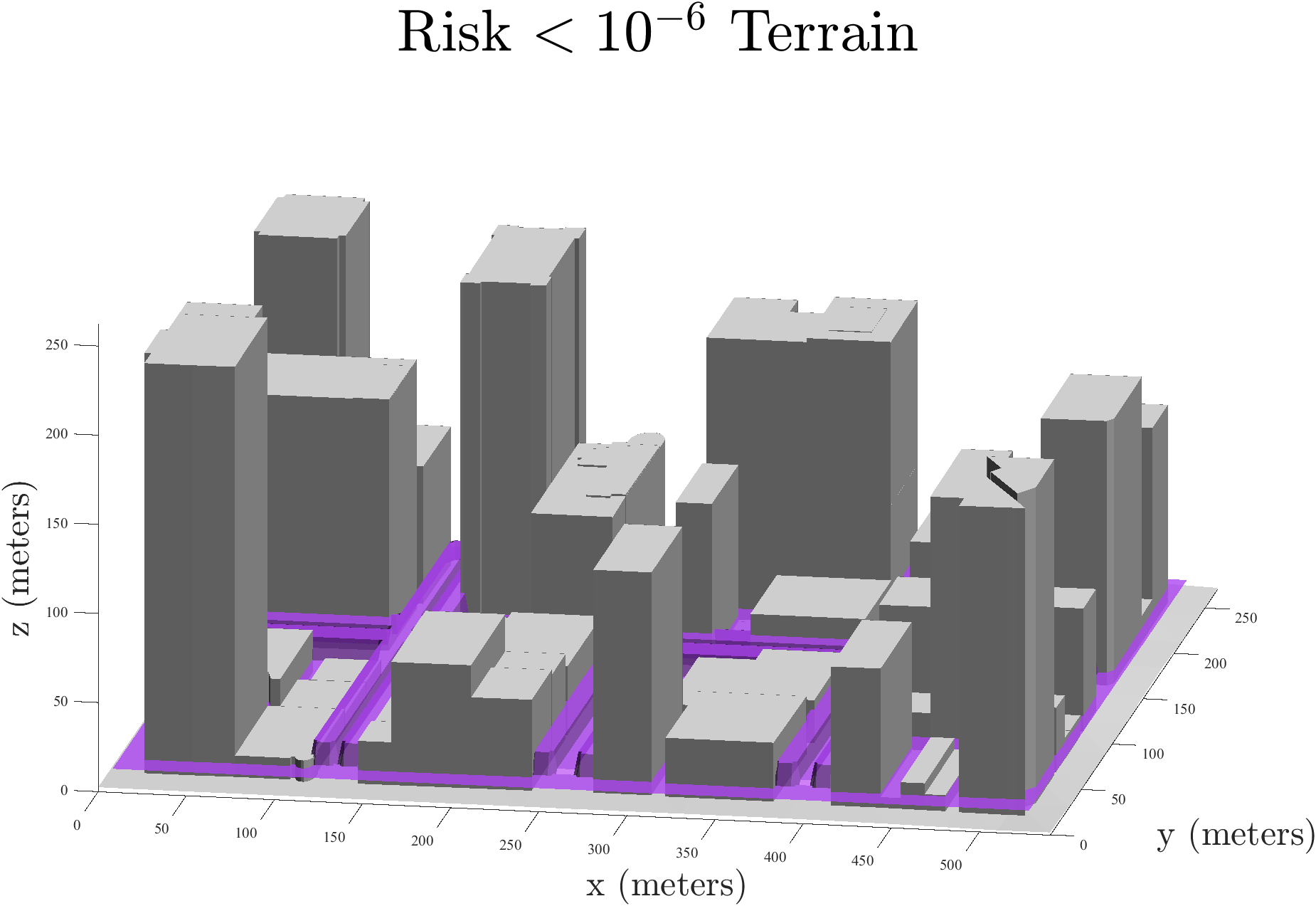}
        \includegraphics[width=0.325\textwidth]{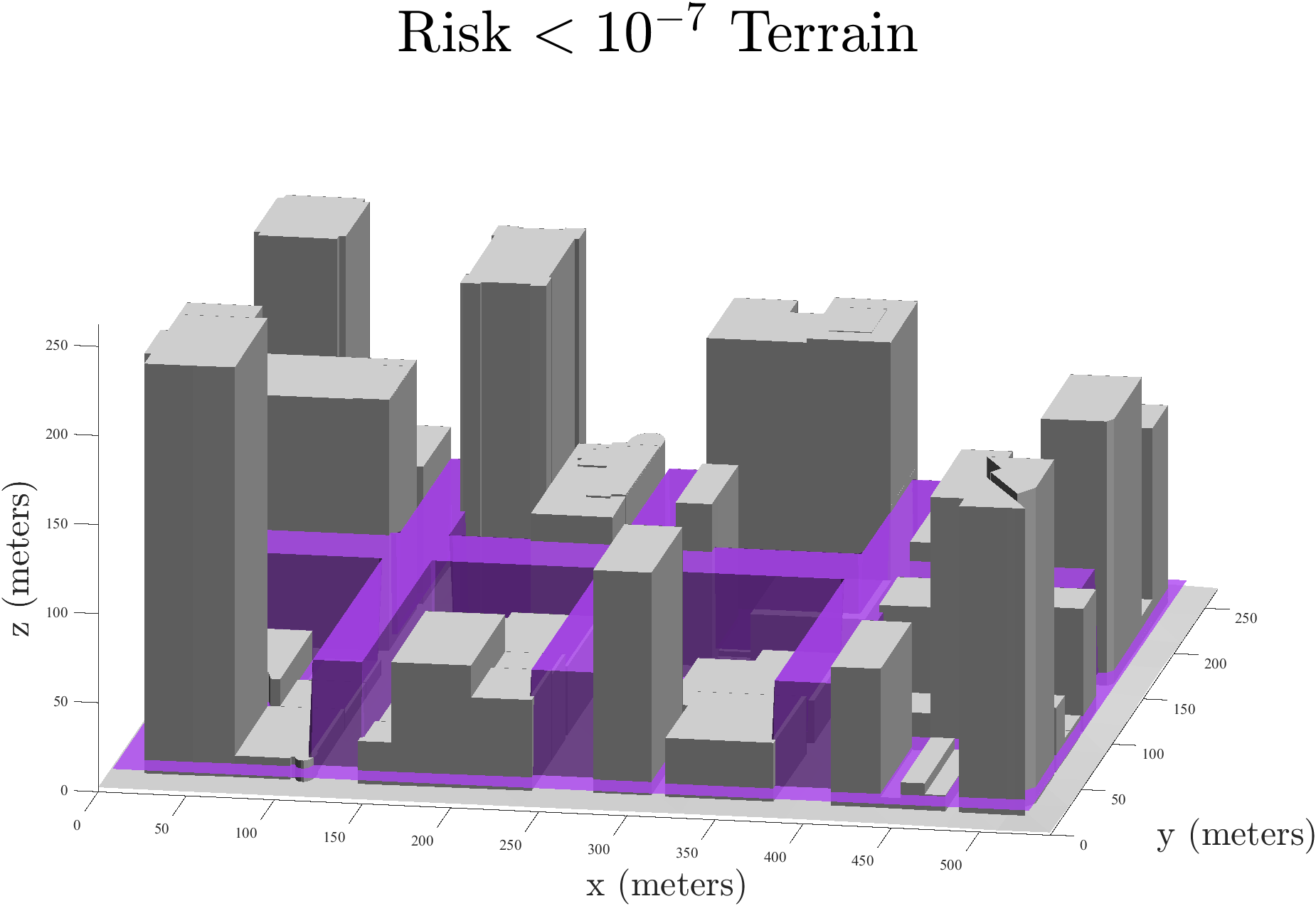}
        \includegraphics[width=0.325\textwidth]{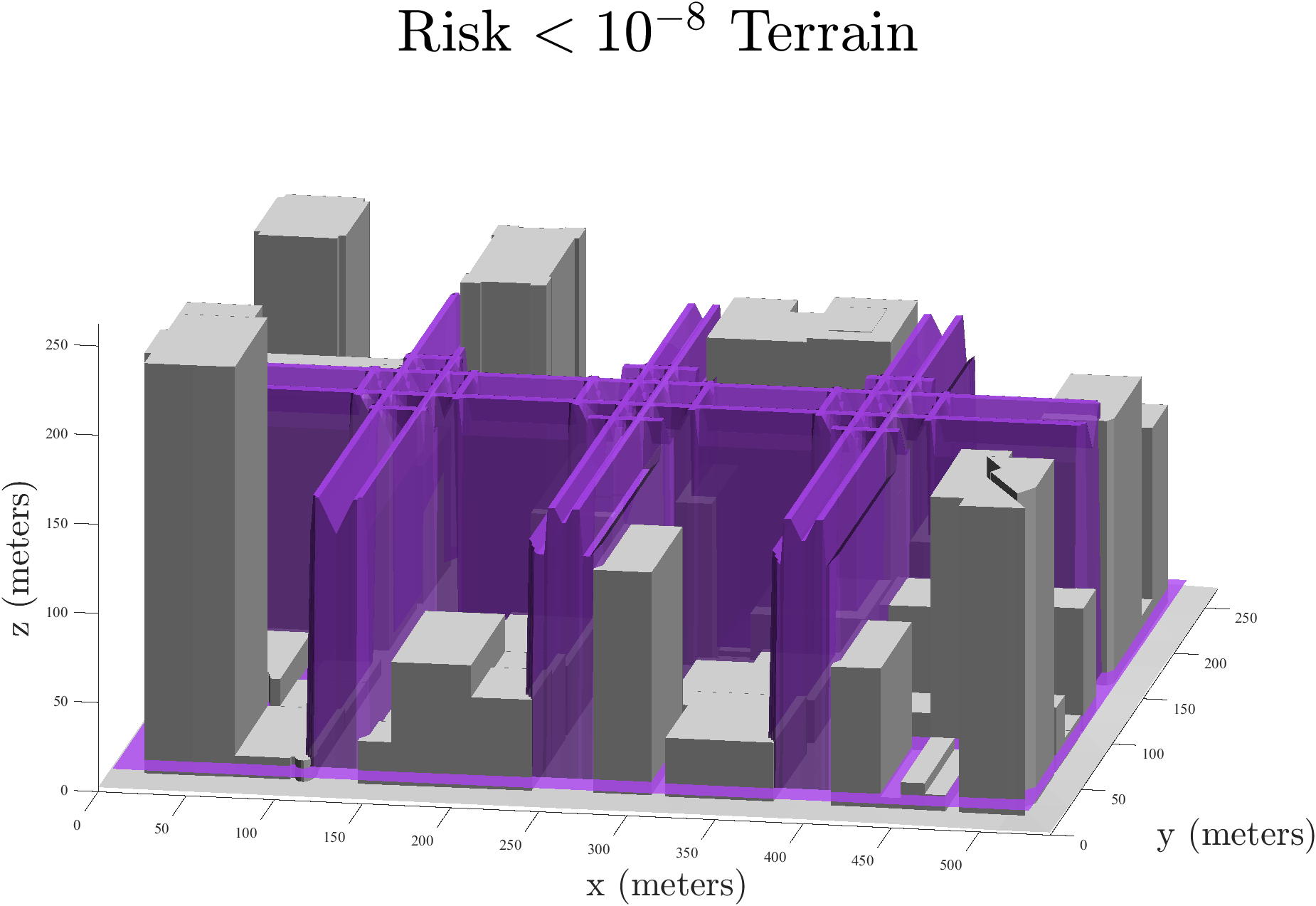}
	\caption{Comparison between virtual risk terrains under Gaussian (top) and Rayleigh (bottom) impact location models.}
	\label{fig:terrains4}
\end{figure}

In the last set of visual comparisons, we observe the differences between the Gaussian impact location model and the Rayleigh impact location model in Figure~\ref{fig:terrains4}. Even though the two impact location models make different assumptions and represent different impact location patterns, their resulting virtual risk terrains have certain features in common, due to the continuous distribution of pedestrians and vehicles on the ground. In Figure~\ref{fig:terrains4}, the overall trends and shapes of the virtual risk terrains are similar between the two impact location models. The two differences are the height and detailed shape of the virtual terrain. At all three risk levels, the virtual terrain under the Rayleigh impact location model is higher than its counterpart under the Gaussian impact location model. And with the Rayleigh impact location model, the surface of the virtual terrain is more complex at the 10$^{-8}$ risk level and smoother at the 10$^{-7}$ risk level.

\subsection{Quantitative Comparisons}


In addition to the visual comparisons, we quantitatively compare between virtual risk terrains under different settings. Among the three risk levels we explored in the last subsection, 10$^{-8}$/flight hour or below is a reference requirement level for policy makers regarding UAS operations in urban environments. Therefore, in this subsection we focus on the 10$^{-8}$ virtual risk terrain. To quantitatively evaluate the magnitude of the virtual risk terrain, we use the minimum clearance altitude (or height) of the virtual terrain. Under the assumption of continuous and uniform EoV distribution on the ground, the minimum clearance altitude is a prominent feature of the virtual risk terrain.

\begin{figure}[h!]
	\centering
        \includegraphics[width=0.425\textwidth]{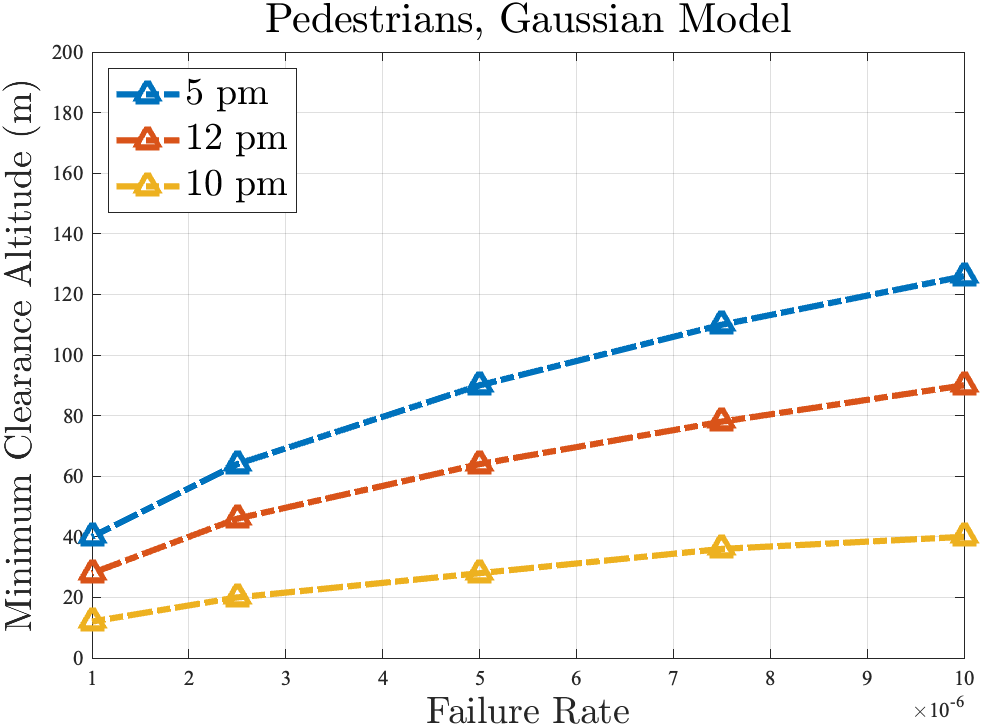}
        \hspace{1cm}
        \includegraphics[width=0.425\textwidth]{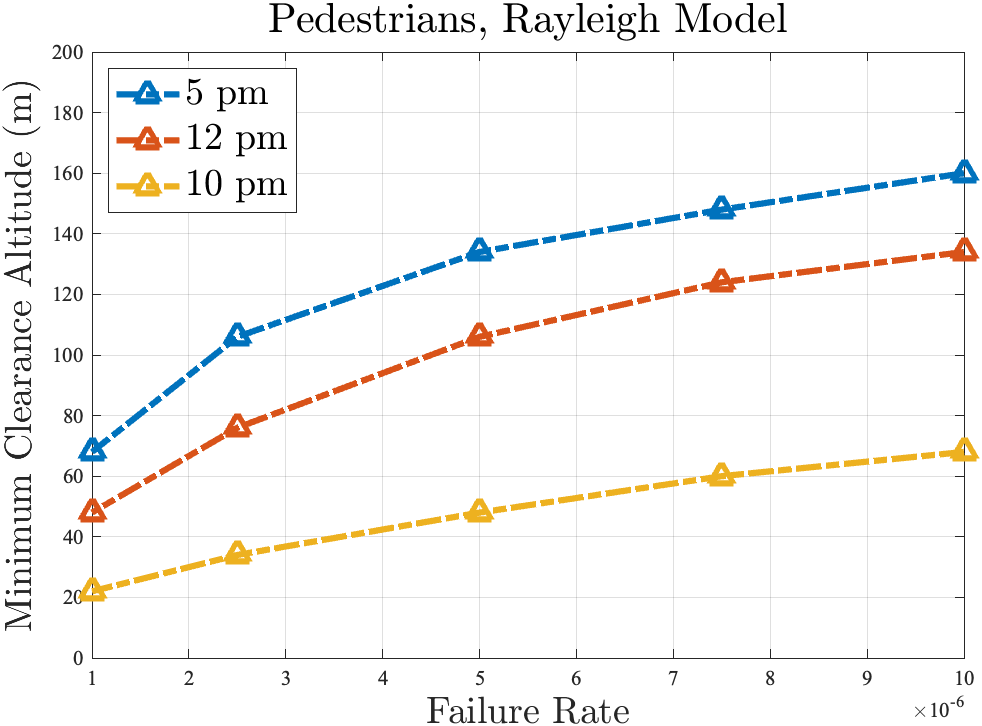}
	\caption{The risk level 10$^{-8}$ minimum clearance altitudes for pedestrians, under Gaussian (left) and Rayleigh (right) impact location models.}
	\label{fig:metrics1}
\end{figure}

\begin{figure}[h!]
	\centering
        \includegraphics[width=0.425\textwidth]{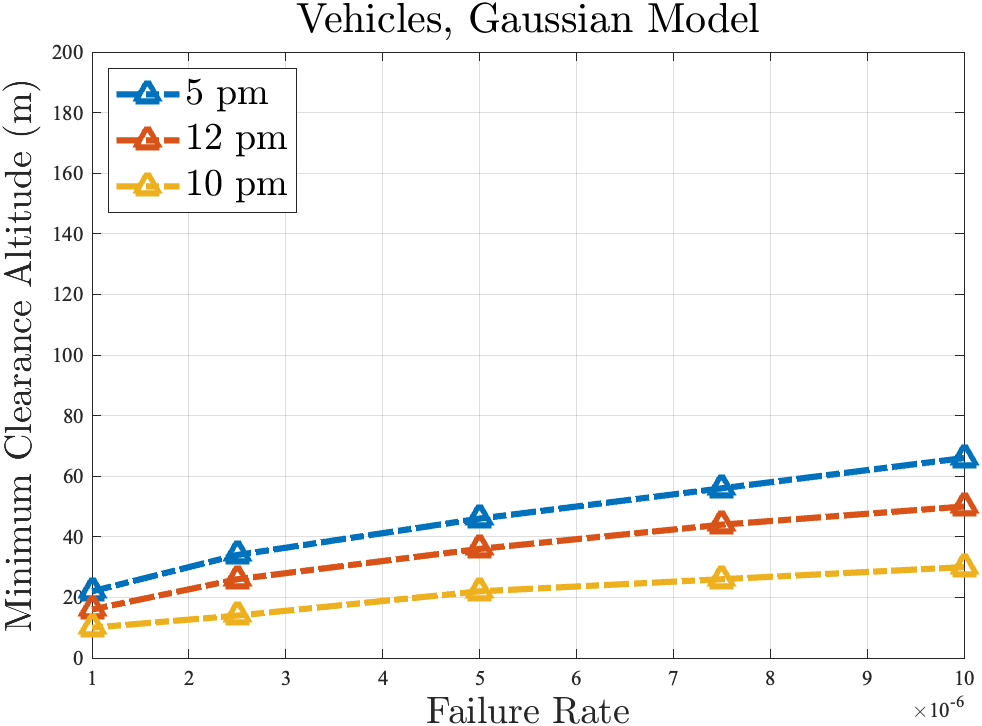}
        \hspace{1cm}
        \includegraphics[width=0.425\textwidth]{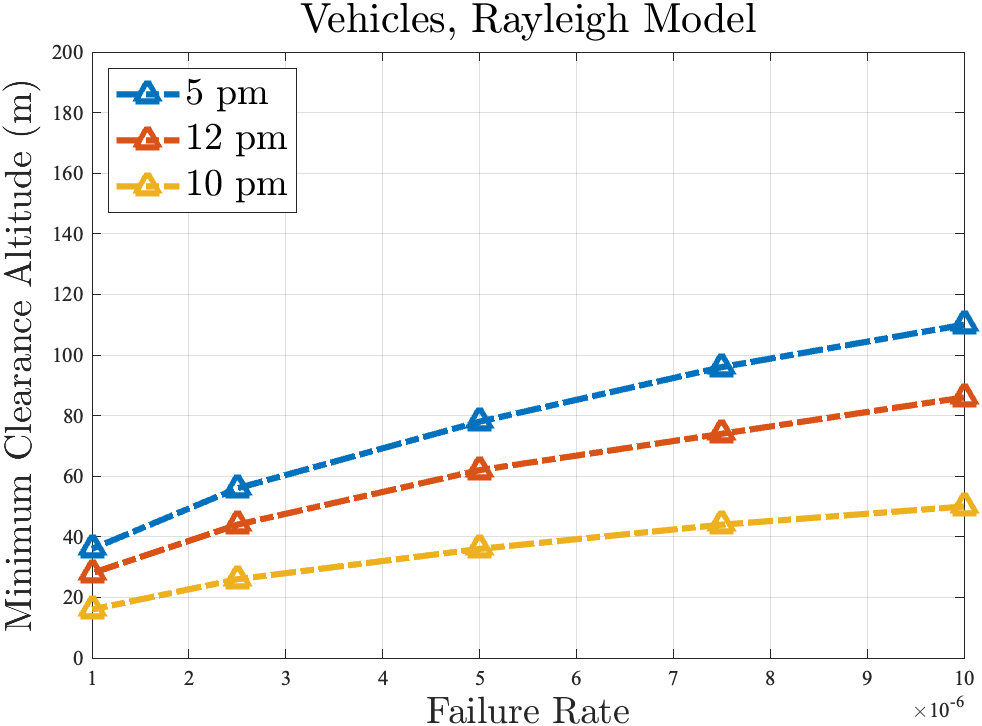}
	\caption{The risk level 10$^{-8}$ minimum clearance altitudes for vehicles, under Gaussian (left) and Rayleigh (right) impact location models.}
	\label{fig:metrics2}
\end{figure}

Figure~\ref{fig:metrics1} shows the patterns of 10$^{-8}$ risk level minimum clearance altitude for pedestrians. The left plot of Figure~\ref{fig:metrics1} shows results under the Gaussian impact location model. In the `worst case' scenario, 10$^{-5}$/flight hour failure rate (right end of the x-axis) and evening rush hour (blue curve), the UAS must fly at around 125 meters above the ground, exceeding the altitude limitation in the current policy. This minimum clearance altitude can be relaxed to below 100 meters if the failure rate is improved to $5 \cdot 10^{-6}$/flight hour or operation takes place at the less congested 12 pm. If the UAS operation were to be allowed at around 40 meters above the ground, that would require one of the following conditions: (1) the failure rate is 10$^{-6}$/flight hour, (2) the time is 10 pm, or (3) the failure rate is $2 \cdot 10^{-6}$/flight hour AND the time is 12 pm. The right plot of Figure~\ref{fig:metrics1} shows results under the Rayleigh impact location model. Like observed in the visual comparisons, the general trends between the two groups of results are similar. Virtual risk terrain under the Rayleigh impact location model has higher minimum clearance altitude, where the difference is between 10 and 40 meters, at every setting. Figure~\ref{fig:metrics2} shows the patterns of 10$^{-8}$ risk level minimum clearance altitude for vehicles. One can observe that the minimum clearance altitudes above vehicles are generally lower than those of pedestrians. For example, under the Gaussian impact location model, the `worst case' minimum clearance altitude for vehicles is one half of that for pedestrians (60 m vs. 120 m). Therefore, in many virtual risk terrain results, such as expressed in the middle plot of Figure~\ref{fig:terrains1}, the terrain is higher above the pedestrian sidewalks and lower above the motor vehicle lanes. These results indicate that, under those conditions, UAS can fly at lower altitudes above motor vehicle lanes and should avoid flying above pedestrians at the same altitudes. Overall, Figures~\ref{fig:metrics1} and \ref{fig:metrics2} provide a (rough) reference for UAS operations planning in a typical urban terrain.

\subsection{The Fusion of Virtual Terrains for Public Acceptance}

In a series of projects to promote the community integration of AAM, a previous work \citep{gao2023noise} has developed a similar concept, the virtual acoustic terrain, for noise-aware flight trajectory planning. The motivation behind the use of virtual acoustic terrain is that the traditional trajectory optimization paradigm requires repetitive computations of a flight operation's noise footprints in complex urban environments and is computationally expensive. By applying acoustic ray tracing and the principle of reciprocity in a complex urban environment, \citep{gao2023noise} convert different noise constraints in the city into 3D exclusion zones which AAM operations should avoid to maintain limited noise impact. The virtual acoustic terrain therefore enables a more efficient non-repetitive noise-aware trajectory optimization paradigm. In this subsection, we present an example which combines the two virtual terrains of different types in the same urban model.

\begin{figure}[h!]
	\centering
        \includegraphics[width=0.325\textwidth]{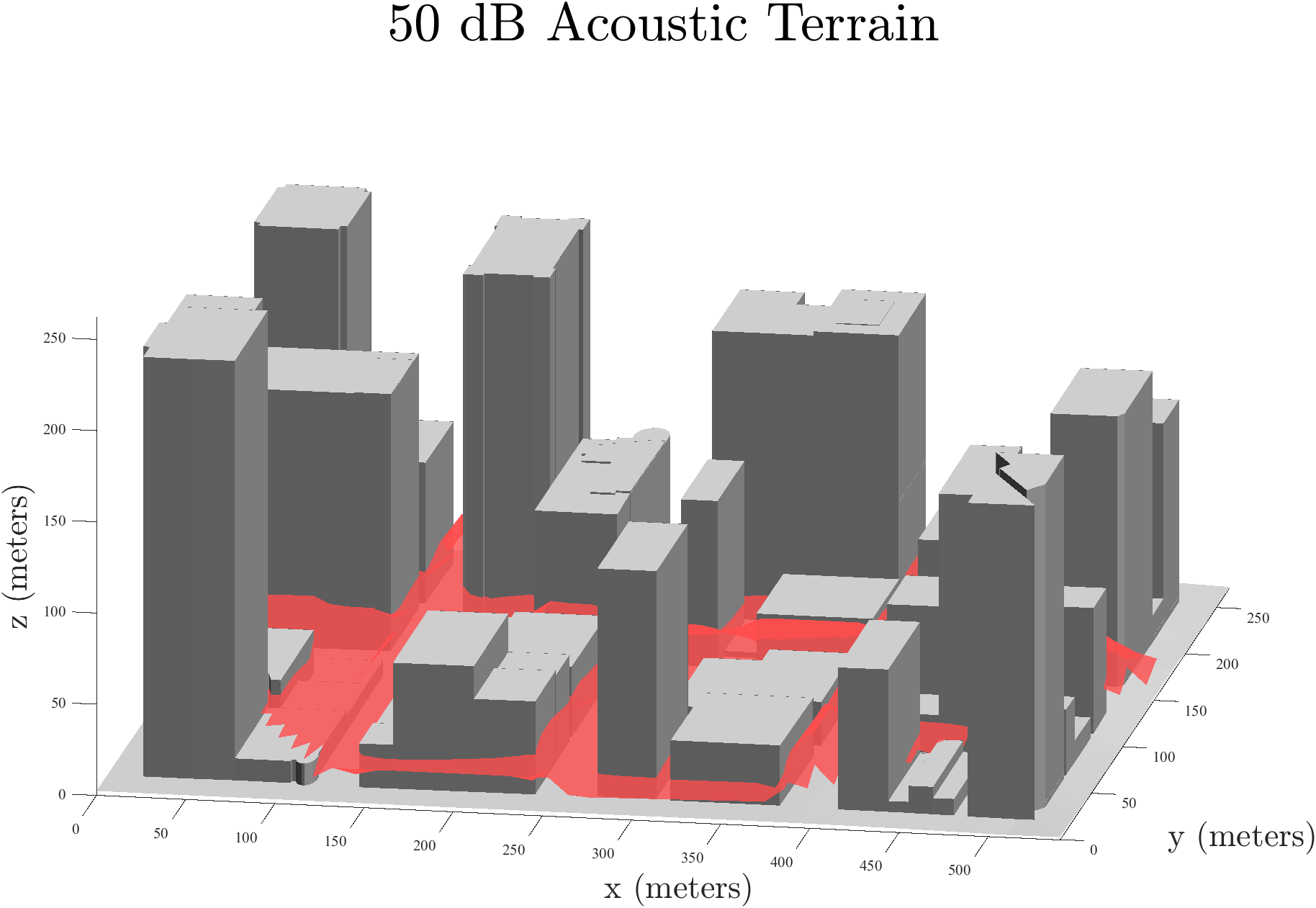}
        \includegraphics[width=0.325\textwidth]{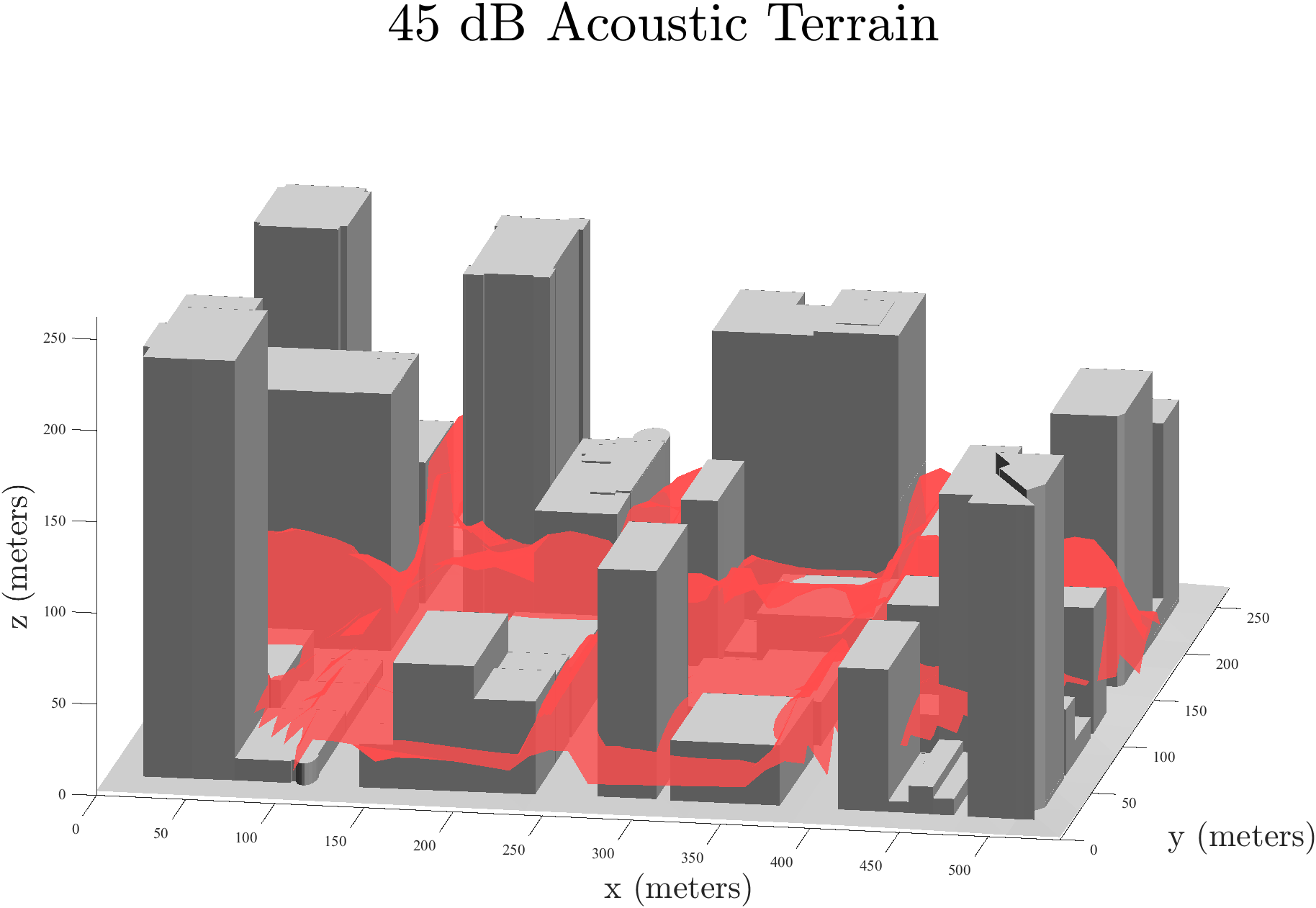}
        \includegraphics[width=0.325\textwidth]{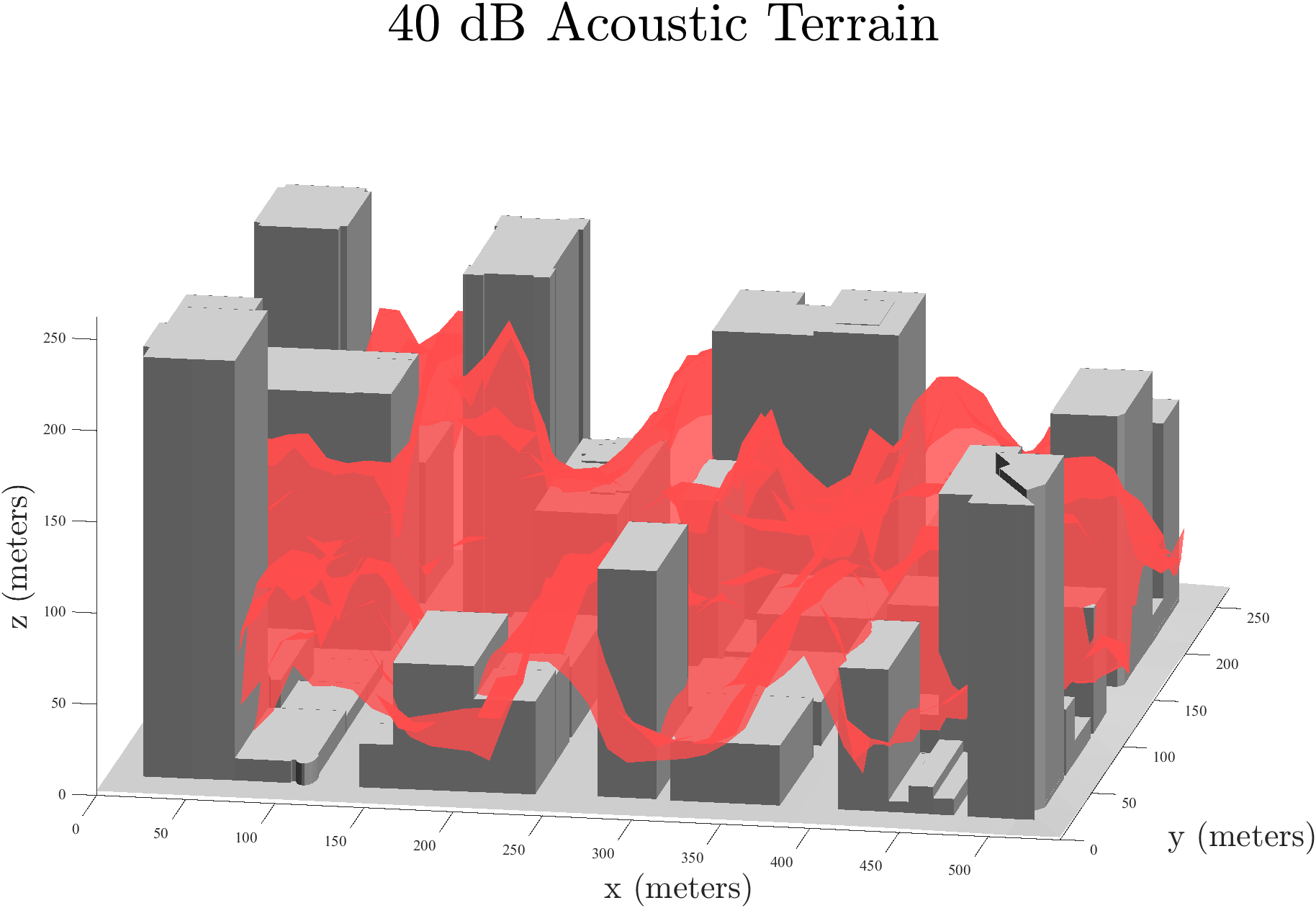}
	\caption{The virtual acoustic terrains with different noise constraint levels: 50 dB (left), 45 dB (middle), and 40 dB (right)}
	\label{fig:Aterrains}
\end{figure}

Figure~\ref{fig:Aterrains} displays a set of virtual acoustic terrains with three noise constraint levels -- 50 dB, 45 dB, and 40 dB. The UAV is assumed to be a monopole sound source with omnidirectional directivity, whose power and frequency are 85 dB and 1,000 Hz, respectively. The acoustic modeling is run with standard day weather and absorption coefficient of 10\% for buildings and the ground. Lke this work, the objective is to limit the UAS operations' noise impact to people on the ground. On the selection of noise constraint levels, 70 dB is often considered noisy in an outdoor urban environment, while 40 dB is considered very quiet. The interpretation of the virtual acoustic terrain is the same: to maintain below a certain noise level on the ground, the UAV must fly above the corresponding virtual acoustic terrain.

\begin{figure}[h!]
	\centering
        \includegraphics[width=0.325\textwidth]{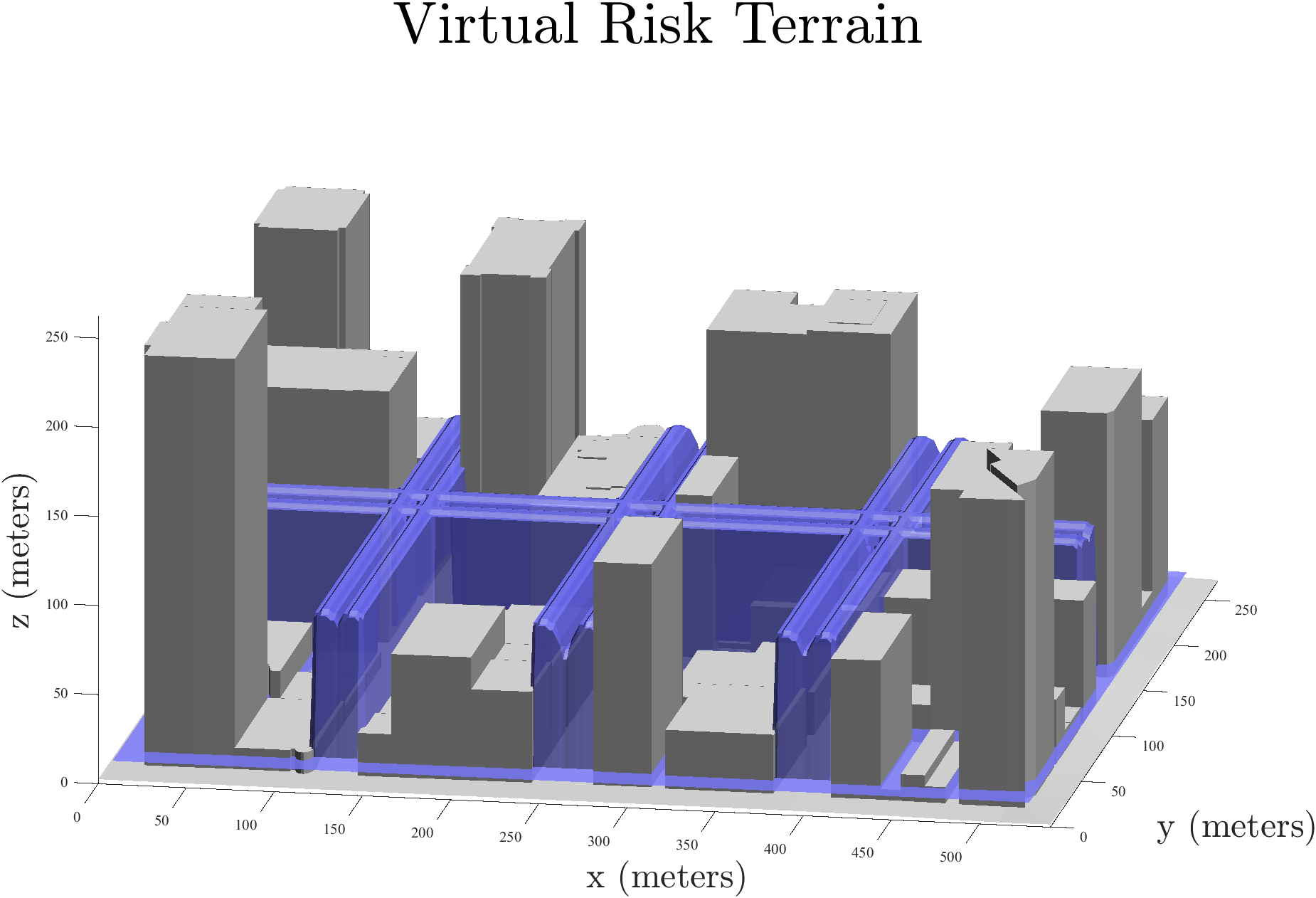}
        \includegraphics[width=0.325\textwidth]{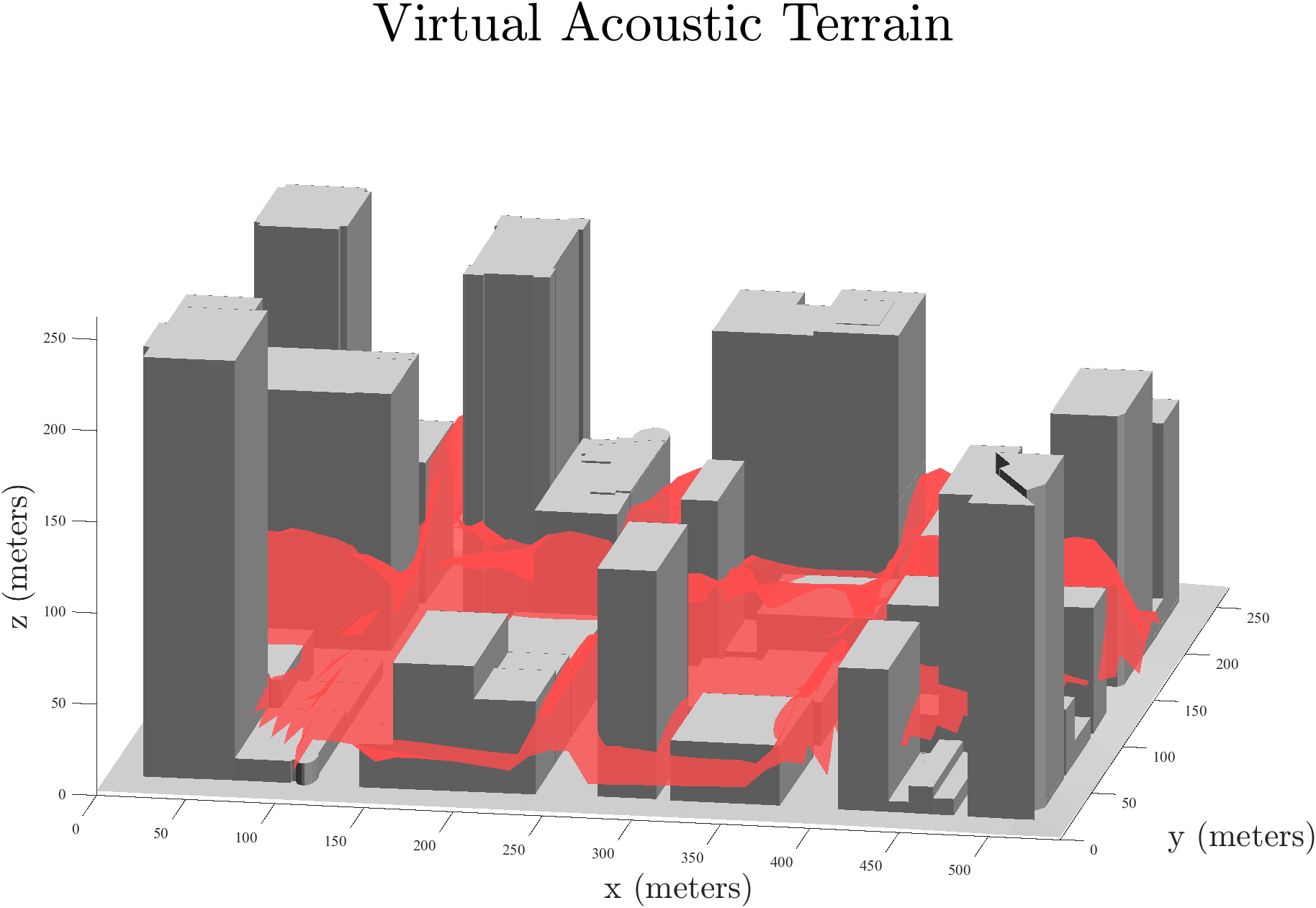}
        \includegraphics[width=0.325\textwidth]{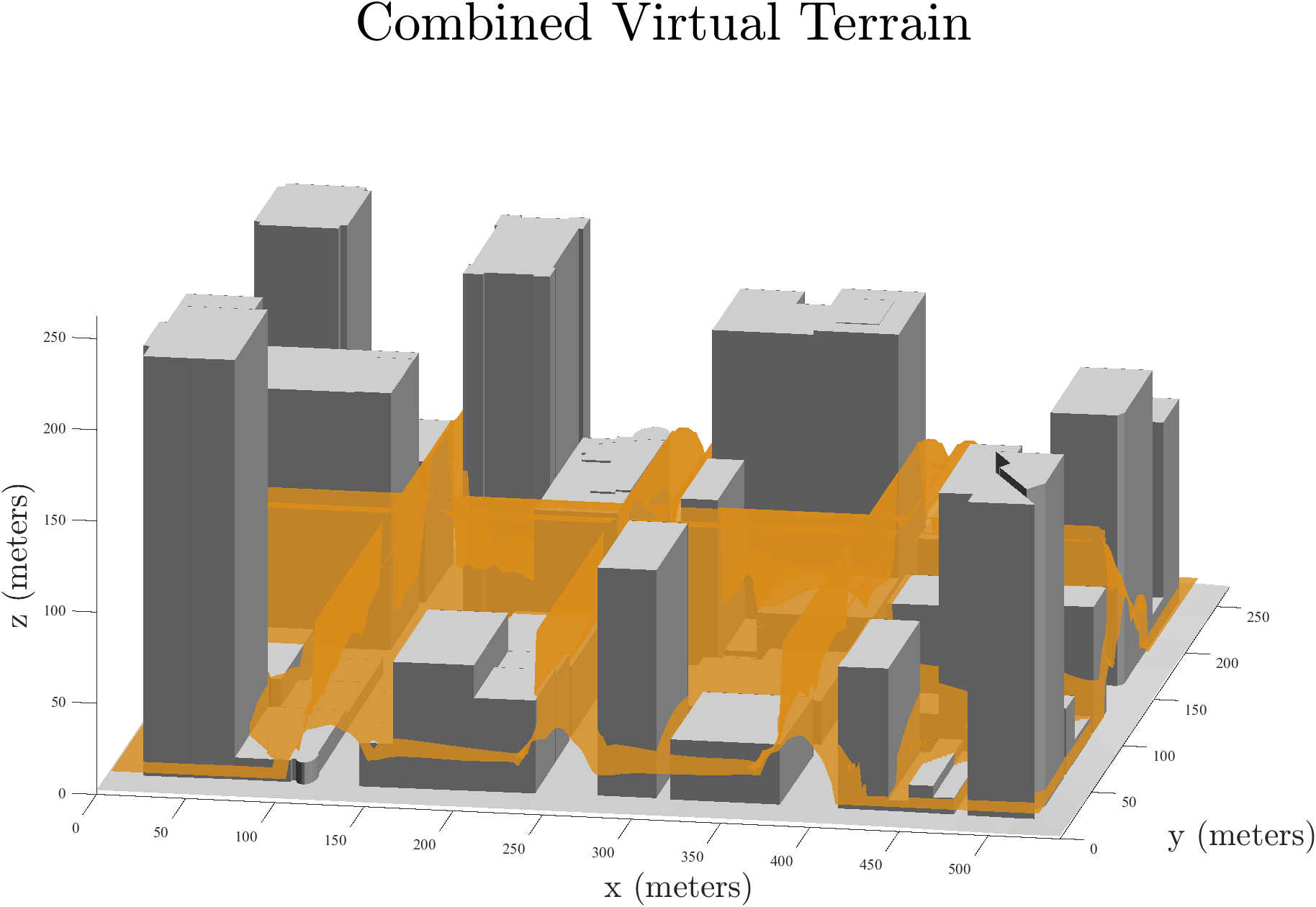}
	\caption{The combination of virtual risk terrain (left) and virtual acoustic terrain (middle) into the combined virtual terrain (right).}
	\label{fig:Cterrains}
\end{figure}

\begin{figure}[h!]
	\centering
        \includegraphics[width=0.825\textwidth]{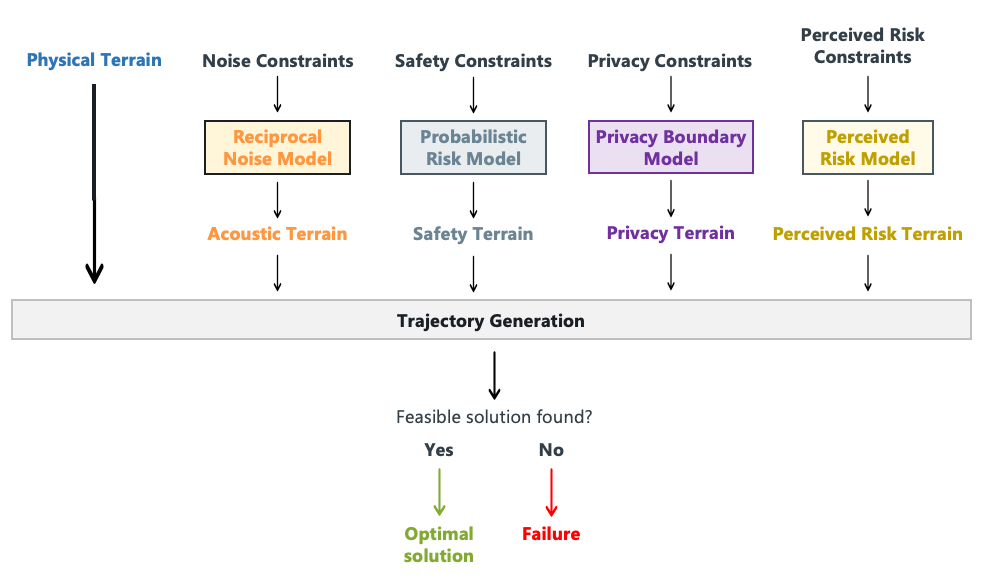}
	\caption{Efficient AAM trajectory optimization enabled by virtual societal impact terrains}
	\label{fig:TrajectoryPlanning}
\end{figure}

In the last numerical example provided in Figure~\ref{fig:Cterrains}, we demonstrate the integration of two types of virtual terrains. The left plot of Figure~\ref{fig:Cterrains} is a representative 3D virtual risk terrain selected from the middle plot of Figure~\ref{fig:terrains2}; the middle plot of Figure~\ref{fig:Cterrains} is a representative virtual acoustic terrain selected from the middle plot of Figure~\ref{fig:Aterrains}. Their combined virtual terrain is shown in the right plot of Figure~\ref{fig:Cterrains}. This combined 3D virtual terrain is the union of both ``no-fly'' zones. When planning UAS operations in this selected area, flight trajectories that avoid this combined virtual terrain can limit the operation's impacts on both ground risk and ground noise. In fact, the virtual societal impact terrains can enable efficient community-aware UAS trajectory planning in an urban environment in a broader sense. In an optimization paradigm depicted in Figure~\ref{fig:TrajectoryPlanning}, societal constraints from four different aspects -- noise, safety, privacy, and perceived risk, can all be converted into 3D virtual terrains. The combination of all virtual terrains with the physical urban terrain defines an overall acceptable fly zone for UAS operations and enables an efficient non-repetitive 3D trajectory optimization process. This will considerably impact and facilitate the community integration of UAS and AAM in urban environments.

\section{Remarks}\label{sec:remarks}

\subsection{Limitations}


The main contributions of this work are the novel concept of 3D virtual risk terrain and the integrated modeling approach outlined in Figure~\ref{fig:Approach}. The framework is a coherent confluence of multiple sub-models for modeling the risks of UAS. For each sub-model, we have adopted either the latest research outcomes in the literature or widely embraced standard practices. Each sub-model belongs to a specialized research field that is progressing rapidly. In this regard, every sub-model of the framework must be continually reinforced to upgrade the overall modeling to incorporate and benefit from future advancements in each field. We expect that the modeling fidelity can greatly benefit from continued research efforts in recovery model, impact location model, and exposure model. Accurate modeling of UAS recovery and impact location remain challenging due to a lack of precise data and a number of uncertain factors; the exposure model can be further refined by using the increasingly available high-resolution mobility data.

In addition, there are some limitations in the present case study. While the settings and parameters were selected to be as representative as possible of a package delivery scene in a complex urban environment, LOC is the `worst case' UAV failure type among the possible scenarios. Therefore, the resulting 3D virtual terrains are also conservative with respect to the type of failure. The choice on LOC is currently constrained by the limited knowledge and models of a UAV's behavior under more complex failure scenarios. On the other hand, the case study has so far only considered ground risks. When operating UAS in a complex urban environment, one extra type of TPR to consider is the surface of the buildings. Although buildings are such a strong sheltering factor that people within the buildings are normally safe from small UAS operations, possible property damage on the buildings can be taken into account. The secondary effects of a UAV collision with the surface of a building can pose additional ground risks.

\subsection{Future Work}


As the very first effort to build 3D virtual risk terrains for UAS operations and urban airspace management, this work will open up many future research avenues for further extensions. Here we briefly mention three opportunities that will enhance the operations planning and community integration of AAM. The first essential avenue is to extend the same concept and apply it to eVTOL aircraft, a key player in on-demand UAM. Because of the differences between UAV, eVTOL aircraft, and their operations, the virtual risk terrain for eVTOL aircraft requires new sub-models in almost every aspect. The ultimate objective is to plan eVTOL aircraft trajectories considering TPR and contingency management (e.g., emergency landing). The second opportunity is to generate the virtual risk terrain on a larger scale. In the case study we generated virtual risk terrains for a portion of the Chicago downtown area. Later on, this virtual terrain will be generated to cover the entire city. This requires the integration of the proposed framework with a GIS capability and can eventually make real-time risk terrain like the traffic map on Google Maps. Third, some ongoing efforts are applying the idea of virtual terrain to other forms of societal impacts/concerns, which include privacy and perceived risk. A fusion of various virtual terrains will provide comprehensive insights into the community integration of AAM and useful references for regulatory policies.

\section{Conclusions}\label{sec:conclusions}

In this paper we introduced virtual risk terrain, a novel concept for UAS operations planning and urban airspace management with risk considerations. By converting public risk constraints in an urban environment into 3D `no-fly' zones, the virtual risk terrain enables efficient UAS trajectory planning and can provide clear guidance to safety regulations for UAS operations in complex urban environments. The computational framework is a conditional probability approach which integrates six sub-models for UAS safety risk and a 3D urban model. We conducted a case study on the Chicago downtown area and generated 3D virtual terrains for the ground risk of multi-rotor UAV cargo delivery operations. We showed how the characteristics of the 3D virtual risk terrain could change under different safety risk levels, UAV reliability levels, impact location models, and times of the day. We also summarized and compared the more general minimum clearance distances/altitudes from EoVs in those scenarios. At the end of the case study, we demonstrated how the virtual risk terrain can also be developed for other societal impact constraints. The amalgamation of all virtual societal impact terrains will advance the operations planning and societal integration of UAS and AAM. We look forward to continuously upgrading this framework with new research outcomes in each sub-model and conducting studies for larger urban models. 

\section*{Acknowledgements}

This work was sponsored by the National Aeronautics and Space Administration (NASA) University Leadership Initiative (ULI) program under project ``Autonomous Aerial Cargo Operations at Scale'', via grant number 80NSSC21M071 to the University of Texas at Austin. The authors are grateful to NASA project technical monitors and project partners for their support. Any opinions, findings, conclusions, or recommendations expressed in this material are those of the authors and do not necessarily reflect the views of the project sponsor. The authors would also like to thank Dr. Mirmojtaba Gharibi for helpful discussions that contributed to aspects of this work.

\newpage

\bibliographystyle{model5-names}\biboptions{authoryear}
\bibliography{main.bib}

\appendix

\end{document}